\documentclass[cmp]{svjour}
\usepackage{amsfonts,amssymb,stmaryrd,limit}
\begin{document}
\title{Limits and Degenerations\\ of Unitary Conformal Field Theories} 
\author{Daniel Roggenkamp \and Katrin Wendland 
}
\titlerunning{Limits and Degenerations\\ of Unitary Conformal Field Theories}
\author{
Daniel Roggenkamp\inst{1,}\inst{2} \and
Katrin Wendland\inst{3}
}
\institute{Physikalisches Institut,
Universit\"at Bonn, Nu\ss allee 12, D-53115 Bonn, Germany.\\
\email{roggenka@th.physik.uni-bonn.de}
\and
Department of 
Mathematics,  King's College London, Strand, London WC2R 2LS, United Kingdom.
\and
Mathematics Institute, University of Warwick,
Coventry CV4-7AL, United Kingdom.\\
\email{wendland@maths.warwick.ac.uk}}
\date{}
\authorrunning{Daniel Roggenkamp and Katrin Wendland}
%
%
%
%
\maketitle
\begin{abstract}
In the present paper,
degeneration phenomena in conformal field theories are studied. For this
purpose, a notion of convergent sequences of CFTs is introduced.
Properties of the resulting limit structure are used to associate geometric
degenerations to degenerating  sequences of CFTs, which, 
as familiar from large volume limits of non-linear sigma models, can be
regarded as commutative degenerations
of the corresponding ``quantum geometries''.

As an application,
the large level limit of the A-series of unitary Virasoro minimal models 
is investigated in detail. In particular, its geometric interpretation 
is determined.
\end{abstract}
\section*{Introduction}
Limits and degenerations of conformal field theories (CFTs)
have occurred in various ways in the
context of compactifications of moduli spaces of  CFTs, in particular
in connection with string theory. For example,  zero curvature or large 
volume limits of CFTs  that
correspond to sigma models are known to give boundary points of the 
respective moduli spaces \cite{agm94a,mo93b}. These limits
provide the connection between string theory  and  classical geometry 
which for instance is  used in the study of D-branes.
In the Strominger/Yau/Zaslow mirror construction \cite{vawi95,syz96,gr99}, 
boundary points play a 
prominent role. 
In fact, 
Kontsevich and Soibelman have proposed a  mirror construction 
on the basis of the Strominger/Yau/Zaslow conjecture
which relies on the 
structure of the boundary of certain CFT moduli spaces \cite{koso00}.

All the examples mentioned above feature interesting degeneration phenomena. 
Namely, subspaces of the  
Hilbert space which are confined to be finite dimensional for a 
well-defined CFT achieve infinite dimensions in the limit. 
In fact, such degenerations 
are expected if the limit is formulated in terms of 
non-linear sigma models, where
at large volume,
the algebra of low energy observables  is expected to yield 
a non-commutative deformation of an algebra 
$\AAA^\infty$ of  functions 
on the target space. The algebra of observables whose energy converges 
to zero then  reduces to
$\AAA^\infty$ at infinite volume. 
An entire non-commutative geometry can be
extracted from the underlying CFT, which approaches the target
space geometry in the limit \cite{frga93}. 
By construction, this formulation should encode
geometry in terms of Connes' spectral triples \cite{co85,co90,co96}. 

By the above, degeneration phenomena are crucial in order to single
out an algebra which encodes geometry in CFTs. 
An intrinsic understanding of limiting processes in CFT 
language is therefore desirable. This will also be necessary
in order to take advantage of the geometric tools mentioned before,
away from those limits. Vice versa, a good understanding of such 
limiting processes in CFTs could allow to take advantage of the 
rich CFT structure in  geometry.

The main aim of the present work is to establish an intrinsic notion
of such limiting processes in pure CFT language and to apply it to 
some interesting examples. 
To this end, we give a definition of convergence for sequences of CFTs,
such that the corresponding limit has the following structure:
There is a limiting pre-Hilbert space $\HHH^\infty$
which carries the action of
a Virasoro algebra, and similar to ordinary CFTs
to each state in $\HHH^\infty$ we assign a tower of modes. 
Under an additional condition the limit even has the structure of
a full CFT on the sphere. This is the case in all known examples, and
in particular, our notion of limiting processes is compatible with
deformation theory of CFTs.

If the limit of a converging sequence of CFTs has the structure of a 
CFT on the sphere, but is not a full CFT, then this is due to a 
degeneration as mentioned above. In particular, the degeneration of
the vacuum sector can be used to read off a geometry
from such a degenerate limit. Namely, in our limits the algebra 
of zero modes assigned to those states in $\HHH^\infty$
with vanishing energy is commutative and  can therefore
be interpreted as algebra of smooth functions on some manifold $M$. 
The asymptotic behaviour of the associated energy eigenvalues allows
to read off a degenerating metric on $M$  and an additional smooth function 
corresponding to the dilaton
as well.
Moreover, being a module of this commutative algebra,  $\HHH^\infty$ can be
interpreted as a space of sections of a sheaf over $M$ as is explained
in \cite{koso00}.
 
Simple examples which we can apply our techniques to are the 
torus models, where our limit structure yields geometric degenerations
of the corresponding target space 
tori \`a la Cheeger-Gromov \cite{chgr86,chgr90}. In this case,
$\HHH^\infty$ is the space of sections of a trivial vector bundle over
the respective target space torus. Similar statements are true
for orbifolds of torus models, only that in this case the fiber structure 
of $\HHH^\infty$ over the respective torus orbifold is non-trivial.
Namely, the twisted sectors contribute  sections of skyscraper sheaves 
localized on the orbifold fixed points.

Our favorite example, which in fact was the starting point of our 
investigations,
is the family of unitary Virasoro minimal models.
Some of their structure constants have a very 
regular behaviour under the variation of the level of the individual models.
We use this to show that the A-series of 
unitary Virasoro minimal models constitutes a 
convergent sequence of CFTs. All fields in its limit theory at infinite level
can  be constructed in terms of operators in the 
$\fs\fu(2)_1$ WZW model.
The sequence degenerates, and the limit  has a geometric
interpretation in the above sense on the interval $[0,\pi]$ equipped
with the (dilaton-corrected) metric $g(x)=\inv[4]{\pi^2}\sin^4\!x$ (in fact,
the $x$-dependent contribution is entirely due to the dilaton). 
This also allows us to read off the geometry of D-branes in these models.
Though this means that the vacuum sector of our limit is well understood, it 
remains an interesting open problem to investigate the full fusion rules
in detail, in particular whether an appropriate limiting 
S-matrix can be found.

A different limit for the A-series of 
unitary Virasoro minimal models at infinite level was proposed in 
\cite{grw01,ruwa01,ruwa02}. It is described by a 
well-defined non-rational CFT of central charge one,
which bears some resemblance
to Liouville theory. In particular, its spectrum is continuous, but 
degenerations do not occur. Our techniques can also be used to describe 
this latter limit. The relation between the two different limit structures 
is best 
compared to the case of  a free boson, 
compactified on a circle of large radius, where apart from the degenerate 
limit described above
one can also obtain the decompactified free boson.
While the limit investigated in this article  
has the advantage that it leads to a consistent geometric
interpretation, the one which corresponds to the decompactified free boson
gives a new well-defined non-rational CFT.

This work is organized as follows: 
In Sect.\ \ref{gcft} we explain how 
non-commutative geometries can be extracted from CFTs, 
after giving a brief overview of some of the basic concepts. 
Sect.\ \ref{lim} contains our definitions of sequences, convergence, 
and limits, and is the technical heart of this paper. Moreover, the
geometric interpretations of degenerate limits are discussed.
Sect.\ \ref{exes} is devoted to the study of torus models and 
orbifolds thereof, where we exemplify our techniques. 
In Sect.\ \ref{kinfinity} we present our results on the A-series
of unitary Virasoro minimal models. We end with a discussion in 
Sect.\ \ref{disc}. Several appendices contain background material and
lengthy calculations.
\subsubsection*{Acknowledgments}
It is a pleasure to thank Gavin Brown, Jarah Evslin, 
Jos\'e Figueroa-O'Farrill, 
Matthias Gaberdiel, Maxim Kontsevich,
Werner Nahm, Andreas Recknagel, Mi\-cha\-el R\"osgen,
Volker Schomerus, G\'erard Watts and the referee for helpful comments or
discussions. 
We also wish to thank the ``abdus salam international center for theoretical
physics'' for hospitality, since part of this work was performed there.

D.\ R.\  was supported by DFG Schwerpunktprogramm 1096 and 
by the Marie Curie Training Site 
``Strings, Branes and Boundary Conformal Field Theory'' at King's 
College London, under EU grant HPMT-CT-2001-00296. 
K.\ W.\ was partly supported under U.S. DOE grant DE-FG05-85ER40219, TASK A,
at the University of North Carolina at Chapel Hill.
\section{From geometry to conformal field theory,
and back to geometry}\label{gcft}
String theory establishes a natural map which associates CFTs to certain,
sometimes degenerate geometries. Conversely, one can associate a 
\textsc{geometric interpretation} to certain CFTs, and the latter construction
is made precise by using Connes' definition of spectral triples and
non-commutative geometry.

In Sect.\ \ref{sptr}
we very briefly remind the reader of \textsc{spectral triples},
explaining how they encode geometric data. Somewhat relaxing the conditions
on spectral triples we define \textsc{spectral pre-triples} which will be 
used in  Sect.\ \ref{cfttripel}. There,
we recall the basic structure of CFTs and show how to
extract spectral pre-triples from them. If the spectral pre-triple defines
a spectral triple, then this will generate a non-commutative
geometry from a given CFT. In Sect.\ \ref{comgeom} we explain how
in favorable cases we can generate commutative geometries from CFTs. 
In the context of string theory, this prescription gives back the original
geometric data of the compactification space.

Much of this  Sect.\ \ref{gcft} consists of a summary of known results
\cite{co90,frga93,co96,re99,koso00},
but it also serves to introduce our notations.
\subsection{From Riemannian geometry to spectral triples}\label{sptr}
For a compact 
Riemannian manifold $(M,g)$, which for simplicity we 
assume to be smooth and connected, the spectrum of the associated 
Laplace-Beltrami operator 
$\Delta_g\colon C^\infty(M)\longrightarrow C^\infty(M)$ encodes 
certain geometric
data of $(M,g)$. However, in general one cannot hear the shape of a drum,
and more information than the set of eigenvalues of $\Delta_g$ is needed
in order to recover $(M,g)$. 

By the Gel'fand-Naimark theorem, the point set topology of $M$ is completely
encoded in $C^0(M)=\ol{C^\infty(M)}$: 
We can recover each point $p\in M$ from the ideal of functions which
vanish at $p$. In other words, given the structure of $C^\infty(M)$ as a
$\CC^\ast$-algebra
and its completion $C^0(M)$, 
$M$ is homeomorphic to the set of closed points of 
$\spec(\OOO_M)$, where $\OOO_M$ is the sheaf of regular functions on $M$.
Connes' dual prescription 
uses $\CC^*$-algebra homomorphisms $\chi\colon C^\infty(M)\longrightarrow \CC$,
instead, such that $p\in M$  corresponds to 
$\chi_p\colon C^\infty(M)\longrightarrow \CC$ with $\chi_p(f):=f(p)$;
the Gel'fand-Naimark theorem ensures that for every commutative $\CC^*$-algebra
$\ol\AAA$ there exists a Hausdorff space $M$ with $\ol\AAA=C^0(M)$.
$M$ is compact if $\ol\AAA$ is unital.
\bex{circ}
Let $R\in\RR^+$, then
$M=\sone^1_R=\RR^1/\sim$ with coordinate $x\sim x+2\pi R$
has the Laplacian $\Delta=-{d^2\over dx^2}$. Its eigenfunctions
$|m\ket_R, m\in\Z$, obey
\beq\el{circleeigen}
\begin{array}{lrcl}
\fa m\in\Z\colon&
|m\ket_R\colon\; x\mapsto e^{i m x/R}\; ;\quad \quad \quad 
\inv{2}\Delta|m\ket_R&=&\inv[m^2]{2R^2}|m\ket_R;\\[4pt]
\fa m,m^\prime\in\Z\colon&
|m\ket_R\cdot|m^\prime\ket_R&=&|m+m^\prime\ket_R,
\end{array}
\eeq
and they form a basis of $C^0(M)$ and $C^\infty(M)$ with respect to the
appropriate norms. Any smooth
manifold  is homeomorphic to $\sone^1_R$, equipped with
the Zariski topology, if its algebra of continuous functions has a basis 
$f^m,\; m\in\Z$,
which obeys the multiplication law $f^m\cdot f^{m^\prime}=f^{m+m^\prime}$. 
\eex
To recover the Riemannian metric $g$ on $M$ as well, we consider the
\textsc{spectral triple} $\left( \HH:=L^2(M,\dvol_g), H:={1\over2}\Delta_g,
\AAA:=C^\infty(M)\right)$, where $H$ is viewed as self-adjoint operator
which is densely defined 
on the Hilbert space $\HH$, and $\AAA$ is interpreted as algebra of 
bounded operators which act on elements of
$\HH$ by pointwise multiplication. Following
\cite{co90,frga93,co96}, we can define a distance functional 
$d_g$ on the topological
space $M$ by considering
\beqns
\FFF:=\bigl\{ f\in\AAA \bigr.
&\bigm|& 
G_f:= [f,[H,f]] =
- \left( f^2\circ H + H\circ f^2\right) + 2f\circ H\circ f\\
&&\bigl.
\quad\quad\;\;
\mbox{ obeys }\;\fa h\in C^\infty(M)\colon\;\; |G_f h|\leq |h| \bigr\}.
\eeqns
One now sets 
\beq\el{metric}
\fa x,y\in M\colon\quad
d_g(x,y):=\sup_{f\in\FFF} \left| f(x)-f(y) \right|.
\eeq
In Ex.\ \ref{circ} with $M=\sone^1_R$  one checks that for all 
$f,h\in C^\infty(M)\colon G_f h=(f^\prime)^2h$, and in general
$G_f h=g(\nabla f,\nabla f)h$. In fact, by definition \cite[Prop. 2.3]{bgv92},
any second-order differential operator $O$ satisfying 
$[f,[{1\over2}O,f]]=g(\nabla f,\nabla f)$ is a \textsc{generalized 
Laplacian}.
Using the time coordinate of a geodesic from
$x$ to $y$ and truncating and smoothing it appropriately one checks that
\eq{metric} indeed gives back the geodesic distance between $x$ and $y$
which corresponds to the metric $g$. In other words, $(M,g)$ can be recovered
from the spectral triple $(\HH,H,\AAA)$. 

More generally, consider a spectral triple $(\HH,H,\AAA)$ with 
$\HH$ a Hilbert space, $H$ a self-adjoint positive
semi-definite operator, which on $\HH$ is densely defined
with $\HHH_{0,0}:=\ker(H)\cong\CC$, and
$\AAA$ a $\CC^\ast$-algebra of bounded operators acting on $\HH$. In fact, 
in the above let us assume that $M$ is spin and
replace $H={1\over2}\Delta_g$ by the 
corresponding Dirac operator $\DDD$
and $\HH=L^2(M,\dvol_g)$ by the Hilbert space $\HH^\prime$ of 
square-integrable sections of the spinor bundle on $M$. 
Note that $H$  can be calculated from $\DDD$, see \eq{nabla} and \eq{lapl}.
Moreover, we assume that $(\HH^\prime,\DDD,\AAA)$ obeys the 
\textsc{seven axioms of non-commutative geometry} 
\cite[p.159]{co96}. Roughly speaking, these axioms ensure that 
the eigenvalues of $H$ have the correct growth behaviour \eq{growth}, 
that $\DDD$ defines a map $\nabla$ on $\AAA$ with
\beq\el{nabla}
\fa f\in\AAA:\quad
\nabla f:=[\DDD,f]\colon\,\HH^\prime\rightarrow\HH^\prime; \quad\quad
\fa h\in\AAA:\quad
[\nabla f,h]=0,
\eeq
where in the above examples
$\nabla f$ acts on $\HH^\prime$ by Clifford multiplication,
and that $\AAA$ gives smooth coordinates on an ``orientable geometry''; 
furthermore,
there are finiteness and reality conditions as well as a type of Poincar\'e  
duality
on the K-groups of $\AAA$. If all these assumptions hold, then 
by \eq{metric} the triple
$(\HH^\prime,\DDD,\AAA)$ defines a non-commutative geometry \`a la Connes
\cite{co85,co90,co96}. If the algebra $\AAA$ is commutative, then  
the triple $(\HH^\prime,\DDD,\AAA)$ in fact defines a unique ordinary
Riemannian geometry $(M,g)$ \cite[p.162]{co96}.
The claim that the  differentiable  and the spin structure of
$(M,g)$ can be fully recovered is detailed in\footnote{We thank 
Diarmuid Crowley for bringing this paper to our attention.} \cite{re99}.

Following \cite{frga93}, instead of studying \textsc{spectral triples}
$(\HH^\prime,\DDD,\AAA)$,
we will be less ambitious and mainly focus on  triples
$(\HH,H,\AAA)$, somewhat relaxing the defining conditions:
\bdefi{pretrip}
We call $(\HH,H,\AAA)$ a \textsc{spectral pre-triple} if $\HH$ is 
a pre-Hilbert space over $\CC$, $H$ is a self-adjoint positive semi-definite
operator on $\HH$ with $\HHH_{0,0}:=\ker(H)\cong\CC$, and 
$\AAA$ is an algebra 
of operators acting on $\HH$. Since $\HHH_{0,0}\cong\CC\ni1$, we can view
$\AAA\hookrightarrow\HH$ by $A\mapsto A\cdot1$.

If additionally 
the eigenvalues of $H$ have the appropriate growth behaviour,
i.e.\  for some $\gamma\in\RR$ and $V\in\R$\mb{:}
\beq\el{growth}
N(E):=\dim_\CC\left( \bigoplus_{\lambda\leq E} \left\{ \varphi\in\HH
\bigm| H\varphi=\lambda\varphi \right\} \right), 
\quad
N(E)\stackrel{E\rightarrow\infty}{\sim} V\cdot E^{\gamma/2},
\eeq
then $(\HH,H,\AAA)$ is called a \textsc{spectral pre-triple of dimension 
$\gamma$}.

If there exists an operator $\DDD$ which is densely defined
on a Hilbert space $\HH^\prime$ that carries an action of
$\AAA$ with
\mb{\eq{nabla}} such that 
\beq\el{lapl}
\fa f,h\in\AAA\colon\quad
\langle\nabla f,\nabla h\rangle_{\HH^\prime} = 2\langle f,H h\rangle_\HH
\eeq
and such that $(\HH^\prime, \DDD,\AAA)$ obeys 
the seven axioms of non-commutative
geometry, then we call $(\HH, H,\AAA)$ a \textsc{spectral triple} or a
\textsc{non-commutative geometry
of dimension $\gamma$}.
\edefi
\vspace*{-1.5em}
\brem{dilaton}
Note that our condition \req{lapl} for the operator $H$ does not 
imply $H={1\over2}\Delta_g$ on $L^2(M,\dvol_g)$. In fact, $H$ will 
in general be 
a generalized Laplacian with respect to a metric $\wt g=(\wt g_{i j})$ 
in the conformal class of $g$. More precisely, we 
will have $\dvol_g=e^{2\Phi}\dvol_{\wt g}$ with $\Phi\in C^\infty(M)$, and 
with $\wt g^{-1}=(\wt g^{i j})$,
\beq\el{genlaplace}
2H=-e^{-2\Phi} \sqrt{\det\wt g^{-1}} 
\sum_{i,j}\, \partial_i e^{2\Phi} \sqrt{\det\wt g}\;\wt g^{i j}\, \partial_j
\eeq
with respect to local coordinates, in accord with \req{lapl}. We call
$g$ the \textsc{dilaton corrected metric} with \textsc{dilaton} $\Phi$.
Note that $\wt g$ is easily read off from the symbol of $H$, allowing to 
determine $\Phi$ from $\dvol_g=e^{2\Phi}\dvol_{\wt g}$.
\erem
A generalization of Connes' approach, which is natural from our point of 
view, is given in \cite{lo00}. There, the Dirac operator on the 
spinor bundle of $M$ is replaced by the Dirac type operator $\DDD=d+d^*$
on $\HH^\prime=\Lambda^*(T^*M)$. Since $\Delta_g=\DDD^2$, \eq{nabla}--\eq{lapl}
remain true, but the list of axioms reduces considerably
to the definition of a \textsc{Riemannian non-commutative
geometry} \cite[III.2]{lo00}. However, our main emphasis lies on the 
recovery of the metric structure $(M,g)$ rather than the
differentiable structure.
Similarly, in \cite{koso00} the main emphasis lies on triples
$(M,\RR^+\!g,\varphi)$, where $\varphi\colon\,M\longrightarrow\MMM$ is
a map into an appropriate moduli space $\MMM$ of CFTs.

It will be easy to associate a spectral pre-triple to every CFT. Using
degenerations of CFTs in the spirit of \cite{koso00}, one can often
associate spectral pre-triples of dimension $\gamma=c$ to a CFT with central 
charge $c$. A general theorem, however, which allows to associate 
non-commutative geometries to arbitrary CFTs seems out of reach.
In all cases we are aware of 
where a non-commutative geometry is obtained from CFTs, this is in fact
proven by deforming an
appropriate  commutative geometry. 
In Sect.\ \ref{kinfinity}, we present a non-standard example of this
type which should lead to interesting non-commutative geometries 
by deformation.
\subsection{Spectral triples from CFTs}\label{cfttripel}
We do not attempt to give a complete  definition of CFTs in 
this section; the interested reader may consult, e.g.,
\cite{bpz84,mose88b,gi88b,mose89,dms96,gaga00}. Some further
properties of CFTs that are 
needed in the main text are collected in App.\ \ref{cfts}.

A \textsc{unitary two-dimensional
conformal field theory} (CFT) is specified by the 
following data:
\begin{itemize}
\item
a $\CC$--vector space $\HHH$ of \textsc{states}
with scalar product $\bra\cdot|\cdot\ket$. This
scalar product is positive definite,
since we  restrict our discussion to \textit{unitary} CFTs;
\item
an anti-$\CC$-linear involution $*$ on $\HHH$, often called \textsc{charge 
conjugation};
\item
an action of two commuting copies   
$\vir_c$, $\ol{\vir}_c$ of a Virasoro 
algebra \eq{vir}
with central charge\footnote{As a matter of 
convenience, we always assume left
and right handed central charges to agree.} 
$c\in\RR$ on $\HHH$, with generators  
$L_n,\, \ol L_n,\, n\in\Z$,
which\footnote{The 
indexing of all modes 
below corresponds to energy, not to its negative.}  
commutes with $\ast$. The Virasoro generators
$L_0$ and $\ol{L}_0$ 
are diagonalizable on $\HHH$, such that $\HHH$ 
decomposes into eigenspaces\footnote{In this work, we
restrict our investigations to bosonic CFTs.} 
\beq\el{deco}
\HHH=\bigoplus_{\stackrel{\scriptstyle h,\ol{h}\in\RR,}{\scriptstyle
h-\ol h\in\ZZ}}\HHH_{h,\ol{h}}\,,
\eeq
and we set $\HHH_{h,\ol{h}}:=\{0\}$ if $h-\ol{h}\not\in\Z$.
The decomposition \eq{deco} is orthogonal with respect to
$\bra\cdot|\cdot\ket$;
\item
a \textsc{growth condition} for the eigenvalues $h,\,\ol h$ in \eq{deco}:
For  some $\nu\in\RR^+$ and $V\in\R$:
\beq\el{cftgrowth}
\fa E\in\RR^+\colon\quad
\infty \quad>\quad
\dim \left( \bigoplus_{(h+\ol{h})^\nu\leq E}\HHH_{h,\ol{h}}\right)
\stackrel{E\rightarrow\infty}{\sim}
\exp\left( V \sqrt{E}\right).
\eeq
In particular, for all $h,\,\ol h\in\RR$ we  have 
$\HHH_{h,\ol{h}}^*\cong\HHH_{h,\ol{h}}$, and we define
$$
\check\HHH^*:=\bigoplus_{h,\ol{h}\in\RR}\HHH_{h,\ol{h}}^*\;;
$$
\item
a unique $*$-invariant \textsc{vacuum} $\Omega\in\HHH_{0,0}\cong\CC$,
as well as a dual $\Omega^*\in\check\HHH^*$ 
characterized by \eq{vacuum};
\item
a map $C:\check\HHH^*\otimes\HHH^{\otimes 2}\longrightarrow\CC$ that encodes 
the \textsc{coefficients of the operator product expansion (OPE)},
such that 
\beq\el{pairing}
C(\cdot,\Omega,\cdot):\check\HHH^*\otimes\HHH\longrightarrow\CC,
\quad (\Psi,\chi)\longmapsto \Psi(\chi),
\eeq
i.e.\  the induced map is the canonical pairing.
The OPE-coefficients $C$ obey  \eq{cc} and  \eq{trimetric} and 
can be used to define an isomorphism
\begin{equation}\el{adj}
\begin{array}{rcl}
\HHH&\longrightarrow&\check\HHH^*\\
\psi&\longmapsto&\psi^*,
\end{array}
\;\mbox{ s.\ th.\ }\; \forall\,\chi\in\HHH:\quad
\psi^*(\chi)= C(\psi^*,\Omega,\chi)
=\bra\psi|\chi\ket.
\end{equation}
\end{itemize}
There are many properties of the map $C$, 
like the sewing relations, that have to be 
fulfilled for reasons of consistency, 
and which we will not indulge to list
explicitly. Some properties of CFTs that follow from these consistency
conditions should be kept in mind, however:
\begin{itemize}
\item
$\varphi\in\HHH$ is a lowest weight vector (lwv) with respect to the
action of $\vir_c$, $\ol{\vir}_c$, i.e.\ a \textsc{primary state},
iff for all\footnote{We agree on
$0\in\NN$, as argued in \cite[IV.4.1,R.6.2]{bo70}.}
$n\in\NN-\{0\}\colon\; L_{-n}\varphi=0$, $\ol L_{-n}\varphi=0$.

For any $\ZZ$--graded algebra 
$\LLL=\smash{\bigoplus\limits_{n\in\ZZ}}\LLL_n$ we define
\beqn\el{prime}
\LLL^\pm &:=& \bigoplus_{\pm n>0}\LLL_n, \\
\HHH^\LLL := \quad \ker\LLL^-
&=& \left\{ \varphi\in\HHH \bigm| \fa n\in\NN-\{0\}, \,
\fa w\in\LLL_{-n}\colon\; w\varphi=0\right\}.\nonumber
\eeqn
In other words, setting $\LLL=\vir=\vir_c\oplus\ol\vir_c$
by abuse of notation, $\HHH^{\vir}$ denotes the 
\textsc{subspace of primary states in $\HHH$}.
\item
The OPE, which we encode in the map $C$ as introduced above, allows
to associate to each $\varphi\in\HHH$ a tower $\varphi_{\mu,\ol\mu}$,
$\mu,\ol\mu\in\RR$,
of linear operators $\varphi_{\mu,\ol\mu}\colon\, \HHH_{h,\ol h}\longrightarrow
\HHH_{h+\mu,\ol h+\ol\mu}$, called \textsc{(Fourier) modes}, see
\eq{mode}. 
In particular, the elements $L_n,\, \ol L_n,\, n\in\Z$,
of $\vir_c$, $\ol{\vir}_c$ can be interpreted as the
Fourier modes of the 
holomorphic and antiholomorphic parts $T\,,\ol{T}$
of the \textsc{energy-momentum tensor}.
Moreover, $\Omega_{0,0}$ acts as identity on $\HHH$, and all other modes
of $\Omega$ act by multiplication with zero.
By abuse of notation we write $T=L_2\Omega\in\HHH_{2,0}$,
$\ol T=\ol L_2\Omega\in\HHH_{0,2}$ for the \textsc{Virasoro states}
in $\HHH$.
\end{itemize}
A sextuple 
$\CCC=(\HHH,\,\ast,\,\Omega,\,T,\,\ol T,\,C)$ with
$\HHH,\,\ast,\,\Omega,\,T,\,\ol T,\,C$ as above   
specifies a CFT. Two CFTs $\CCC=(\HHH,\,\ast,\,\Omega,\,T,\,\ol T,\,C)$
and $\CCC^\prime=(\HHH^\prime,\,\ast^\prime,\,\Omega^\prime,\,T^\prime,\,
\ol T^\prime,\,C^\prime)$ are \textsc{equivalent}\label{equivalent}, 
if there exists a
vector space homomorphism $I:\HHH\longrightarrow\HHH^\prime$, such that
$I\colon(\Omega,\,T,\,\ol T)\mapsto(\Omega^\prime,\,T^\prime,\,\ol T^\prime)$
and $\ast^\prime=I\circ\ast$, $C^\prime=C\circ(I^\ast\otimes I\otimes I)$.

Instead of primary states in $\HHH^{\vir}$,
below,  we will be interested in primary states with respect
to a larger algebra than $\vir$, namely the 
(generic) \textsc{holomorphic and antiholomorphic
W-algebra} $\WWW^*\oplus\ol\WWW^*$, see \eq{walg}.
By \eq{prime} the primary states with respect to a subalgebra $\WWW$ of 
$\WWW^*\oplus\ol\WWW^*$ are 
$$
\HHH^{\WWW}=\;
\ker\WWW^-=\left\{ \varphi\in\HHH \bigm| \fa n\in\NN-\{0\}, \,
\fa w\in\WWW_{-n}\colon\; w\varphi=0\right\}.
$$
To \textsc{truncate the OPE 
to primaries} note that by \eq{cftgrowth} for given 
$\varphi\in\HHH$ and $\chi\in\HHH_{h,\ol h}$, we have
$\varphi_{\mu,\ol\mu}\chi\neq0$ for a discrete set of $(\mu,\ol\mu)\in\R^2$.
Hence, whenever  the set 
$$
I_\WWW(\varphi,\chi):=\left\{ (\mu,\ol\mu)\in\RR^2 \bigm|
\exists\psi\in \HHH^{\WWW}\colon\,
\psi^*(\varphi_{\mu,\ol\mu}\chi) \neq0\right\}
$$  
is finite, we can define the truncated OPE
$\varphi\boxast\psi$ as
the orthogonal projection of
$\sum\limits_{(\mu,\ol\mu)\in I_\WWW(\varphi,\chi)}\!\!\!\!\!
\varphi_{\mu,\ol\mu}\chi$
onto $\HHH^\WWW$:
\beqn\el{trunc}
&&\ho
:=\left\{ \varphi\in\HHH^\WWW\bigm|\fa\chi\in\HHH^\WWW\colon\,
\left| I_\WWW(\varphi,\chi) \right| <\infty \right\};\nonumber\\
&&\quad\quad\forall\,\varphi\in\ho,\fa\chi\in\HHH^\WWW:\\
&&\quad\quad\quad\quad\varphi\boxast\chi\in\HHH^\WWW \mbox{ s.th. }
\fa\psi\in\HHH^\WWW\colon\;\;
\psi^*(\varphi\boxast\chi) = C(\psi^*,\varphi,\chi).\nonumber
\eeqn
Let us remark that the above definition of $\boxast$ may well be too
restrictive: By introducing appropriate (partial) completions of 
the relevant vector spaces one can attempt to replace our finiteness
condition in \req{trunc} by a condition on normalizability and thereby get
rid of the restriction to $\ho$. Although in most of our examples we find
$\ho=\HHH^\WWW$, for the orbifolds discussed in Sect.\ \ref{orbex},
$\HHH^\WWW-\ho$ contains all twisted ground states. The latter do not
enter into the discussion of the zero mode algebra,
which is relevant for finding geometric interpretations
(see Sect.\ \ref{geomintlim}). Summarizing, our definition
of $\boxast$, above, is well adapted to our purposes, though it may be
too restrictive in general. By construction,
%
$$
\fa\varphi\in\ho\colon\quad
\varphi\boxast\Omega=\Omega\boxast\varphi=\varphi.
$$
Let us extract a spectral pre-triple from a CFT 
$\CCC=(\HHH,\,\ast,\,\Omega,\,T,\,\ol T,\,C)$.
By definition of the adjoint (see \eq{adjoint}, \eq{viradj}),  
$L_0$ acts as self-adjoint
operator on $\HHH$, and $L_1^\dagger=L_{-1}$.
Moreover, $2L_0=[L_1^\dagger,L_1]$ shows that $L_0$ is positive
semi-definite, and similarly
for $\ol L_0$. Therefore, to associate a spectral pre-triple to a 
CFT $\CCC$, we will always use $H:=L_0+\ol L_0$, which by the uniqueness
of $\Omega$ obeys $\ker(H)=\HHH_{0,0}\cong\CC$. 
Following \cite{frga93}, we let
$$
\wt\HH:=\HHH^{\WWW}=\ker\WWW^-
$$
denote the space of primaries in $\HHH$ with respect to an appropriate
subalgebra $\WWW$ of the holomorphic and antiholomorphic W-algebras.  
Moreover, to every $\varphi\in\ho$ 
we associate an operator $A_\varphi$
on $\wt\HH$ 
which acts by the truncated  OPE $\varphi\boxast$ as in  \eq{trunc}.
The operators $A_\varphi,\, \varphi\in\ho$, generate our algebra $\wt\AAA$
with the obvious vector space structure and with
composition of operators as multiplication:
\beq\el{algebra}
\fa\varphi\in\ho\colon\;\;
A_\varphi\colon\, \wt\HH\longrightarrow\wt\HH, \;\;
A_\varphi(\chi):=\varphi\boxast\chi;\quad
\wt\AAA := 
\left\langle \vphantom{\sum}
A_\varphi  \right|
\left. \vphantom{\sum}\varphi\in\ho \right\rangle.
\eeq
It is not hard to see that $(\wt\HH,H,\wt\AAA)$ obeys 
Def.\ \ref{pretrip} thus defining
a spectral pre-triple. As a word of caution, we remark that in general for
$\varphi,\chi\in\ho$, 
$A_\varphi\circ A_\chi\neq A_{\varphi\boxast\chi}$.

Several other attempts to associate an algebra to a CFT can be
found in the literature. 
\textsl{Truncation of the OPE to leading terms}, as suggested in
\cite[2.2]{koso00},   gives a straight-forward
algebra structure but seems not to capture
enough of the algebraic information encoded in the OPE: On the 
one hand, if all  states in $\HHH^{\WWW}$ are given by simple currents,
e.g.\ for the toroidal theories focused on in \cite{koso00},
then truncation of the OPE to leading terms is equivalent to our truncation
\eq{trunc}. On the other hand, for the example that we  present
in Sect.\ \ref{kinfinity}, it is not, and we show how our truncation \eq{trunc}
gives a convincing geometric interpretation.
For holomorphic vertex operator algebras, 
\textsc{Zhu's commutative algebra} is 
a commutative associative
algebra which can  be constructed using the normal ordered product by modding
out by its associator (see \cite{zh96,brna99,gane00}), and it
is isomorphic to the zero-mode algebra \cite{brna99}.
Although to our knowledge Zhu's commutative algebra
has not been generalized to the non-holomorphic
case, it  is very likely that such a generalization
would yield the same geometric interpretations for degenerate CFTs 
that we propose below; this is related to the fact that
$A_\varphi\circ A_\chi= A_{\varphi\boxast\chi}$  holds for 
the relevant states in these degenerate
CFTs, see Lemma \ref{commcond} and Prop.\ \ref{commutativity}.
Summarizing, our truncation \eq{trunc}, which goes back to
\cite{frga93}, seems to unite the useful aspects
of both the truncation of the OPE to leading terms and Zhu's algebra.

An application of Tauber's theorem known as Kawamata's theorem
\cite[Thm. 4.2]{wi41} shows that the growth 
condition \eq{cftgrowth} ensures  the
eigenvalues of $H$ to obey \eq{growth} for   $\gamma=\nu$.
In general, $\gamma$
will not coincide with the central charge $c$ of the CFT, but in many 
examples with integral $c$ we find 
$\gamma=2c$, see e.g.\  Ex.\ \ref{circle} below. 
So far, we have shown:
\bprop{sptfromcft}
To any CFT $\CCC=(\HHH,\,\ast,\,\Omega,\,T,\,\ol T,\,C)$ of central
charge $c$ which obeys \mb{\eq{cftgrowth}} 
with $\gamma=\nu\in\RR$, after the choice
of a subalgebra $\WWW$ of the holomorphic and antiholomorphic 
W-algebras, we associate a triple
$$
\left( \wt\HH:=\HHH^\WWW=\ker\WWW^-,\, H:=L_0+\ol L_0,\,
\wt\AAA:=\langle A_\varphi\bigm| \varphi\in\ho\rangle\right).
$$
Then $(\wt\HH,H,\wt\AAA)$ is
a spectral pre-triple of dimension $\gamma=\nu$ as in
Def.\ \mb{\ref{pretrip}}.
\eprop
The operator
$\nabla:=L_1+\ol L_1\colon\;\wt\HH\rightarrow\HHH$ 
obeys \eq{lapl} as well
as a Leibniz rule. 
However, for general CFTs we are unable to show that
$(\wt\HH,H,\wt\AAA)$ gives a spectral triple of a specific dimension, i.e.\  a
non-commutative geometry according to  
Def.\ \ref{pretrip}. 
$\wt\AAA$ need not, in general, act by bounded operators,
and we are unable to check all seven axioms of Connes' or their reduction in
\cite{lo00}, including the fact that $\wt\AAA$ is a $\CC^*$-algebra. 
Neither are  we aware
of any attempt to do so in the literature, see also
\cite{fgr95} for a discussion of some unsolved problems that this
approach poses. 

For toroidal CFTs, the above construction indeed gives a 
$\CC^*$-algebra $\wt\AAA$ of bounded operators  \cite{frga93}. 
We illustrate this by
\bex{circle}
Let $\CCC_R, R\in\RR^+$, denote the \textsc{circle theory at radius $R$}, 
i.e.\ the CFT with central charge $c=1$ 
that describes a boson compactified on a circle\footnote{Our 
normalizations are such that the
boson compactified on a circle of radius $R=1$ is described by the
$\fs\fu(2)_1$ WZW model.} of radius
$R$. All $\CCC_R$ possess a subalgebra $\WWW=\fu(1)\oplus\ol{\fu(1)}$
of the holomorphic and antiholomorphic W-algebra\footnote{To clear notations,
our symbol $\fg$ always 
denotes the loop algebra associated to the Lie group $G$
with Lie algebra $g$, and $\fg_k$ denotes its central extension of level $k$.}
(see App.\ \ref{c1}), and the pre-Hilbert space
$\HHH_R$ of $\CCC_R$ decomposes into irreducible representations
of $\WWW$. The latter can be labeled by left- and right handed
dimension and charge
$h_R,\, Q_R$  and $\ol h_R,\, \ol Q_R$ of their lwvs, where 
$h_R=\inv{2} Q_R^2$, $\ol h_R = \inv{2} \ol Q_R^2$.
The space of primaries of 
$\CCC_R$ with respect to $\WWW$ is
\beqns
\wt\HH\;:=\;\HHH^\WWW
&=&\span_\CC\left\{ \textstyle
|Q_R;\ol Q_R\ket:=
\hwv[{{Q_R^2\over2}}]{Q_R}\otimes\hwv[{{\ol Q_R^2\over2}}]{\ol Q_R}
\right|\\
&&\quad\quad\quad\;\left.
\exists\; m,n\in\ZZ:\;
Q_R = \inv{\sqrt2}\left(\inv[m]{R}+n R\right),
\; \ol Q_R = \inv{\sqrt2}\left(\inv[m]{R}-n R\right)
\right\},
\eeqns
see \eq{circlespec}. To obtain the spectral pre-triple associated to $\CCC_R$
by Prop.\ \ref{sptfromcft} from $\wt\HH=\HHH^\WWW$ we need to
consider the truncated OPE \eq{trunc}. By \eq{algebra}
and \eq{circletrunc}, orthonormalizing the $|Q_R;\ol Q_R\ket$ 
as in \eq{u1primenorm}, we have
\beqn\el{product}
A_{|Q_R;\ol Q_R\ket} \circ
A_{|Q_R^\prime;\ol Q_R^\prime\ket}
&=& (-1)^{(Q_R+\ol Q_R)(Q_R^\prime-\ol Q_R^\prime)/2}
A_{|Q_R+Q_R^\prime;\ol Q_R+\ol Q_R^\prime\ket}\nonumber\\
&=& A_{|Q_R;\ol Q_R\ket\boxast |Q_R^\prime;\ol Q_R^\prime\ket}.
\eeqn
We see that $\ho=\HHH^\WWW$ and
$\wt\AAA:=\langle A_\varphi\bigm| \varphi\in\HHH^\WWW\rangle$
is generated by the $A_\varphi$ with $\varphi\in\HHH^\WWW$ as a 
vector space, 
i.e.\  $\boxast$ defines an (associative!) product on $\HHH^\WWW$,
which simplifies the situation considerably in comparison to the
general case. The algebra $\widetilde{\AAA}$ is clearly non-commutative. It
is a straight-forward non-commutative extension of the
product \eq{circleeigen} of the algebra of smooth functions on the 
circle, taking winding \textit{and} 
momentum modes  into account. In fact, 
$\widetilde{\AAA}$ is the twisted
group algebra $\CC_\eps[\Gamma]$ of the $\fu(1)\oplus\ol{\fu(1)}$-
charge lattice 
\beq\el{u1chargelattice}
\Gamma=\left\{\left.(Q_R;\ol Q_R)= (Q_{m,n};\ol Q_{m,n})=
\inv{\sqrt2}\left(\left(  \inv[m]{R}+n R\right);
\left(\inv[m]{R}-n R\right)\right)
\right| m,n\in\ZZ\right\}
\eeq
(see \req{circlespec}),
twisted by the two-cocycle $\eps$ of \req{circletrunc}, yielding
a non-commutative generalization of the algebra of smooth functions 
on $\sone^1_R\times \sone^1_{1/R}$.
Moreover, one checks that $(\wt\HH,H,\wt\AAA)$ 
is a spectral pre-triple of 
dimension $2=2c$, and  we have
\beqn\el{eigencheck}
Q_R= \inv{\sqrt2}\left(\inv[m]{R}+n R\right),\;
\ol Q_R= \inv{\sqrt2}\left(\inv[m]{R}-n R\right)\hphantom{Q_R}\\
\Longrightarrow\quad
H|Q_R;\ol Q_R\ket=(h_R+\ol h_R)|Q_R;\ol Q_R\ket
&=& \left( \inv[m^2]{2R^2}+\inv[n^2R^2]{2}\right) |Q_R;\ol Q_R\ket,
\nonumber
\eeqn
in perfect agreement with \eq{circleeigen}.
\eex
\vspace*{-1.5em}
\subsection{Commutative (sub)-geometries}\label{comgeom}
By Prop.\ \ref{sptfromcft}, there is a spectral pre-triple associated
to every CFT. However, this construction is not very satisfactory. Namely, it
depends on the choice of 
a W-subalgebra $\WWW$, and it does not allow us to extract a
non-commutative geometry \`a la Connes in a straight-forward manner.
Moreover, if we start e.g.\  with the one-dimensional Riemannian geometry
$(\sone^1_R,g)$ discussed in Ex.\ \ref{circ}, from
its associated CFT we read off  a spectral pre-triple
$(\wt\HH, H, \wt\AAA)$ of dimension $2$ in Ex.\ \ref{circle}. 
The original one-dimensional spectral pre-triple 
$\left( \HH=L^2(\sone^1_R,\dvol_g), H={1\over2}\Delta_g,
\AAA\right.$ $\left. =C^\infty(\sone^1_R)\right)$ can 
of course be obtained from $(\wt\HH, H, \wt\AAA)$
by restriction:
\beqns
\HH&=&\span_\CC\left\{ |m\ket_R \bigm| m\in\ZZ \right\}\\
&\cong& \span_\CC\left\{ |Q_R;\ol Q_R\ket\in\wt\HH
\bigm| Q_R = \ol Q_R =\inv[m]{\sqrt2 R},  m\in\ZZ \right\}
\;=\; \ker\langle \WWW^-, j_0-\ol\jmath_0\rangle,
\eeqns
where $j_0, \ol\jmath_0$ denote the zero modes of generators $j, \ol\jmath$ of
$\fu(1),\ol{\fu(1)}$ as in \eq{heisenberg}. 
In \eq{eigencheck} we have checked that $H$ has the
correct eigenvalues on the generators of $\HH$. 
Also, by \eq{product}, $\boxast$ is associative and
commutative on $\HH\cong\ker\langle \WWW^-, j_0-\ol\jmath_0\rangle$,
and 
$\AAA\cong\langle A_\varphi\bigm|\varphi\in\HH\rangle$. This motivates
\bdefi{geomint}
Let $\CCC$ denote a CFT with central charge $c$, $\WWW$ a subalgebra
of its holomorphic and antiholomorphic W-algebras, and 
$(\wt\HH=\ker\WWW^-, H, \wt\AAA)$ the associated spectral pre-triple
of dimension $\gamma$ as in Prop.\ \mb{\ref{sptfromcft}}. A spectral pre-triple
$(\HH, H, \AAA)$ of dimension $c$ is called a \textsc{geometric interpretation
of $\CCC$} if $\HH\subset\wt\HH$, 
$\AAA=\langle A_\varphi\big|_\HH\bigm| \varphi\in\HH\rangle$ 
is commutative, and if there are appropriate completions 
$\ol\HH,\,\ol\AAA$ of $\HH,\,\AAA$ such that
$(\ol\HH, H, \ol\AAA)$ is a spectral triple of dimension $c$, i.e.\ 
$\ol\HH=L^2(M,\dvol_g)$, $H={1\over2}\Delta_{\wt g}$,
$\ol\AAA=C^\infty(M)$ for some Riemannian manifold
$(M,g)$ of dimension $c$ and $\dvol_g=e^{2\Phi}\dvol_{\wt g}$,
$\Phi\in C^\infty(M)$.
\edefi
One checks that each CFT $\CCC_R,\, R\in\RR^+$, of Ex.\ \ref{circle}
has precisely two geometric interpretations 
$(\HH^\pm_R, H, \AAA^\pm_R)$
with $\ol{\HH^\pm_R}=\ol{\ker\langle\WWW^-, j_0\mp\ol\jmath_0\rangle}\cong
L^2(\sone^1_{R^{\pm1}})$. The ambiguity is not a problem but a well-known
virtue, since the CFTs $\CCC_R$ and $\CCC_{R^{-1}}$ are equivalent
according to the definition given on p.~\pageref{equivalent}.
On the other hand, the non-linear 
sigma model construction with target the geometric interpretation 
$\sone^1_{R^{\pm1}}$ of $\CCC_R$ gives back the  CFT $\CCC_R$,
yielding the notion of \textsc{geometric interpretation} introduced in
Def.\ \ref{geomint} very natural.
For general $(M,g)$, however, a rigorous construction of a non-linear
sigma model with target $M$ is problematic: Renormalization is necessary,
and perturbative methods give good approximations only for $M$ with large
volume. Hence we can only expect those properties of Ex.\ \ref{circle}
to generalize which characterize $\CCC_R=\CCC_{R^{-1}}$ at $R^{\pm1}\gg0$. 
In fact, 
$\HH^{\pm}_R$ is generated by the 
$|Q_{R^{\pm1}};\ol Q_{R^{\pm1}}\ket\in\wt\HH$
with $Q_{R^{\pm1}} = \pm \ol Q_{R^{\pm1}} =\inv[m]{\sqrt2 R^{\pm1}}$, 
$m\in\ZZ$. Hence by \eq{eigencheck}
for $R^{\pm1}\gg0$  
the geometric interpretation $(\HH^{\pm}_R, H, \AAA^{\pm}_R)$
is obtained from $(\wt\HH, H, \wt\AAA)$ by restriction to those states 
in $\wt\HH$ which at large volume retain bounded energy:
\bdefi{pgeomint}
Let $\CCC$ denote a CFT \mb{(}or a limit of a sequence of CFTs, 
see Def.\ \mb{\ref{convergent})}
with associated spectral pre-triple
$(\wt\HH, H, \wt\AAA)$ and geometric interpretation 
$(\HH\subset\wt\HH, H, \AAA\subset\wt\AAA)$. Assume that there is an 
$\eps>0$, $\eps\ll1$, such that
$$
\AAA_0:=\langle A_\varphi\big|_\HH\bigm| \varphi\in\wt\HH,\;
|H\varphi|\leq\eps|\varphi| \rangle
\quad\mb{ obeys }\quad
\HH=\span_\CC\left( \AAA_0\left( \HHH_{0,0} \right)\right).
$$
Then $\CCC$ is called \textsc{degenerate} and $(\HH, H, \AAA)$ or
$(M,g)$ with $\ol\HH=L^2(M,\dvol_g)$, $H={1\over2}\Delta_{\wt g}$ and
$\ol\AAA=C^\infty(M)$ as in Def.~\mb{\ref{geomint}}
is called \textsc{preferred geometric interpretation
of $\CCC$}. 
\edefi
The preferred geometric interpretations are exactly those geometric 
interpretations proposed and studied in \cite{frga93}. It is believed
that a degenerate CFT $\CCC$ with preferred geometric interpretation $(M,g)$
in fact yields a \textit{degenerate Riemannian geometry} $(M,g)$.
More precisely, in \cite{koso00}, families $\CCC_\eps$ of degenerate CFTs
with $\eps>0$ as in Def.\ \ref{pgeomint} are studied as
$\eps\rightarrow0$, where the preferred geometric interpretations
$(M_\eps,g_\eps),\, \eps>0$, all
yield the same topological manifold $M_\eps\cong M$.
Then $(M,g_\eps)_{\eps\rightarrow0}$ is believed to describe a 
Gromov-Hausdorff limit of a metric on $M$, where some
cycles  collapse while 
keeping the curvature bounded. Such  limits of metrics have been
studied  in \cite{chgr86,chgr90}. In the physics literature,
the limiting geometries which
arise from degenerate CFTs are sometimes referred to as
\textsc{large volume limits}, see \cite{mo93b} for a useful account.

Since each collapse of cycles $(M,g_\eps)_{\eps\rightarrow0}$  in 
\cite{chgr86,chgr90} gives a boundary point of the moduli space
of Riemannian metrics on $M$, it is natural to use sequences 
$(\CCC_\eps)_{\eps\rightarrow0}$ as above to construct corresponding boundary points
of moduli spaces of CFTs. In a more general context, such a possibility
was alluded to in \cite{koso00}. It presumes the definition of topological
data on the families of CFTs under consideration:
\bdefi{modsp}
A \textsc{CFT-space} is given by the following data:
A sheaf $\SSS$ over a 
topological Hausdorff space $\MMM$, such that for each $p\in\MMM$, 
$\CCC_p$ is a CFT with associated pre-Hilbert space
$\HHH_p=\SSS_p$.
Furthermore $\Omega,T,\ol T$ are global sections of $\SSS$, and 
all CFT-structures as e.g.\ OPE-coefficients, evaluated on local
sections of $\SSS$,  are  continuous. If $\MMM$ is a $D$-dimensional
variety, then $D$  is called the \textsc{dimension} of the CFT-space
$\SSS$.
\edefi
If $\SSS$ is obtained as a deformation space of CFTs
in the sense of conformal deformation theory, then $\MMM$
comes equipped
with a metric, the Zamolodchikov metric, which induces a
standard topology on $\MMM$
as well as flat connections on $\SSS$ \cite{sc58,ka79,rsz94}. This is
in particular true for the family $(\CCC_R)_{R\in\RR^+}$ studied in 
Ex.\ \ref{circle}. 

Intuitively, it is now clear how boundary points of CFT-spaces 
$\SSS$ over $\MMM$ could be constructed: One considers continuous
paths $p:[0,\infty)\rightarrow \MMM$ giving rise to one-dimensional
CFT-subspaces $\SSS_{|p}$ of $\SSS$ with 
$(\SSS_{|p})_t=\HHH_{p(t)}$ for $t\in[0,\infty)$.
If for $t\rightarrow\infty$ the CFT-structures of the $\CCC_{p(t)}$
converge in a suitable sense,
e.g.\ as specified in Sect.\ \ref{lim}, then the limit structure gives 
rise to a boundary point of the CFT-space  $\SSS_{|p}$. If
$\lim_{t\rightarrow\infty}p(t)=\ol p\in\MMM$, 
then the CFT-structures converge to the corresponding structures of
$\CCC_{\ol p}$, and the boundary point of the CFT-subspace $\SSS_{|p}$
just corresponds to this CFT. If however 
$\ol p\in\ol{\MMM}-\MMM$, then the boundary point of the CFT-subfamily
$\SSS_{|p}$ can be considered as boundary point of $\SSS$.

Moreover, assume that
to  each $\CCC_{p(t)}$ we can associate a spectral pre-triple
$(\HH_{p(t)}, H,$ $\AAA_{p(t)})$ obtained from an appropriate subspace of
constant sections along $p$.
If there is an $N\in\NN$ such that for every constant section 
$\varphi$ from such a subspace
one has $|H\varphi_{p(t)}|/|\varphi_{p(t)}| 
\stackrel{t\rightarrow\infty}{=} O(t^{-N})$,
and if all structure constants of $(\HH_{p(t)}, H, \AAA_{p(t)})$
converge for $t\rightarrow\infty$, then 
we obtain a limiting spectral pre-triple $(\HH^\infty, H,\AAA^\infty)$.
The above assumption that all eigenvalues
of $H$ on $\HH_{p(t)}$ converge  with the same speed $O(t^{-N})$
allows to define $H^\infty:=\lim_{t\rightarrow\infty}t^NH$ and should allow to
read off a non-degenerate Riemannian geometry from 
$(\HH^\infty, H^\infty,\AAA^\infty)$.

For now, instead of considering CFTs, 
let us stay in the regime of function spaces and 
inspect limits of 
commutative geometries in terms of spectral triples. This serves as a 
motivation for  Sect.\ \ref{lim} and also leads to some
ambiguities which should be kept in mind.
\bex{circlelim}
We consider possible limiting procedures 
for the spectral triples 
$\left(\HH_R=L^2(\sone^1_R)\right.$, 
$\left.H={1\over2}\Delta, \AAA_R=C^\infty(\sone^1_R)\right)$
as $R\rightarrow\infty$. 
By \eq{circleeigen}, each $\HH_R$ is
generated by $\oplus_{m\in\ZZ} V_m^R$ with $V_m^R:=\span_\CC\{ |m\ket_R\}$
an eigenspace of $H$ with eigenvalue ${m^2\over2R^2}$. 
It is therefore natural to choose constant sections 
$\Phi=\{ \varphi^{m}\bigm| m\in\ZZ\}$ of the sheaf $\SSS$ over
$\RR^+$ with
$\SSS_R=\HH_R$ by $\varphi^{m}_R:=|m\ket_R$. Since the sections are
constant with respect to the inner product on the Hilbert spaces 
$\HH_R$, $R\in\RR^+$, they are in particular  compatible
with the Hilbert space structure, which allows us to formally
define a limiting Hilbert space
$$
\HH_{(1)}^\infty:=\ol{\bigoplus_{m\in\ZZ}
\span_\CC\left\{ |m\ket_\infty \right\}}
\;\mbox{ with }\;
|m\ket_\infty := \{ |m\ket_R\bigm| R\in\RR^+\}.
$$
By \eq{circleeigen},
$\varphi^{m}\cdot\varphi^{m^\prime}=\varphi^{m+m^\prime}$ for all
$m,m^\prime\in\ZZ$, so we are lead to set $A_m |m^\prime\ket_\infty
:=|m+m^\prime\ket_\infty$ and thus obtain a commutative algebra of 
bounded operators $\AAA_{(1)}^\infty:=\langle A_m \bigm| m\in\ZZ\rangle$ on
$\HH_{(1)}^\infty$. The $H$-eigenvalues of all $|m\ket_R$ converge
to zero with the same speed as $R\rightarrow\infty$, hence we can naturally 
define $H^\infty_{(1)}|m\ket_\infty:=\inv[m^2]{2}|m\ket_\infty$ to obtain 
the commutative geometry 
$(\HH^\infty_{(1)}\cong L^2(\sone^1_1), H^\infty_{(1)},$
$\AAA^\infty_{(1)}\cong C^\infty(\sone^1_1))$ in the limit. 

Mathematically, having $|m\ket_\infty$ represent the sequence 
$\{ |m\ket_R\bigm| R\in\RR^+\}$ means that 
$\span_\CC\left\{|m\ket_\infty\right\}$ is the \textsc{direct
limit} (see, e.g.\ \cite{do72})
of the $\left\{V_m^R,\,R\in\RR^+\right\}$,
where for $R,R^\prime\in\R^+$, we use 
$f_{R,R^\prime}\colon\,\sone^1_{R^\prime}\rightarrow\sone^1_R$
with $f_{R,R^\prime}(x):=x\cdot{R\over R^\prime}$ to construct a 
\textsc{direct system} $(\HH_R, f_{R,R^\prime}^\ast)$. 
Then, $\HH^\infty_{(1)}$
is the \textsc{direct limit} of $(\HH_R, f_{R,R^\prime}^\ast)$.

We have used the category $I_1$ with objects $\Ob(I_1)\cong\R^+$
the circles of radii $R\in\RR^+$ and morphisms the diffeomorphisms
between the circles. Note that there is precisely one diffeomorphism
$f_{R,R^\prime}$ for every pair of circles\footnote{We use oriented 
circles with base points to get rid of the translations and reflections.} 
$(\sone^1_{R^\prime},\sone^1_R)$. The limit 
$(\HH^\infty_{(1)}, H^\infty_{(1)}, \AAA^\infty_{(1)})$ is the 
inductive limit of the functor $F_1\colon\, I_1\rightarrow\Vect$
which on objects maps $\sone^1_R\mapsto C^\infty(\sone^1_R)$, and
on morphisms maps $f_{R,R^\prime}\mapsto f_{R,R^\prime}^\ast$.

Instead of $I_1$, there is another quite obvious category $I_2$ we could 
have chosen, namely with objects $\Ob(I_2)=\Ob(I_1)\cong\R^+$
the circles of radii $R\in\RR^+$ and morphisms the isogenies (i.e.\  local
isometries) between circles. That is, there exists a morphism
$g_{R,R^\prime}\colon\,\sone^1_{R^\prime}\rightarrow\sone^1_R$
with $g_{R,R^\prime}(x):=x$ precisely if ${R^\prime\over R}\in\NN$. 
The inductive limit of the functor $F_2\colon\, I_2\rightarrow\Vect$
which on objects maps $\sone^1_R\mapsto C^\infty(\sone^1_R)$, and
on morphisms maps $g_{R,R^\prime}\mapsto g_{R,R^\prime}^\ast$
is 
\beqns
\HH^{(2)}_\infty&:=&\ol{
\bigoplus_{\varrho\in\RR/\QQ}
\span_\CC\left\{ |\varrho\bullet0\ket_\infty \right\}
\oplus
\bigoplus_{r\in\RR^\ast}
\span_\CC\left\{ |r\ket_\infty \right\}}\\
&&\quad\mbox{ with }\; 
\fa \varrho\in\RR/\QQ\colon\;
|\varrho\bullet0\ket_\infty 
:= \{ |0\ket_{N\varrho}\bigm| N\in\ZZ\},\\
&&\quad\hphantom{\mbox{ with }}\; 
\fa r\in\RR^\ast\colon\;\quad\quad\;\,
|r\ket_\infty := \{ |n\ket_{n/r}\bigm| n\in\ZZ\}.
\eeqns
Here, $\RR/\QQ$ denotes classes of real numbers which are commensurable
over $\QQ$. We have
$|\varrho\bullet0\ket_\infty\equiv1$ for all $\varrho\in\RR/\QQ$,
and for all $n\in\ZZ-\{0\}\colon$ 
$|n\ket_{n/r}\colon\, x\mapsto e^{i x r}$. Hence we naturally define 
$H^\infty_{(2)}|\varrho\bullet0\ket_\infty:=0$ for $\varrho\in\RR/\QQ$, and
$H^\infty_{(2)} |r\ket_\infty := {r^2\over2} |r\ket_\infty$ for $r\in\RR$.
This again yields a degenerate limit, but we cannot rescale $H^\infty_{(2)}$
as before. 
Namely, to interpret the $|r\ket_\infty$
in terms of sections $\varphi^r,\, r\in\RR$ of $\SSS$ over $\RR^+$,
$\SSS_R=\HH_R$
(where the label
$r=0$ is replaced by $r=\varrho\bullet0$,
$\varrho\in\RR/\QQ$), 
we have to set 
$\varphi^r_R:=|r R\ket_R$ iff $r R\in\ZZ$ 
($\varphi^{\varrho\bullet0}_R:=|0\ket_R$ iff $R=\varrho$ in $\RR/\QQ$)
and $\varphi^r_R:=0$, otherwise. 
To yield the $\varphi^r_R$ continuous, we need to introduce a discrete
topology on $\RR^+$. Then we can also naturally define a 
spectral triple 
$(\HH^\infty_{(2)}, H^\infty_{(2)}, \AAA^\infty_{(2)})$,
with $A_r\in \AAA^\infty_{(2)}$, $r\in\RR$, acting by
$A_r  |r^\prime\ket_\infty = |r+r^\prime\ket_\infty$,
$A_r  |r^\prime\bullet0\ket_\infty = |r\ket_\infty$  iff 
$r$ and $r^\prime$ are commensurable over $\QQ$, 
i.e.\ $r=r^\prime$  in $\RR/\QQ$, and
$A_r  |r^\prime\ket_\infty =0$, $A_r  |r^\prime\bullet0\ket_\infty = 0$,
otherwise. Similarly, for $\varrho\in\RR/\QQ$ we have 
$A_{\varrho\bullet0}\in \AAA^\infty_{(2)}$ acting by
$A_{\varrho\bullet0} |r^\prime\ket_\infty = |r^\prime\ket_\infty$,
$A_{\varrho\bullet0} |r^\prime\bullet0\ket_\infty 
= |r^\prime\bullet0\ket_\infty$  iff
$\varrho=r^\prime$  in $\RR/\QQ$ and
$A_{\varrho\bullet0}  |r^\prime\ket_\infty =0$, 
$A_{\varrho\bullet0}  |r^\prime\bullet0\ket_\infty = 0$,
otherwise. In other words, $A_{\varrho\bullet0}$ acts as a projection, and
$\left\{ A_{\varrho\bullet0}\mid \varrho\in\RR/\QQ\right\}$ defines a 
``partition of identity'', $A_0:=\sum_{\varrho\in\RR/\QQ} A_{\varrho\bullet0}$.
This indeed 
gives a commutative geometry, namely $\RR$ with
the flat metric and an interesting topology.
\eex
Summarizing, Ex.\ \ref{circlelim} 
motivates the use of direct limits for the construction of limits 
of spectral
pre-triples and CFTs. 
Moreover, as a word of caution, we have found
two different limiting geometries for the spectral triples
$\left(\HH_R=L^2(\sone^1_R)\right.$, 
$\left.H={1\over2}\Delta, \AAA_R=C^\infty(\sone^1_R)\right)$ as
$R\rightarrow\infty$, depending on the choice of the constant sections
of $\SSS$ over $\RR^+$ with
$\SSS_R=\HH_R$. 
Both limits are natural in their own right.
$(\HH^\infty_{(1)}, H^\infty_{(1)}, \AAA^\infty_{(1)})$
is motivated by the approach of \cite{frga93,koso00}, whereas
$(\HH^\infty_{(2)}, H^\infty_{(2)},$ $\AAA^\infty_{(2)})$
corresponds to a  decompactification of $\sone_R^1$ as 
$R\rightarrow\infty$ equipped with a discrete topology.
Similarly, the definition of limits for CFTs
that we  propose in Sect.\ \ref{lim} will incorporate some ambiguity.
\brem{diffapp}
We do not claim that
direct limits yield the only sensible construction for limits 
of algebras or spectral triples as in  Ex.\ \ref{circlelim}.
There, we have already performed a generalization 
from direct limits of ordered systems to direct limits of merely 
partially ordered systems. 
However, an ordered set $(\AAA_i,\bullet_i,\bra.,.\ket_i)_{i\in I}$ 
of algebras with non-degenerate bilinear forms   need not be
a direct system at all in order to make sense of its
``limit''. 

Since we  mainly focus on
the more natural direct limit construction, below,
we do not give a formal definition
of the more general one, here. The main idea, however, is to regard 
a vector space $\AAA$ as limit of the ordered set $(\AAA_i)_{i\in I}$ 
if for every $i\in I$ there is an 
epimorphism $f_i:\AAA\rightarrow\AAA_i$, 
such that for each $\varphi\in\AAA-\{0\}$ there exists an 
$N\in I$ with $f_i(\varphi)\neq 0$ for all $i>N$. If the respective limits,
below, exist, then we can equip $\AAA$ 
with a limit bilinear form and algebra structure by setting
$$
\bra \varphi,\chi \ket
:=\lim_{i}\bra f_i(\varphi),f_i(\chi)\ket_i\,,\quad\quad
\bra \psi,\varphi\bullet\chi\ket
:=\lim_i\bra f_i(\psi),f_i(\varphi)\bullet_if_i(\chi)\ket_i\,.
$$
Note that this only defines an algebra structure on $\AAA$ if
$\bra\cdot,\cdot\ket$ is non-degenerate. 

As an example let us discuss the limit of the algebras 
$C^\infty(\sone^1_R)$ of Ex.\ \ref{circlelim}, equipped with the Hermitean 
form
$$
\bra \varphi,\chi\ket_R
={1\over 2\pi R}\int_0^{2\pi R} \ol{\varphi(x)} {\chi(x)}dx\,.
$$
The radii $R\in\RR^+$ of 
the circles $\sone^1_R$ constitute the ordered index set $I$.
As limit space $\AAA$ we choose the space $C^\infty_c(\RR)$ 
of compactly supported smooth functions on $\RR$. Then, we define
\beqns
f_R:C^\infty_c(\RR)&\longrightarrow& C^\infty(\sone^1_R)\\
\varphi(x)&\longmapsto&
f_R(\varphi)(y):={1\over\sqrt{R}}\sum_{m\in\ZZ}\varphi(\inv[m]{R})e^{i R y/m},
\eeqns
which is a discrete version of a Fourier transform.
Indeed, $(C^\infty(\sone^1_R),f_R)_{R\in\RR^+}$ 
fulfills all the conditions mentioned above, 
and the limit algebra structure on $\AAA=C^\infty_c(\RR)$,
corresponding to the
ordinary product of the Fourier transformed functions,
is the convolution product 
$$
\varphi\bullet\chi(x)=\int_\RR  \varphi(x-y) \chi(y)dy\,.
$$
This construction
can be extended to a limit of spectral triples as in Ex.\ \ref{circlelim},
and the limit geometry is $\RR$ with the standard topology.\hspace*{\fill}
\erem
\vspace*{-0.5em}
\section{Limits of conformal field theories: Definitions}\el{lim}
This section gives our main definitions and 
is the technical heart of the paper. As explained above, 
our construction is motivated by the ideas of
\cite{frga93,koso00}. The guiding example is that of the circle
theories discussed in Exs.\ \ref{circ}, \ref{cftgrowth}, \ref{circlelim},
or more generally the toroidal CFTs discussed in Sect.\ \ref{torex}, since
these models as well as their large volume limits are well understood.
Further motivation arises from the observation that the family of unitary
Virasoro minimal models $\MMM(m,m+1)$, $m\in\NN-\{0,1\}$, 
can be treated by our 
techniques, too, as detailed in Sect.\ \ref{kinfinity}. 

Sect.\ \ref{seli} is devoted to the definition of sequences of CFTs
and their limits; we propose a list of conditions which ensure that
the limit possesses enough structure in order to realize some of the ideas
of \cite{frga93,koso00}. In Sect.\ \ref{geomintlim} we explain how
our limits can give rise to geometric interpretations.
\subsection{Sequences of CFTs and their limits}\el{seli}
In Ex.\ \ref{circlelim} we have given a motivation for
our general approach to limiting processes for CFTs, which
uses direct systems and direct limits\footnote{Although not
spelled out in this language, in \cite[\S6]{mose88b} a
notion of classical limits of CFTs in terms of direct limits was 
introduced.}. 
We  recall the
basic definitions below but refer the reader to the 
literature for a more detailed exposition, see e.g.\ \cite{do72}.
We start by defining sequences of CFTs:
\bdefi{series}
Let $(\CCC^i)_{i\in\NN}=(\HHH^i,\,*^i,\,\Omega^i,
\,T^i,\,\ol{T}^i,\,C^i)_{i\in\NN}$ 
denote a family of CFTs with left and right central charges
$c_i$. Given vector space homomorphisms $f_i^j$
such that
\begin{eqnarray}\label{defseries}
&&\forall\,i,\,j\in\NN,\,i\leq j:\;
f_i^j:\HHH^i\longrightarrow\HHH^{j}, \quad\mbox{ and }\\[2pt]
&&\left.
\begin{array}{rcl}
f_i^j(\Omega^i)&=&\Omega^{j},\\
f_i^j(T^i)&=&T^{j},\\ 
f_i^j(\ol T^i)&=&\ol T^{j},
\end{array}
\right\}
\mbox{ and }
\forall\, i,\,j,\,k\in\NN,\, i\leq j\leq k:\;
\left\{
\begin{array}{rcl}
f_i^i&=&\mbox{id}_{\HHH^i},\\[2pt]
f_j^k f_i^j&=&f_i^k,\\[2pt]
*^{j}f_i^j&=&f_i^j*^i,
\end{array}
\right.\nonumber
\end{eqnarray}
we call $(\CCC^i,f_i^j)$  
a \textsc{sequence of conformal field theories}. 
\edefi
Note that we do not demand any 
further CFT-structure to be preserved by the morphisms $f_i^j$, which therefore
are not morphisms of CFTs.  Hence a sequence
of CFTs cannot be regarded as a direct system of CFTs.
However, \req{defseries} by definition gives a
direct system of  vector spaces $(\HHH^i,f_i^j)$. It allows us to define
a direct limit vector space \cite{do72}
$$
\KKK^\infty:=\lim_{\longrightarrow}\HHH^k 
\simeq
\left[\bigoplus_{k\in\NN}\HHH^k\right]\Big\slash
{\rm span}_\CC\left\{\varphi^i-f_i^j(\varphi^i)\,\mid 
i,\,j\in\NN,\,i\leq j,\,\varphi^i\in\HHH^i\right\}\,,
$$
where by abuse of notation for $i\in\NN$
we have omitted the inclusion homomorphisms
$\imath^i:\HHH^i\hookrightarrow\bigoplus_{k\in\NN}\HHH^k$.
The above definition of $\KKK^\infty$ means that 
for each $\varphi\in\KKK^\infty$ there exist 
$k\in\NN$ and $\varphi^k\in\HHH^k$
such that $\varphi$ is represented by $\varphi^k$, 
i.e. $\varphi=[\varphi^k]=[f_k^l(\varphi^k)]$ for all 
$l\geq k$. In the following, $\varphi^k$ will always denote a 
representative of this form for $\varphi\in\KKK^\infty$.
By
$$
\forall\, i\in\NN:\quad
f_i^\infty:\quad
\HHH^i
\quad\stackrel{\imath^i}{\hookrightarrow}\quad
\bigoplus_{k\in\NN}\HHH^k
\quad\stackrel{\rm proj}{\longrightarrow}\quad
\KKK^\infty
$$ 
we 
denote the homomorphisms given by the composition of inclusion and projection,
with $f^\infty_j\circ f_i^j=f^\infty_i$ for $i,\, j\in\NN,\,
i\leq j$. With the above notations, 
$\varphi=f_k^\infty(\varphi^k)\in\KKK^\infty$.

Similarly, for $i,\,j\in\NN$, $i\leq j$ and $\psi\in\HHH^i$ we define
$(f_i^j)^*(\psi^*):=(f_i^j(\psi))^*$. This gives a direct system
$\left((\check\HHH^*)^i, (f_i^j)^*  \right)$. Its direct limit is denoted
$(\check\KKK^*)^\infty$, and we have 
projections $(f_i^*)^\infty = (f_i^\infty)^*$ as above.

By \eq{defseries}, the  limits $\KKK^\infty$ and
$(\check\KKK^*)^\infty$  possess special 
elements $\Omega,\,T,\,\ol T$ and $\Omega^*$, 
and an involution $*$. However,
the definition of CFT-like structures on the limit vector space $\KKK^\infty$
requires some more conditions on a sequence of CFTs, 
which we shall
discuss now. In particular, we need a notion of convergence. 

In the following, let $(\CCC^i, f_i^j)$ denote a sequence of CFTs.
\bcond{opeconv}
The OPE-coefficients $C^i$ of $\CCC^i$
\textsc{converge with respect to the $f_i^j$}, i.e.
\beqns
&&\forall \varphi\in\HHH^i, \chi\in\HHH^j, \psi\in\HHH^k: \\
&&\quad\quad
C^{m}\left( (f_k^m(\psi))^*, f^m_i(\varphi), f^m_j(\chi) \right)
\;\stackrel{m\rightarrow\infty}{\longrightarrow} \;
C\left((f_k^\infty)^*(\psi^*),f^\infty_i(\varphi), 
f^\infty_j(\chi)\right)\in\CC\,.
\eeqns
\vspace{-1em}
\econd
The limits $C$ of the OPE-coefficients only
depend on elements of the direct limits $\KKK^\infty$, 
$(\check\KKK^*)^\infty$
and are trilinear. Thus a sequence of CFTs $(\CCC^i,f_i^j)$ 
fulfilling Cond.\ \ref{opeconv} gives rise to a trilinear function
$$
C^\infty:
(\check\KKK^\infty)^*\otimes\KKK^\infty\otimes\KKK^\infty\longrightarrow\CC\,.
$$
$C^i,\, *^i$, and the map $\HHH^i\rightarrow(\check\HHH^i)^*,\,
\psi\mapsto\psi^*$ with \req{pairing}, \req{adj}
determine the Hermitean structure of
$\HHH^i$. Since the homomorphisms $f_i^j$ are compatible with this structure,
if Cond.\ \ref{opeconv} holds, then
the vector space $\KKK^\infty$ inherits a limiting bilinear form
$\bra\cdot|\cdot\ket_\infty=C^\infty(\cdot^*,\Omega,\cdot)$, which 
may be degenerate, though.
Define $\NNN^\infty\subset\KKK^\infty$ to be the space of 
\textsc{null vectors} of $\bra\cdot|\cdot\ket_\infty$
in $\KKK^\infty$, i.e.
\beq\label{nulldef}
\NNN^\infty
:=\{\nu\in\KKK^\infty\,| C^\infty(\nu^*,\Omega,\nu)=0\,
\}\,.
\eeq
Since the Cauchy-Schwarz inequality is valid for all 
$\bra\cdot|\cdot\ket_i=C^i(\cdot^*,\Omega^i,\cdot)$, Cond.\ \ref{opeconv} 
implies that $C^\infty(\cdot^*,\Omega,\cdot)$
defines a non-degenerate bilinear form on 
\beq\el{kdef}
\HHH^\infty:=\KKK^\infty/\NNN^\infty \quad\mbox{ with }\quad
\pi^\infty:\KKK^\infty\longrightarrow\HHH^\infty
\mbox{ the  projection}.
\eeq
In the following, we will frequently use elements $\varphi\in\KKK^\infty$
to represent a class in $\HHH^\infty$ and by abuse of notation write
$\varphi\in\HHH^\infty$.
Note that $\pi^\infty$ is compatible with $C^\infty$ 
only if the following condition holds:
\bcond{null}
All OPE-constants involving null vectors $\nu\in\NNN^\infty$ as in 
\mb{\req{nulldef}} vanish in the limit, i.e.\ the following conditions hold:
\beqns
\forall\,\nu\in\NNN^\infty,
\;
\forall\,\varphi,\chi\in\KKK^\infty:
\quad
C^\infty(\chi^*,\nu,\varphi) &=& 0\\
C^\infty(\chi^*,\varphi,\nu)&=&0\;=\;C^\infty(\nu^*,\varphi,\chi).
\eeqns
By \mb{\eq{cc}}, the latter two conditions are equivalent.
\econd
Cond.\ \ref{null} implies that $C^\infty$ descends to a well-defined map
$$
C^\infty:
(\check\HHH^\infty)^*\otimes\HHH^\infty\otimes\HHH^\infty\longrightarrow\CC\,.
$$
Though short and elegant, Cond.\ \ref{null} seems not to be very 
convenient to check in our applications. See 
Rem.\ \ref{simcon}.\ref{simconii} for
a simplification and note that in our Def.\ \ref{fulconv} we avoid
this difficulty.

In order to recover a CFT-like structure in the limit, we will 
introduce a direct limit of the decomposition \req{deco} 
on $\HHH^\infty$. To this end, we will need
\bcond{bigrading}
There are decompositions of the vector spaces $\HHH^i$ into common
$L_0^i$- and $\ol{L}_0^i$-ei\-gen\-spa\-ces, 
which are preserved by the $f_i^j$, i.e.
\begin{eqnarray*}
&&\HHH^i=\bigoplus_{\alpha\in\III_i}\HHH^i_{\alpha}\,,
\qquad L_0^i\big|_{\HHH^i_{\alpha}}
=h^i_\alpha\,{\rm id}_{\HHH^i_\alpha}\,,\quad
\ol{L}_0^i\big|_{\HHH^i_\alpha}
=\ol{h}^i_\alpha\,{\rm id}_{\HHH^i_\alpha}\,,\\
&&\quad\quad\quad
\forall i,j\in\NN\,,\,\forall\,\alpha\in\III_i\;\;\exists\beta\in\III_j:\;
f_i^j(\HHH_\alpha^i)\subset\HHH_\beta^j\,;\quad
f_i^j(\alpha):=\beta.
\end{eqnarray*}
\vspace{-1em}
\econd
In fact, the induced maps $f_i^j:\III_i\longrightarrow\III_j$ 
defined by Cond.\ \ref{bigrading} constitute a direct system on the 
index sets, whose direct limit will be called 
$\III_\infty:=\lim\limits_{\longrightarrow} \III_i$. 
The preservation of the decompositions by the $f_i^j$  
guarantees the existence of a decomposition 
\beq\el{kdec}
\KKK^\infty=\bigoplus_{\alpha\in\III_\infty}\KKK^\infty_\alpha\,.
\eeq
Cond.\ \ref{bigrading} even guarantees that if Cond.\ \ref{opeconv} is  
satisfied as well, 
then \eq{kdec} imposes an analogous decomposition of $\NNN^\infty$
and therefore 
results in  
\beq\el{hdec}
\HHH^\infty=\bigoplus_{\alpha\in\III_\infty}\HHH^\infty_\alpha\,.
\eeq
For $\varphi\in\HHH^\infty_\alpha$ with $\alpha=[\alpha^k]$,
$\varphi\neq0$,
$$
h_{\alpha^k}=h_{\varphi^k}
:= {C^k((\varphi^k)^*,T^k,\varphi^k)
\over C^k((\varphi^k)^*,\Omega^k,\varphi^k)}\,,\qquad
\ol{h}_{\alpha^k}=\ol{h}_{\varphi^k}
:={C^k((\varphi^k)^*,\ol{T}^k,\varphi^k)
\over C^k((\varphi^k)^*,\Omega^k,\varphi^k)}\,
$$
give the $(L_0^k,\ol L_0^k)$ eigenvalues of $\varphi^k$.
By Conds.\ \ref{opeconv} and \ref{null} all limits
$$
h_\alpha:=\lim_{k\rightarrow\infty}h_{\alpha^k}^k\,,\qquad
\ol{h}_\alpha:=\lim_{k\rightarrow\infty}\ol{h}_{\alpha^k}^k
$$
exist. Therefore, we can define the following operators on $\HHH^\infty$:
$$
L_0\big|_{\HHH^\infty_\alpha}
:=h_\alpha\,{\rm id}_{\HHH^\infty_\alpha}\,,
\qquad
\ol{L}_0\big|_{\HHH^\infty_\alpha}
:=\ol{h}_\alpha\,{\rm id}_{\HHH^\infty_\alpha}\,.
$$
These give rise to a coarser decomposition of $\HHH^\infty$ than
\eq{hdec} into 
$(L_0,\ol L_0)$ ei\-gen\-spa\-ces,
$$
\HHH^\infty=\bigoplus_{\stackrel{\scriptstyle h,\ol{h}\in\RR,}{\scriptstyle
h-\ol h\in\ZZ}} \HHH^\infty_{h,\ol{h}}\,.
$$
In particular, as opposed to a well-defined CFT, it is not guaranteed that 
all $\HHH^\infty_{h,\ol h}$ are
finite dimensional. Indeed, the $\HHH^\infty_{h,\ol h}$ will be infinite 
dimensional for some of the examples studied in Sects.\ \ref{exes} and 
\ref{kinfinity}.
In order to nevertheless
allow a definition of modes analogously to \req{mode},
we will therefore need
\bcond{finitedim}
For all $\alpha\in\III_\infty$ and all $\varphi\in\HHH^\infty$, 
$\chi\in\HHH^\infty_\alpha$,
$$
\III_{\mu,\ol{\mu}}(\varphi,\chi)
:=
\left\{ \beta\in\III_\infty \Big|
\HHH_\beta^\infty\subset\HHH_{h_\alpha+\mu, \ol h_\alpha+\ol\mu}^\infty,\;\;
\exists\psi\in\HHH_\beta^\infty\colon\;
C^\infty(\psi^*,\varphi,\chi)\neq0 \right\}
$$
is finite, such that
$$
V_{\mu,\ol{\mu}}(\varphi,\chi):=
\bigoplus_{\beta\in\III_{\mu,\ol{\mu}}(\varphi,\chi)} \HHH_\beta^\infty
$$
is finite dimensional. 
\econd
Cond.\ \ref{finitedim} can also be derived from a version
of uniform convergence on the $C^i$ which we discuss in 
Rem.\ \ref{simcon}.\ref{simconiv}.

To summarize, a sequence of CFTs  which obeys 
Conds.\ \ref{opeconv}--\ref{finitedim}
gives rise to a 
limit vector space $\HHH^\infty$ 
with non-degenerate bilinear form 
$\bra\cdot|\cdot\ket_\infty=C^\infty( \cdot^*,\Omega,\cdot)$
and an OPE-like structure, which assigns modes
to each vector in this vector space analogously to \eq{mode}:
\beqn\el{limitmode}
&&
\forall\,\varphi\in\HHH^\infty,\;
\forall\,\mu,\,\ol\mu, \in\RR, \; \forall\,\alpha\in\III_\infty,\;
\forall\,\chi\in\HHH^\infty_{h_\alpha,\ol h_\alpha}:\\
&&\quad
\varphi_{\mu,\ol\mu}\chi \in 
V_{\mu,\ol\mu}(\varphi,\chi)\;
\mbox{ s.th. } \;
\forall\,\psi \in V_{\mu,\ol\mu}(\varphi,\chi):
\;
\psi^* (\varphi_{\mu,\ol\mu}\chi) = C^\infty(\psi^*, \varphi, \chi). \nonumber
\eeqn
Recall that for a well-defined CFT, the modes of specific subsectors
form closed algebras, like the holomorphic and antiholomorphic W-algebras.
However,  we need  
additional  conditions which ensure that this algebra structure
is preserved in \req{limitmode}. We first specify
\bdefi{wstable}
For a sequence $(\CCC^i,f_i^j)$ of CFTs, let $\wt{\WWW}^i$ denote a 
sequence of finite subsets of $\ker\ol L_0^i\oplus\ker L_0^i$ with
$T^i,\,\ol T^i\in \wt{\WWW}^i$ and $f_i^j(\wt{\WWW}^i)=\wt{\WWW}^j$,
which generate subalgebras 
$$
\WWW^i:=\left\langle \varphi_{n,0},\, \varphi_{0,n} \Big|
\varphi\in\span_\CC\wt{\WWW}^i, \; n\in\ZZ\right\rangle
$$
of the holomorphic and antiholomorphic W-algebras. Assume that the 
$\WWW^i$ are all of the same type, i.e.\ they differ only by their structure 
constants with respect to the elements of $\wt{\WWW}^i$. 
Then the family 
$\WWW^i\supset\vir_{c_i}\oplus\ol{\vir}_{c_i}$ 
is called the \textsc{stable W-algebra}, and the elements of 
$\wt{\WWW}^i$ are called \textsc{W-states}.
\edefi
By definition, the Virasoro algebra is stable in every sequence of CFTs,
and we denote
$$
\wt\WWW^\infty:= \pi^\infty f_i^\infty(\wt{\WWW}^i).
$$
To guarantee that \req{limitmode} induces the action of a W-algebra on
$\HHH^\infty$, the stable W-algebras have to obey 
the following two conditions:
\bcond{wprimaries}
The $f_i^j$ preserve the primaries of the $\WWW^i$, which never become null:
$$
\fa i,j\in\NN\colon\quad
f_i^j\left( (\HHH^i)^{\WWW^i} \right) \subset (\HHH^j)^{\WWW^j},
\quad\mbox{ and }\quad
f_i^\infty\left( (\HHH^i)^{\WWW^i}-\{0\} \right)\cap\NNN^\infty=\emptyset.
$$
\vspace*{-2.5em}
\econd
\bcond{modeconv}
For every holomorphic
W-state $w$ and $\chi\in\KKK^\infty$,
$n\in\ZZ$, the sequence $w_{n,0}^i\chi^i$ \textsc{converges weakly to 
$w_{n,0}\chi$} as defined by \mb{\req{limitmode}}, that is:
\beqn\el{Cconv}
&&\forall\, \chi,\psi\in\KKK^\infty,\, 
\forall\, w,\wt w\in\wt\WWW^\infty,\, 
\forall\, n\in\ZZ:\nonumber\\
&&\quad\quad\quad\quad\;
C^\infty( \psi^*, \wt w, w_{n,0}\chi)
\;\;=\;\; 
\lim_{i\rightarrow\infty} C^i( (\psi^i)^*, \widetilde w^i, w^i_{n,0}\chi^i),
\eeqn
and analogously for antiholomorphic W-states $w$.
\econd
Indeed, a sequence
of CFTs with stable W-algebras $\WWW^i$ which obeys Conds.\ \ref{opeconv} -- 
\ref{modeconv} features a
W-algebra action of $\WWW^\infty$ on $\HHH^\infty$, generated by
the modes of all W-states in $\wt\WWW^\infty$,
and with structure constants 
obtained as limits of the structure constants of the $\WWW^i$.
The stable W-algebras $\WWW^i$ are non-trivial by definition, since 
at least $\vir_{c_i}$ and $\ol{\vir}_{c_i}$ are 
stable. Hence, for example
$$
c:= 2\; C^\infty( \Omega^*, T,
L_2\Omega)
\reeq{Cconv} 2 \lim_{i\rightarrow\infty}
C^i( (\Omega^i)^*, T^i,L^i_2\Omega^i)
= \lim_{i\rightarrow\infty} c_i
$$
gives the central charge of the limiting Virasoro algebras
$\vir_c,\,\ol\vir_c\subset\WWW^\infty$. 
Analogously, as expected, 
for $\varphi\in\HHH_{h,\ol h}^\infty,\, \varphi\neq0$,  we have
\beq\el{weight}
h = \inv{2} C^\infty( \varphi^*, T, L_1\varphi)/|\varphi|^2, \quad\quad
\ol h = \inv{2} C^\infty( \varphi^*, \ol T, \ol L_1\varphi)/|\varphi|^2.
\eeq
Finally, in order to introduce a limit of the truncated OPE
\req{trunc} for all states that are relevant for our geometric interpretations
in Sect.\ \ref{geomintlim}, we will need
\bcond{truncconv}
For all $\alpha, \beta\in\III_\infty$ and all $\varphi\in\HHH^\infty_\alpha,
\chi\in\HHH^\infty_\beta$, let 
$$
I(\varphi,\chi):=\left\{ (\mu,\ol\mu)\in\RR^2 \bigm|
\exists\psi\in f_i^\infty\left( (\HHH^i)^{\WWW^i}\right)\colon\,
\psi^*(\varphi_{\mu,\ol\mu}\chi) \neq0\right\},
$$
as in \mb{\eq{trunc}}. If $\HHH_\alpha^\infty\subset
\HHH_{0,0}^\infty$, then $|I(\varphi,\chi)|<\infty$.
\econd
We would like to point out that Cond.\ \ref{truncconv} 
is required to ensure the finiteness condition used in the
definition of the truncated OPE $\boxast$ \req{trunc} in the limit.
As mentioned there, this finiteness condition however seems to be
very restrictive and can probably be replaced by
appropriate normalizability of the vectors
$\sum\limits_{(\mu,\ol\mu)\in I_\WWW(\varphi,\chi)}\!\!\!\!\!
\varphi_{\mu,\ol\mu}\chi$ for $\varphi,\,\chi\in\HHH^\WWW$.
Then Cond.\ \ref{truncconv} would be dispensable.

We are now ready to give a definition of convergence for CFTs which we
find natural:
Any sequence of CFTs obeying Conds.\ \ref{opeconv} -- 
\ref{truncconv} will carry an OPE-like structure on $\HHH^\infty$ with
an action of the limiting W-algebra $\WWW^\infty$ on $\HHH^\infty$
which is compatible with the mode construction.
However, so far,
we have used an ``algebraic'' approach to CFTs 
as described at the beginning of Sect.\ \ref{cfttripel} and in 
App.\ \ref{cfts}.
This approach has the drawback that the ``analytic'' structure
of CFT $n$-point functions, also explained in App.\ \ref{cfts},
is encoded in a rather complicated way. It
makes some of the above conditions quite intricate, but
does not ensure the existence of $n$-point functions in the limit. 
On the other hand,  some of these conditions 
follow from convergence of four-point functions
on the sphere. In fact, Conds.\ \ref{opeconv} -- \ref{truncconv} alone turn
out to be problematic in view of our aim to find geometric interpretations
of the limit, similar to Def.\ \ref{pgeomint}: We have not succeeded to 
derive commutativity of the relevant algebra $\AAA^\infty$ from them, 
see Prop.\ \ref{commutativity}.
Therefore,  we formulate a notion of convergence of sequences of CFTs 
which poses stronger conditions by
incorporating the ``analytic'' structure of CFTs:
\bdefi{fulconv}
Let $(\CCC^i,f_i^j)$ denote a sequence of CFTs with stable W-algebras $\WWW^i$,
whose four-point functions on $\PP^1$ converge with respect to the 
$f_i^j$ as real analytic functions outside the partial diagonals,
with the standard behaviour near the singularities \mb{(}see App. 
\mb{\ref{cfts}}\mb{)}. In other words, 
for all $i\in\NN$, $\varphi,\chi,\psi,\omega\in\HHH^i$,
$$
(z,\ol z)\longmapsto
\lim_{j\rightarrow\infty}
\bra f_i^j(\psi)|f_i^j(\varphi)(1,1)f_i^j(\chi)(z,\ol{z})|f_i^j(\omega)\ket_j
$$
exists as real analytic function of 
$z,\ol z\in\CC\backslash\{0,1\}\cong\PP^1\backslash\{0,1,\infty\}$ with 
expansions \mb{\eq{fpsing}} around the points $0,1,\infty$.
If the sequence moreover fulfills 
Conds.\ \mb{\ref{bigrading} -- \ref{truncconv}}, then it
is called \textsc{fully convergent}.
\edefi
As a word of caution we remark that in general, the limits of four-point 
functions do not descend to well-defined objects on $(\HHH^\infty)^{\otimes4}$.
\bprop{anconv}
Let $(\CCC^i,f_i^j)$ denote a fully convergent sequence 
of CFTs with stable W-algebras $\WWW^i$. Then this sequence obeys 
Conds.\ \mb{\ref{opeconv} -- \ref{truncconv}}.
\eprop
\bpr
We need to show that Conds.\ \ref{opeconv} and \ref{null} follow from
the convergence of four-point functions on the sphere and 
Conds.\ \mb{\ref{bigrading} -- \ref{truncconv}}.

Indeed, Cond.\ \ref{opeconv} is an immediate consequence, 
since for $\varphi,\chi,\psi\in\HHH^i$: 
$$
\bra \psi|\varphi(1,1)\Omega^i(z,\ol{z})|\chi\ket_i
\equiv\bra \psi|\varphi(1,1)|\chi\ket_i
\reeq{C3} C^i(\psi^*,\varphi,\chi),
$$ 
and $f_i^j(\Omega^i)=\Omega^j$.

To see that Cond.\ \ref{null} is satisfied, first assume 
that there are vectors 
$\nu\in\NNN^\infty_{a,\ol{a}}$,
$\varphi\in\KKK^\infty_{b,\ol{b}}$, 
$\chi\in\KKK^\infty_{c,\ol{c}}$ 
such that $C^\infty(\nu^*,\varphi,\chi)\neq 0$, and choose a sequence
$\{\wt\psi^i_j\}_j$ 
of  orthogonal bases of the $\HHH^i$, 
$\wt\psi^i_j\in\HHH^i_{h^i_j,\ol h^i_j}$,
which converges weakly
to an orthogonal basis of $\KKK^\infty$.
Using \eq{cc}, we can expand the following
four-point function around $z=0$ as in \eq{fpexp}:
\beq\el{fpprop}
\bra\chi^i|(\varphi^i)^\dagger(1,1)\varphi^i(z,\ol{z})|\chi^i\ket_i
=\sum_j \Big|{C^i((\wt\psi_j^i)^*,\varphi^i,\chi^i)\over 
|\wt\psi^i_j|^i}\Big|^2 
z^{h_j^i-b^i-c^i}
\ol{z}^{\ol{h}_j^i-\ol{b}^i-\ol{c}^i}.
\eeq
In particular, we can choose $\wt\psi^i_1:=\nu^i$
and obtain a contradiction to the convergence of four-point functions.
By \req{cc}, this also contradicts the existence of $\nu,\,\varphi,\,\chi$
as above with $C^\infty(\chi^*,\varphi,\nu)\neq0$.

Finally, \req{trimetric} shows that $C^\infty(\psi^*,\nu,\chi)=0$ 
follows for all
$\nu\in\NNN^\infty$, $\chi\in\KKK^\infty$, and primary $\psi\in\KKK^\infty$.
But all three-point functions on the sphere which involve descendants
can be obtained from those involving the corresponding primaries by
application of differential operators. Hence the convergence 
of four-point functions
on the sphere together with Conds.\ \ref{wprimaries}, \ref{modeconv} ensure
that this suffices to prove the claim.
\epr
Although   in the limit of a sequence of CFTs we get 
well-defined OPE-coefficients by construction,
we cannot expect convergence of all correlation functions, as e.g.\ 
torus partition functions. 
This means that the limit of a converging sequence of CFTs is not a CFT, 
in general. 
It may be possible to pose more restrictive conditions on the notion of 
convergence in order to ensure the limit to be described 
by a well-defined CFT. 
However, we are   interested in   certain 
degeneration phenomena, which for example occur in the large volume limits
mentioned in Exs.\ \ref{circle}, \ref{circlelim}. 
These limits do not have a well-defined
torus partition function. So from our viewpoint it is not even
desirable to have a CFT as limit of a converging sequence, in general.
We wish to emphasize that a  limit in the above sense severely depends 
on the choice of the homomorphisms $f_i^j$ in our sequence 
$(\CCC^i, f_i^j)$ of CFTs.
This is analogous to the ambiguity described in Ex.\ \ref{circlelim}.
\vspace{1em}
\brem{simcon}
\vspace{-1.5em}
\begin{enumerate}
\renewcommand\theenumi{\roman{enumi}}
\item\el{simcongra}
Since for $\psi\in\HHH_{h,\ol{h}}^i$
in a convergent sequence of CFTs
$\bra \Omega^i|\Omega^i(1,1)\psi(z,\ol{z})|\psi\ket_i$ 
$=C^i((\Omega^i)^*,\psi,\psi)z^{-2h}\ol{z}^{-2\ol{h}}$,
the convergence of the $(L_0,\ol L_0)$ eigenvalues 
for states with non-vanishing norm follows
independently from the arguments given after Cond.\ \ref{bigrading}.
\item\el{simconii}
The crucial step in the proof of Prop.\ \ref{anconv} is the observation that 
each coefficient in \req{fpprop} remains bounded
in the limit. The latter is equivalent to
the following condition:
\beqn\el{normconverges}
&&\mbox{For } \chi,\psi\in\KKK^\infty \mbox{ set }
\mu^i:=h_{\psi^i}-h_{\chi^i},\,
\ol\mu^i:=\ol h_{\psi^i}-\ol h_{\chi^i};\\ 
&&\quad\quad\mbox{ then } \forall\,\varphi\in\KKK^\infty:\quad
|\varphi^i_{\mu^i,\ol\mu^i}\chi^i| \quad \mbox{ is bounded as }
i\rightarrow\infty.\nonumber
\eeqn
Hence \req{normconverges} implies Cond.\ \ref{null} and is equivalent to
the convergence of four-point functions 
$\langle\chi^i|(\varphi^i)^\dagger(1,1)\varphi^i(z,\ol z)|\chi^i\rangle_i$.
\item\el{simconnull}
In general, null vectors of a representation of the Virasoro algebras
$\vir_{c}$, $\ol{\vir}_{c}$ are defined to be states
that descend from a lowest weight vector but vanish under the action of
each $L_n,\,\ol L_n$ with $n<0$. Although our definition
\req{nulldef} of null vectors is different, a fully convergent sequence of CFTs
has stable Virasoro algebras and allows us to define the action of
$\vir_{c}$, $\ol{\vir}_{c}$ on $\HHH^\infty$ such that null
vectors in this conventional sense are not present, either.
\item\label{simconi}
Our definition of a fully convergent sequence of CFTs with stable
W-algebras $\WWW^i$ simplifies greatly if $\NNN^\infty$
as in \req{nulldef} reduces to $\{0\}$. Then Cond.\ \ref{null} is void.
Moreover, Cond.\ \ref{modeconv} follows from the convergence of 
four-point functions, since the limiting four-point functions are 
well-defined on $\HHH^\infty$ and
the factorization properties \req{fpexp},
\req{confbl} -- \req{cross} of four-point functions remain valid in the limit.
As in ordinary CFTs, this also implies associativity of the OPE in the 
limit, and 
the existence of all $n$-point functions on $\PP^1$.
\item\el{simconiv}
It is not hard to show that Cond.\ \ref{finitedim} is equivalent to a 
version of uniform convergence
of the OPE-constants:
\beqns
\forall\,\varphi\in\HHH^\infty ,\,
\fa\chi\in\HHH^\infty_\alpha,\; \alpha\in\III_\infty,\;
\forall\,\mu,\ol\mu\in\RR,
\forall\,\eps>0\quad\exists I\in\NN:\quad\quad\quad\quad\nonumber\\
\forall\,i\geq I,\;\forall\,\psi\in V_{\mu,\ol\mu}(\varphi,\chi)\,
\mbox{ with }
C^\infty(\psi^*,\varphi,\chi)\neq0:\quad\quad
\hphantom{C^\infty(\psi^*,\varphi,\chi)}
\\
\left| C^i( (\psi^i)^*,\varphi^i,\chi^i) 
- C^\infty(\psi^*,\varphi,\chi)\right|
< \eps\left| C^\infty(\psi^*,\varphi,\chi)\right|.
\hphantom{C^\infty(\psi^*,\varphi,\chi)}\nonumber
\eeqns
\vspace*{-3em}
\end{enumerate}
\erem
The above notion of full convergence turns out to be too 
restrictive
for our purposes. In fact, we would like to allow for 
diverging conformal weights and other structure constants
 in decoupled sectors of the CFTs. 
This happens for example in the large radius limit
of the free boson on the circle, where the winding modes get 
infinitely massive as $R\rightarrow\infty$, see \eq{eigencheck}.
As motivated by Def.\ \ref{pgeomint},
in these cases we should restrict our considerations to the
closed sectors with converging conformal weights:
\bdefi{convergent}
We call a sequence $(\wih\CCC^i,\wih f_i^j)$ of CFTs 
$\wih\CCC^i=(\wih\HHH^i,\,\wih*^i,\,\Omega^i,\,T^i,\,\ol T^i,\,\wih C^i)$
\textsc{convergent}, if the  following holds:

For every $i\in\NN$, the  subspace  $\HHH^i\subset\wih\HHH^i$ 
consisting of those vectors whose conformal weights converge 
under the $\wih f_i^j$  is closed under the OPE. Moreover, 
$$
\forall\, i,\,j\in\NN, \; i\leq j:\quad
f_i^j := \wih f_i^j{}_{\mid\HHH^i}, \quad
C^i := 
\wih C^i{}_{\mid(\check\HHH^*)^i\otimes\HHH^i\otimes\HHH^i}, 
\quad
*^i:=\wih *^i_{\mid\HHH^i}
$$
defines a fully convergent system $(\HHH^i,C^i, f_i^j)$ with 
stable W-algebras $\WWW^i\supset\vir_{c_i}\oplus\ol{\vir}_{c_i}$.
The corresponding 
direct limit
$$
\CCC^\infty:=
(\HHH^\infty:=(\lim_{\longrightarrow}\HHH^i)/\NNN^\infty,\,
*^\infty,\,\Omega,\, T,\,\ol T, \,C^\infty)
$$
is called \textsc{limit of the sequence $(\wih\CCC^i,\wih f_i^j)$ of CFTs}.
The stable $\WWW$-algebras are called \textsc{preserved W-algebras}.
\edefi
\vspace*{-1em}
\brem{famconv}
The discussion of convergence of sequences of CFTs generalizes
to one-di\-men\-sio\-nal CFT-spaces $\SSS$
(see Def.\ \ref{modsp}). Instead of
the homomorphisms $f_i^j$ we specify a connection (i.e.\ the parallel sections)
on the  sheaf $\SSS$. 
Sequences can then be defined on local trivializations of $\SSS$ by
parallel transport. If such a sequence converges  
in the sense of Def.\ \ref{convergent}, then the limit structures 
discussed
above give rise to a boundary point of the CFT-space.

For a general CFT-space 
$\SSS$ over $\MMM$ with
non-compact  $\MMM$,  equipped 
with a flat connection (e.g.\ obtained 
from deformation theory),  we can then construct boundary points 
as limits of  convergent sequences which come from one-dimensional 
CFT-subspaces as above. We will discuss such a boundary of the CFT-space of 
toroidal CFTs\footnote{In fact, CFT-spaces of toroidal models, more 
generally of WZW- and coset-models (see \cite{foro03}) or of orbifolds thereof,
and discrete sequences of CFTs
are the only well known examples of CFT-spaces.
Although the moduli space $\MMM$ of $N=(4,4)$ SCFTs on $K3$ is known
\cite{asmo94}, the corresponding CFT-space $\SSS$ over $\MMM$
has not yet been constructed.} 
in Sect.\ \ref{torex}.
\erem
\vspace*{-1.5em}
\subsection{Geometric interpretations}\el{geomintlim}
As mentioned above, our notion of convergence admits the occurrence of 
degeneration phenomena. One of them is the \textsc{vacuum degeneracy}, 
i.e.\ the degeneration of the subspace of 
states with vanishing conformal weights. While this subspace is 
one-dimensional in a well-defined CFT, it may become 
higher-dimensional, and even infinite-dimensional, in the limit of CFTs. 
In Def.\ \ref{pgeomint} we have introduced \textsc{preferred geometric
interpretations} of CFTs; in this section we will argue that limits
of CFTs with an appropriate vacuum degeneracy can be expected to allow
such geometric interpretations. Similar approaches have
been proposed in \cite[\S6]{mose89} as well as \cite{frga93,koso00}, but
with no general
definitions of sequences and limits of CFTs at hand.

In the following, let $(\CCC^i)_{i\in\NN}=(\HHH^i,\,*^i,\,\Omega^i,
\,T^i,\,\ol{T}^i,\,C^i)_{i\in\NN}$ denote a convergent sequence of CFTs. 
As in Def.\ \ref{convergent}, its limit is
denoted $\CCC^\infty
=(\HHH^\infty:=\KKK^\infty/\NNN^\infty,\,*^\infty$, $\Omega$, $T,\,\ol T, 
\,C^\infty)$. By Cond.\ \ref{wprimaries} we can set 
$$
\wt\HH^\infty
:=f_i^\infty \left( (\HHH^i)^{\vir}\right)\subset\HHH^\infty, 
\;\;\mbox{ and }\;\;
\HH^\infty:=\ker\left(L_0\right)\cap\ker\left(\ol{L}_0\right)
=\HHH^\infty_{0,0}
\subset\HHH^\infty.
$$
Note that $\HH^\infty\subset\wt\HH^\infty$, since descendants cannot
have vanishing dimensions.
To every $\varphi\in\HH^\infty$ we associate an operator $A_\varphi$ on
$\wt\HH^\infty$ by truncation of the OPE, as before: By 
Conds.\ \ref{finitedim} and \ref{truncconv}, for $\chi\in\wt\HH^\infty$
we can copy \eq{trunc} verbatim to define $\varphi\boxast\chi$. 
Then $A_\varphi(\chi):=\varphi\boxast\chi$. Let $\AAA^\infty$ denote the
algebra generated by all these operators:
\beqns
\fa\varphi\in\HH^\infty\colon\;\;
A_\varphi\colon\, \wt\HH^\infty\longrightarrow\wt\HH^\infty, \;\;
A_\varphi(\chi)&:=&\varphi\boxast\chi;\quad
\AAA^\infty := 
\left\langle \vphantom{\sum}
A_\varphi  \right|
\left. \vphantom{\sum}\varphi\in\HH^\infty \right\rangle,\\
\mbox{where }\;
\fa\psi\in \wt\HH^\infty\colon\;\;
\psi^*(\varphi\boxast\chi) &=& C^\infty(\psi^*,\varphi,\chi).
\eeqns
We first collect some properties of $\AAA^\infty$:
\blem{modeaction}
In the limit of a sequence of CFTs,
for every state $\varphi\in\HH^\infty$ of vanishing conformal weights
one has $L_1\varphi=\ol L_1\varphi=0$.
Moreover, $A_\varphi$ preserves weights, i.e.\ with 
$\wt\HH^\infty_{h,\ol h}:= \wt\HH^\infty\cap  \HHH^\infty_{h,\ol h}$,
for all $h,\ol h\in\RR$ we find $A_\varphi\left( \wt\HH^\infty_{h,\ol h}\right)
\subset \wt\HH^\infty_{h,\ol h}$. In particular, $\AAA^\infty$ acts on 
$\HH^\infty$.
\elem
\bpr
Fix $\varphi\in\HH^\infty$. 
Note that
$L_1^i\varphi^i$ converges weakly to $L_1\varphi$ by Cond.\ \ref{modeconv}.
Using this, we first show that in 
$\KKK^\infty=\smash{\lim\limits_{\longrightarrow}}\HHH^i$,
$L_1\varphi$ is a null vector, i.e.\ $L_1\varphi\in\NNN^\infty$. Indeed,
$$
C^\infty(  (L_1\varphi)^\ast, \Omega, L_1\varphi) 
\reeq{limitmode}
C^\infty( (L_1\varphi)^* ,T, \varphi) \\
\reeq{cc} \ol{C^\infty( \varphi^* ,T, L_1\varphi)}
\reeq{weight}0,
$$
which by definition \req{nulldef} proves $L_1\varphi\in\NNN^\infty$.
Similarly, $\ol L_1\varphi\in\NNN^\infty$.

Using weak convergence, Cond.\ \ref{null},
and \eq{l1action}, since $\varphi$ has vanishing conformal weights, we find
$$
\fa \psi\in\HHH^\infty_{h_\psi,\ol h_\psi},\;
\chi\in\HHH^\infty_{h_\chi,\ol h_\chi}\colon\quad
(h_\psi-h_\chi)\,C^\infty(\psi^*, \varphi, \chi)=
C^\infty(\psi^*, L_1\varphi, \chi)=0.
$$
Hence $C^\infty(\psi^*,\varphi,\chi)=\psi^*(A_\varphi\chi)\neq0$
only if $h_\chi=h_\psi$, and similarly $\ol h_\chi=\ol h_\psi$.
This proves the claim.
\epr
By the above, the only non-trivial mode of each $\varphi\in\HH^\infty$ 
is $\varphi_{0,0}$. This motivates
\bdefi{zeromodes}
In the limit $\CCC^\infty$ of a converging sequence of CFTs we set 
$\HH^\infty:=\HHH^\infty_{0,0}$ and call $\AAA^\infty:=
\left\langle \vphantom{\sum}
A_\varphi  \right|
\left. \vphantom{\sum}\varphi\in\HH^\infty \right\rangle$
the \textsc{zero mode algebra}.
\edefi
To fix notations, we now choose an orthonormal basis 
$\left\{\psi_j\right\}_{j\in\NN}$ of $\HH^\infty$ such that
$\ast(\psi_j)=\psi_j$ for all $j\in\NN$ and 
$\psi_j^i\in\HHH_{h_j^i, \ol h_j^i}^i$
where as always $\psi_j^i$ is a representative of $\psi_j$, i.e.
$\psi_j=f_i^\infty(\psi_j^i)$ in $\HHH^\infty$.
We set
\beq\el{onb}
\fa a,b,c\in\NN\colon\quad
C^c_{ab} := C^\infty( \psi_c^*, \psi_a,\psi_b)
\;\reeq{cc}\; C^b_{ac}\;\reeq{trimetric}\; C^b_{ca}.
\eeq
Following \cite{frga93}, we expect the zero mode algebra of
a limit of CFTs to give rise to a spectral triple which defines
a commutative geometry. 
In fact, 
\blem{commcond}
The zero mode algebra $\AAA^\infty$ of the limit of a sequence of CFTs
is commutative if and only if 
$$
\fa\varphi,\chi\in\HH^\infty\colon\quad
A_\varphi\circ A_\chi = A_{\varphi\boxast\chi}.
$$
\vspace*{-1.5em}
\elem
\bpr
With respect to the orthonormal basis $\left\{\psi_j\right\}_{j\in\NN}$
chosen before \eq{onb}, we have
$$
\fa a,b\in\NN\colon\quad
\psi_a\boxast\psi_b = \sum_j C^j_{ab} \psi_j.
$$
One therefore checks:
\beqn\el{commeq}
\fa a,b\in\NN\colon\quad\quad\quad\quad
A_{\psi_a}\circ A_{\psi_b}&=&A_{\psi_b}\circ A_{\psi_a}\nonumber\\
\stackrel{\mbox{\scriptsize\req{onb}}}{\Longleftrightarrow} \quad
\sum_j C^j_{ad}C^j_{b c} &=& \sum_j C^j_{b d}C^j_{ac}\quad \fa c,d\in\NN\\
\stackrel{\mbox{\scriptsize\req{onb}}}{\Longleftrightarrow} \quad\;
A_{\psi_a}\circ A_{\psi_b}&=&A_{\psi_a\boxast\psi_b}.\nonumber
\\[-2em]\nonumber
\eeqn
\vspace{-1em}
\epr
\bprop{commutativity}
The zero mode algebra $\AAA^\infty$ of the limit of a convergent 
sequence of CFTs is commutative.
\eprop
\bpr
By the proof of Lemma \ref{commcond}, the claim is equivalent to
\req{commeq}. This equation follows from the relations imposed on the 
OPE-constants by crossing symmetry. Namely, for all $a,\,b,\,c,\,d\in\NN$,
both sides of
\beq\el{limitcross}
\bra\psi_d^i| \psi_c^i(1,1) \psi_a^i(z,\ol z)|\psi_b^i\ket_i
= \bra\psi_b^i| \psi_c^i(1,1) \psi_a^i(z^{-1},\ol z^{-1})|\psi_d^i\ket_i\;
z^{-2h^i_a}\ol z^{\,-2\ol h^i_a}
\eeq
converge to real analytic functions on $\CC-\{0,1\}\cong\PP^1-\{0,1,\infty\}$
with power series expansions in $z,\,\ol z;\,z^{-1},\,\ol z^{-1}$, 
respectively. Since by 
Cond.\ \ref{wprimaries} the sum over primaries in \req{beta} does not contain
contributions from null vectors, we can use \req{beta} -- \req{cross} to
analyze the structure of \req{limitcross}: Both sides converge to formal
power series in $z,\,z^{-1}$, respectively, with non-negative integer 
exponents, only. Hence both sides must be constant, receiving only 
contributions from the leading terms in the conformal blocks. Then 
\req{cross} shows $\sum_j C_{ab}^j C_{c j}^d=\sum_j C^j_{ad}C^b_{c j}$
which by \req{onb} is equivalent to \req{commeq}.
\vspace{-1em}
\epr
\brem{nocomm}
\vspace*{-1.5em}
\begin{enumerate}
\renewcommand\theenumi{\roman{enumi}}
\item
Similarly to Rem.\ \ref{simcon}.\ref{simconi}, the proof of 
Prop.\ \ref{commutativity} simplifies considerably if null vectors are 
not present in $\KKK^\infty$. Then the proof of Lemma \ref{commcond}
shows $L_1\psi_a$=0 as an element of $\KKK^\infty$
for all $\psi_a\in\HH^\infty$, such that for all
$\psi_a,\,\psi_b,\,\psi_c,\,\psi_d$ $\in\HH^\infty$:
$$
0 = \langle \psi_d | \psi_c(1,1) L_1\psi_a(z,\ol z) | \psi_b \rangle
= {\partial\over\partial z} 
\langle \psi_d | \psi_c(1,1) \psi_a(z,\ol z) | \psi_b \rangle.
$$
In other words, all conformal blocks are constant,  
and crossing symmetry can be used directly to show 
Prop.\ \ref{commutativity}.
\item\el{zhu}
Our definitions easily generalize to the case where the central charges of 
the left and right handed  Virasoro algebras do not coincide. Then
the situation greatly simplifies if all CFTs under consideration are 
chiral: One immediately identifies $A_\varphi(\chi)$,
$\varphi,\chi\in\HH^\infty$, with the normal ordered product of $\varphi$
and $\chi$.
Since $\HHH^\infty=\KKK^\infty/\NNN^\infty$ with $\NNN^\infty$ containing
the ideal generated by $L_1\HH^\infty$, $\HH^\infty$ belongs to 
Zhu's commutative
associative algebra \cite{zh96,brna99,gane00}, which is known to be 
isomorphic to the zero-mode algebra \cite{brna99}. It would be desirable
to generalize the notion of Zhu's algebra to non-chiral theories, and
it would be interesting to know if such a notion can reproduce 
$\AAA^\infty$ in the limit
of convergent sequences of CFTs.
\end{enumerate}\vspace*{-1em}
\erem
By Prop.\ \ref{commutativity},
limits of CFTs are naturally
expected to possess preferred geometric interpretations:
\bdefi{limgeomint}
Let $\CCC^\infty$ denote the limit of a convergent sequence of 
CFTs with limiting central charge $c$ and
zero mode algebra $\AAA^\infty$. Let $N\in\NN$ be maximal such that for all 
$\varphi\in\HH^\infty$ with $\varphi=f_i^\infty(\varphi^i)$,
$\varphi^i\in\HHH^i_{h_i,\ol h_i}$: 
$$
\lambda_\varphi^N 
:= \lim_{i\rightarrow\infty} i^N (h_i+\ol h_i) <\infty,
\quad\quad
H^\infty\varphi := \lambda_\varphi^N \varphi.
$$
Then the linear extension of $H^\infty$  is
a self-adjoint operator
$H^\infty\colon \, \HH^\infty\longrightarrow \HH^\infty$. 

If there exist completions $\ol\HH^\infty,\,\ol\AAA^\infty$ of
$\HH^\infty,\,\AAA^\infty$ such that
$(\ol\HH^\infty, H^\infty, \ol\AAA^\infty)$ is a spectral triple of dimension
$c$, then the latter 
is called a \textsc{geometric interpretation of $\CCC^\infty$}.
\edefi 
The above definition may seem artificial, since we cannot prove
a general result  allowing to give geometric interpretations for
arbitrary limits of CFTs. However, below we will see that 
there are interesting
examples which do allow such geometric interpretations, in particular a 
non-standard one which we  present in Sect.\ \ref{kinfinity}.
Moreover, from the viewpoint of non-linear sigma model constructions
and large volume limits of their underlying geometries,
Def.~\ref{limgeomint} formalizes the expected encoding of geometry in
CFTs, see \cite{mose89,frga93,koso00}, which justifies our definition.
\section{Limits of conformal field theories: Simple examples}\el{exes}
This section consists in 
a collection of known examples, where we discuss limits of 
CFTs and their geometric interpretations in the language 
introduced in Sect.\ \ref{lim}.
Sects.\ \ref{torex} and \ref{orbex} deal with toroidal CFTs and orbifolds 
thereof, respectively. We confirm that our techniques
apply to these cases and that they yield the expected results. In particular,
the discussion of toroidal CFTs fits our approach into the picture drawn
in \cite{frga93,koso00}. 
\subsection{Torus models}\el{torex}
As a first set of examples, let us discuss bosonic toroidal CFTs. These are 
$\fu(1)^d$-WZW models, whose W-algebras contain 
$\fu(1)^d\oplus\ol{\fu(1)}^d$-subalgebras generated by
the modes of the respective ${\fu}(1)^d\cong\RR^d$-valued currents. 
That is, \req{heisenberg} generalizes to
\beqns
j^k(z)=\sum_{n\in\ZZ}a_n^k z^{n-1}\,,\quad \ol{\jmath}^k(\ol{z})
&=&\sum_{n\in\ZZ}\ol{a}_n^k\ol{z}^{n-1}\,,
\quad k=1,\ldots d,\\{}
[a^k_n,a^l_m]=m\delta^{kl}\delta_{m+n,0}\,,\qquad
[\ol{a}^k_n,\ol{a}^l_m]
&=&m\delta^{k,l}\delta_{m+n,0}\,,\qquad [a_n^k,\ol{a}_m^l]=0.
\eeqns
Holomorphic and antiholomorphic energy-momentum tensors
can be obtained as $T={1\over 2}\sum_k\!:\!\!j^k j^k\!\!\!:$, 
$\ol{T}={1\over 2}\!\sum_k:\!\!\ol{\jmath}^k\ol{\jmath}^k\!\!\!:$.
Their modes give rise to holomorphic and antiholomorphic
Virasoro algebras with central charges $c=d$.

The pre-Hilbert space $\HHH_\Gamma$
of a toroidal CFT $\CCC_\Gamma$ decomposes into irreducible lowest
weight representations of 
$\fu(1)^d\oplus\ol{\fu(1)}^d$,
which are completely characterized by their holomorphic and antiholomorphic
$\fu(1)^d\oplus\ol{\fu(1)}^d$-charges $(Q;\ol{Q})\in\Gamma\subset\RR^{2d}$:
$$
\HHH_\Gamma\cong\bigoplus_{(Q;\ol{Q})\in\Gamma}
\VV^{\mathfrak{u}(1)^d}_Q\otimes\ol{\VV}^{\mathfrak{u}(1)^d}_{\ol{Q}}\,.
$$
The corresponding norm-$1$ lwvs $|Q;\ol{Q}\ket$ 
have conformal weights\footnote{For $Q\in\RR^d$, $Q^2$
denotes the standard quadratic form on $\RR^d$.} 
$$
h_{|Q;\ol{Q}\ket}=\inv2 Q^2\,,\qquad
\ol{h}_{|Q;\ol{Q}\ket}=\inv2 \ol{Q}^2\,,
$$
and, by definition, the corresponding fields 
$V_{|Q;\ol{Q}\ket}(z,\ol{z})$ (see \req{in}) obey
\beqn\el{u1charge}
j^k(w)V_{|Q;\ol{Q}\ket}(z,\ol{z})
&=&
{Q^k V_{|Q;\ol{Q}\ket}(z,\ol{z})\over (w-z)}+{\rm reg.}\,,\\
\ol{\jmath}^k(\ol{w})V_{|Q;\ol{Q}\ket}(z,\ol{z})
&=&
{\ol{Q}^k V_{|Q;\ol{Q}\ket}(z,\ol{z})\over (\ol{w}-\ol{z})}+{\rm reg.}\nonumber
\eeqn
The $n$-point functions of the 
$V_{|Q;\ol{Q}\ket}(z,\ol{z})$
reduce to products of the respective holomorphic
and antiholomorphic conformal blocks
\beq\label{toruscorr}
\bra 0|V_{|Q_1;\ol{Q}_1\ket}(z_1,\ol{z}_1),\ldots,
V_{|Q_n;\ol{Q}_n\ket}(z_n,\ol{z}_n)|0\ket_{\PP^1}
=\prod_{1\leq i<j\leq n}(z_i-z_j)^{Q_i Q_j}
(\ol{z}_i-\ol{z}_j)^{\ol{Q}_i\ol{Q}_j}\,.
\eeq
Recall that the right hand side gives a well-defined function for
$z_i,\,\ol z_i\in\CC$ away from the partial diagonals, iff all
$Q_i Q_j-\ol Q_i\ol Q_j\in\ZZ$. Indeed, 
the charges $(Q;\ol{Q})$ of lwvs in a toroidal
CFT constitute an even integral selfdual Lorentzian lattice 
$\Gamma\subset\RR^{d,d}$
of signature $(d,d)$, where 
the quadratic form is given by the 
double spin $2(h-\ol{h})$ of the respective state, and
addition corresponds to fusion.
We denote by $U^d$ the unique even integral selfdual Lorentzian lattice of
signature $(d,d)$ and regard $\Gamma$ as image of an embedding\footnote{The 
embedding is only specified up to automorphisms of $U^d$.} 
$\iota:U^d\stackrel{\sim}{\rightarrow}\Gamma\subset\RR^{d,d}$ of $U^d$ 
into $\RR^{d,d}$. 
Making use of the OPE \req{u1charge}
and contour integration, the $n$-point functions of descendants
can also be extracted from  \eq{toruscorr}. This procedure 
only involves derivation and multiplication by charges
$Q^k,\,\ol Q^k$.

To determine the OPE-constants of lwvs from \req{toruscorr} note that 
the overall sign on the right hand side of this equation depends on 
the ordering of the fields $V_{|Q_i;\ol{Q}_i\ket}(z_i,\ol{z}_i)$. 
This is due to the \textsc{cocycle factors} which for 
$d=1$ have the simple form \req{circletrunc}. To state a general
formula, we first have to generalize \req{circlespec}.
Given a toroidal CFT with charge lattice 
$\iota(U^d)=\Gamma\subset\RR^{d,d}$, 
one can always find a lattice $\Lambda\subset\R^d$
of rank $d$ and a linear map 
$B\colon\R^d\longrightarrow(\R^d)^*$, such that with respect to appropriate 
coordinates on $\RR^{d,d}$ the following holds:
Denote by $\Lambda^*\subset(\R^d)^*$ the $\Z$-dual of $\Lambda$ and
identify $(\R^d)^*$ with $\R^d$ by means of the standard scalar product
on $\R^d$. Then $B$ is skew-symmetric and 
\beq\el{lattice}
\Gamma 
= \left\{ 
(Q;\ol Q)=\inv{\sqrt2} \left( \mu-B\lambda+\lambda; 
\mu-B\lambda-\lambda\right) \bigm| (\mu,\lambda)\in\Lambda^\ast\times\Lambda
\right\}.
\eeq
Moreover,
\beqn\el{torusope}
\mb{for }\;
(Q^{(\prime)};\ol Q^{(\prime)}) &=& \inv{\sqrt2}
\left( \mu^{(\prime)}-B\lambda^{(\prime)}+\lambda^{(\prime)}\; ;\; 
\mu^{(\prime)}-B\lambda^{(\prime)}-\lambda^{(\prime)}\right)
\in \Gamma\colon\qquad
\\
&&C\left(|Q+Q^\prime;\ol{Q}+\ol Q^\prime\ket^*,|Q;\ol{Q}\ket,
|Q^\prime;\ol{Q}^\prime\ket\right)
= (-1)^{\mu\lambda^\prime}\,,\nonumber
\eeqn
with all other OPE-constants vanishing. Note that for $d=1$, 
Eq.\ \req{torusope} simplifies to \req{circlespec} -- \req{circleope}
due to the absence of the $B$-field.

By the above, 
a toroidal CFT $\CCC=\CCC_\Gamma$
is completely characterized by its charge lattice $\Gamma$,
and thus the moduli space of these models is the 
\textsc{Narain moduli space}
$$
{\cal M}_{\rm Narain}^d
\cong {\rm O}(d,d,\ZZ)\backslash{\rm O}(d,d)/O(d)\times O(d)
$$
of even integral selfdual Lorentzian lattices in $\RR^{d,d}$ 
\cite{cent85,na86}.

We will discuss sequences  $(\CCC^i:=\CCC_{\Gamma_i},f_i^j)$ 
of toroidal CFTs in ${\cal M}_{\rm Narain}^d$ with stable W-algebra 
$\fu(1)^d\oplus\ol{\fu(1)}^d$ and fixed
$\iota_i:U^d\rightarrow\Gamma_i$ such that
\beqn\el{torussequence}
&&\forall\, i,j,\in\NN\,,\,\lambda\in U^d\,,\,
P,\ol{P}\in\CC[x_1^1,\ldots,x_1^d,x_2^1,\ldots,x_2^d,\ldots]:\\[3pt]
&&\qquad\quad
f_i^j:P\left((a^i)_m^k\right)\,\ol{P}\left((\ol{a}^i)_n^l\right)\,
|\iota_i(\lambda)\ket_i\longmapsto
P\left((a^j)_m^k\right)\,\ol{P}\left((\ol{a}^j)_n^l\right)\,
|\iota_j(\lambda)\ket_j\,.\nonumber\qquad
\eeqn
This choice of sequence is  natural from the point of
view of deformation theory and completion of ${\cal M}_{\rm Narain}^d$,
see Rem.\ \ref{famconv}.
Indeed, the OPE of lwvs is constant under the $f_i^j$ 
defined in \eq{torussequence},
and  OPE as well as all correlation functions of such 
sequences converge if and only if the lattices
$\Gamma_i=\iota_i(U^d)$ converge in $\RR^{d,d}$. In this case,
\eq{torussequence} is fully convergent in the sense of 
Def.\ \ref{fulconv}. Furthermore $\NNN^\infty=\{0\}$, which 
implies that the limiting $n$-point functions have the usual
factorization properties  (c.f.\ Rem.\ \ref{simcon}.\ref{simconi}). 
In fact, the lattices $\Gamma_i$ 
converge within ${\cal M}^d_{\rm Narain}$, such that 
no degeneration phenomena occur. All  $n$-point functions
on surfaces with positive genus converge as well, 
and the limit of such a sequence
is again a toroidal CFT with charge lattice
$\Gamma_\infty=\lim_{i\rightarrow\infty}\Gamma_i$.

On the other hand, ${\cal M}^d_{\rm Narain}$ is not complete, and a sequence
of lattices $\Gamma_i$ may degenerate but still give a convergent sequence
in the sense of Def.\ \ref{convergent}.
Namely, given any 
primitive null-sublattice
$N\subset U^d$ of rank $\delta\in\NN$, one can construct 
convergent sequences such that
the images $\iota_i(N^\perp)$ converge
with $\iota_i(N)$ collapsing to $\{0\}$, while the images of 
lattice points in $U^d-N^\perp$ diverge. In fact, we can split\footnote{Here,
$\oplus$ denotes the direct sum, not the orthogonal direct sum.}
\beq\el{latticedeco}
U^d  = N^\ast\oplus N^\perp= N^*\oplus N\oplus M,
\eeq
such that $N^*$ is null with $N^*\oplus N\cong U^\delta$. Then 
the lattice in the limit,
$$
\Gamma_\infty:=\lim_{i\rightarrow\infty}\iota_i(N^\perp)
=\lim_{i\rightarrow\infty}\iota_i(M),
$$
is again an even integral selfdual Lorentzian lattice in 
$\RR^{d,d}$, however with smaller
rank: $\Gamma_\infty\cong U^{d-\delta}$.

Every sequence \eq{torussequence} showing this kind of degeneration
is  convergent in the sense of Def.\ \ref{convergent}. The limiting
pre-Hilbert space is a  $\fu(1)^d\oplus\ol{\fu(1)}^d$-module
\beq\el{decomp}
\HHH^\infty\simeq\HHH_{\Gamma_\infty}\otimes
\bigoplus_{\lambda\in N}
\VV^{\fu(1)^\delta}_{Q=0}\otimes\ol{\VV}^{\fu(1)^\delta}_{\ol Q=0}\,.
\eeq
As before, we have  $\NNN^\infty=\{0\}$ and therefore the usual 
factorization properties of the limiting $n$-point functions on the sphere.
However, the degeneration of the lattice results in a diverging torus
partition function, and the limit only has the structure of a
CFT on surfaces of genus $0$ with an infinitely degenerate vacuum sector.

In the spirit of Rem.\ \ref{famconv}, 
we regard these degenerate limits as boundary points of the CFT-space 
$\SSS$ over ${\rm O}(d,d)/{\rm O}(d)\times {\rm O}(d)$
of toroidal CFTs. The underlying Hausdorff space 
of such boundary points has a stratification 
$$
\partial^{\rm deg} \left({\rm O}(d,d)/{\rm O}(d)\times {\rm O}(d)\right)
\cong\bigcup_{1\leq \delta\leq d} 
{\rm O}(d-\delta,d-\delta)/{\rm O}(d-\delta)\times 
{\rm O}(d-\delta)\,.
$$
If one furthermore takes into account the possibly different speeds of 
degeneration, then one ends up with the compactification described in 
\cite{koso00}, which also includes higher boundary strata.

As explained in Sect.\ \ref{cfttripel}, we can extract a spectral
pre-triple $(\wt\HH,H,\wt\AAA)$ with  associative algebra
$\wt\AAA$ from every CFT. For a toroidal CFT with charge lattice 
$\Gamma$, \req{torusope} and a direct
generalization of the discussion in Ex.\ \ref{circle} lead us to the twisted
group algebra
\beq\el{grpalg}
\wt\AAA_\Gamma\cong\CC_\eps[\Gamma]\,,\qquad
\eps\colon\left( (Q;\ol Q), (Q^\prime;\ol Q^\prime)\right)
\longmapsto (-1)^{\mu\lambda^\prime}\,,
\eeq
i.e.\
$$
|Q;\ol Q\ket\boxast |Q^\prime;\ol Q^\prime\ket
=(-1)^{\mu\lambda^\prime}\;|Q+Q^\prime;\ol Q+\ol Q^\prime\ket
$$
with notations as in \req{torusope}. 
For any $N$ as above, the lwvs corresponding to elements in $N$ generate a commutative
subalgebra $\wt\AAA_N\cong\CC[N]\subset\wt\AAA_\Gamma$. 
Similarly to the one-dimensional case discussed
in Sect.\ \ref{comgeom}, restriction to $\wt\AAA_N$ gives a commutative
geometry. In fact, $\wt\AAA_N$ is isomorphic to the  
algebra of smooth functions on a $\delta$-dimensional 
torus $T_{\iota(N)}=\RR^\delta/2\pi\Lambda_\delta$, where $\Lambda_\delta$ is a lattice
of rank $\delta$ such that $\iota(N)\subset\Gamma$ in \req{lattice} is described
by restricting to lattice vectors with 
$(\mu,\lambda)=(\mu,0)$, $\mu\in\Lambda_\delta^*\subset\Lambda^*$.

The zero mode algebra 
$\AAA^\infty$ associated to the degenerate vacuum sector of \eq{decomp}
(see Sect.\ \ref{geomintlim})
is isomorphic to $\wt\AAA_N\cong\CC[N]$, whose closure is
the algebra of smooth functions on a topological torus $T_N$. 
More precisely, each $\mu\in\Lambda_\delta^*$ corresponds to a
function $T_{\iota(N)}\ni x\mapsto e^{i \mu(x)}\in\CC$. It is an
eigenfunction of the Laplacian on $T_{\iota(N)}$, equipped 
with the standard metric inherited from $\RR^\delta$,
with eigenvalue $\mu^2$.
The latter goes to zero as $i\rightarrow\infty$. 

Since 
$\NNN^\infty=\{0\}$, $\AAA^\infty$ 
also acts on the entire limiting pre-Hilbert space 
$\HHH^\infty$,
which is a projective module of finite type over $\AAA^\infty$ 
and can therefore be regarded 
as the space
of sections of a vector bundle over $T_N\cong{\rm Spec}(\AAA^\infty)$. 
Thus, in the limit we find a CFT-fibration
over the moduli space $T_N$ of ground states of the limit theory as described
in \cite{koso00}. In fact, with a suitable choice of coordinates
as in \req{lattice},
the limiting structure of such degenerating sequences of toroidal
CFTs can be regarded as \textsc{large volume limit}, where $T_{\iota_i(N)}$ 
is the sequence of subtori obtaining infinite radii, while their duals 
collapse.
It is a geometric interpretation
in the sense of Def.\ \ref{limgeomint}, if $N$ is a maximal null sublattice,
i.e.\ if $\delta=d$. Otherwise, it can be viewed as geometric interpretation
of an appropriate subtheory with central charge $c^\prime=\delta$.
\brem{decomplim}
As described in Ex.\ \ref{circlelim} for the case of
spectral triples of circles, we can also use partially ordered
systems to obtain limits of toroidal CFTs different from the ones
discussed above. For instance, in the case of circle models
$\CCC_R$, $R\in\RR^+$ (see Ex.\ \ref{circle}, App.\ \ref{c1}), we can
define a partially ordered system analogously to the construction
given in Ex.\ \ref{circlelim}. In the
latter case, the limit spectral triple for functions on circles 
corresponds to $\RR$ with a complicated topology. Similarly,  the limit of the
partially ordered system of circle theories decomposes into subsectors
which only couple through the vacuum.

However, as  explained in Rem.\ \ref{diffapp}, one does
not have to use the direct limit construction to define limits of 
spectral triples, and the same is true for limits of CFTs. Let us briefly
discuss this in the case of the \textsc{infinite radius limit of circle 
models}.

To define a limit of the ordered set $(\CCC_R)_{R\in\RR^+}$ of
CFTs without a given direct system, 
we have to find an appropriate Hermitean limit vector space $\HHH^\infty$ and
epimorphisms $f_R:\HHH^\infty\rightarrow\HHH_R$, $R\in\RR^+$,  by hand. 
In fact, using notations as in \req{u1chargelattice},
\beq\el{c1daH}
\HHH^\infty:=C^\infty_c(\RR)\otimes\CC[x_1,x_2,\ldots]^{\otimes 2}
\eeq
together with
\beq\el{c1daf}
f_R(\chi\otimes P\otimes \ol P)
:={1\over \sqrt{R}}\sum_{m,n}\chi(\sqrt2Q_{m,n})
P(a_r)\ol P(\ol a_s)|Q_{m,n};\ol Q_{m,n}\ket
\eeq
satisfy conditions similar to those stated in Rem.\ \ref{diffapp}. 
Since the respective limits exist, we can define limit correlation functions 
by
$$
\bra 0|\varphi_1(z_1,\ol z_1)\ldots \varphi_n(z_n,\ol z_n)|0\ket
:=\lim_{R\rightarrow\infty}
\bra 0|f_R(\varphi_1)(z_1,\ol z_1)\ldots f_R(\varphi_n)(z_n,\ol z_n)|0\ket_R\,,
$$
and similarly obtain limit OPE-constants.

The notion of convergence of sequences of CFTs 
introduced in Defs.\ \ref{fulconv}, \ref{convergent} can be generalized
to such a limit construction of ordered sets of CFTs.
Indeed, the system $(\CCC_R,f_R)_{R\in\RR^+}$ defined 
by \eq{c1daH} -- \eq{c1daf} 
is convergent in this generalized sense. Its limit is a full CFT, namely 
the \textsc{uncompactified free bosonic theory} with pre-Hilbert 
space\footnote{This pre-Hilbert space 
is a closure of $\HHH^\infty$ defined in \eq{c1daH}.}
$$
\HHH=\bigoplus_{Q\in\RR}\VV_Q^{\fu(1)}\otimes\ol\VV_Q^{\fu(1)}\,,
$$
and in particular does not show any degeneration phenomena.
We emphasize that $\HHH^\infty$ and the $f_R$ had to be constructed by hand
and are not compatible with CFT-deformation theory,
which is our reason for preferring the direct limit construction of 
Sect.\ \ref{lim}.
\erem
\vspace*{-1.5em}
\subsection{Torus orbifolds}\el{orbex}
If a given CFT allows an appropriate action of a finite symmetry 
group\footnote{That is, the group acts as group of
automorphisms on the pre-Hilbert
space of our theory leaving the n-point functions invariant, and the
level matching conditions \cite{dhvw86} are obeyed.}, then 
one can construct a new model from these data
by \textsc{orbifolding}, see e.g. \cite{dvvv89}. Since from 
our point of view the main ideas are apparent already in
the simplest examples of torus orbifold
models, namely the $\sone^1/\Z_2$-orbifold theories $\CCC_R^{\ZZ_2}$, 
$R\in\RR^+$, that is
the $\ZZ_2$-orbifolds of the circle models $\CCC_R$  described in 
Ex.\ \ref{circle} 
and App.\ \ref{c1}, we will restrict our discussion to this family.

On the pre-Hilbert space $\HHH_R$ of the circle theory $\CCC_R$, 
the non-trivial element of $\ZZ_2$ acts by
$$
P(a_n)\ol{P}(\ol{a}_n)|Q_R;\ol Q_R\ket\longmapsto
P(-a_n)\ol{P}(-\ol{a}_n)|-Q_R;-\ol Q_R\ket
$$
for $P,\ol{P}\in\CC[x_1,x_2,\ldots]$. 
The pre-Hilbert space of the resulting orbifold model $\CCC_R^{\ZZ_2}$
consists of the $\ZZ_2$-invariant part of $\HHH_R$
and additional twisted sectors. Each sector decomposes into 
lowest-weight representations of the generic orbifold W-algebra
$\WWW=W(2,4)\oplus\ol{W(2,4)}\subset\fu(1)\oplus\ol{\fu(1)}$ 
as detailed in \cite{na96}:
In the untwisted sector, there are 
norm-$1$ lwvs 
$\inv{\sqrt2}|Q_R;\ol Q_R\ket^{\ZZ_2}=\inv{\sqrt2}|-Q_R;-\ol Q_R\ket^{\ZZ_2}$ 
of conformal weights $h={1\over 2}Q_R^2$, 
$\ol h={1\over 2}\ol{Q}_R^2$ for 
each $\ZZ_2$-equivalence class of charges \eq{circlespec} 
appearing in the original circle theory.
An additional norm-$1$ lwv $|\Theta_R\ket$ of conformal weights $h=\ol h=1$
occurs in the basic 
$\fu(1)\oplus\ol{\fu(1)}$ representation with $Q_R=\ol Q_R=0$.
Furthermore, in the twisted sector, 
there are
four lwvs $|\sigma_R^l\ket$, $|\tau_R^l\ket$, $l\in\{0,1\}$, 
with $h=\ol h={1\over 16}$, $h=\ol h={9\over 16}$, respectively.
 
The OPE in the circle theories is invariant under the $\ZZ_2$-action,
and the OPE in the orbifold models respects the $\ZZ_2$-grading on the 
pre-Hilbert spaces. Hence the
correlation functions and OPE of states in the invariant sectors of the 
orbifold models
coincide with the respective data in the circle 
theories.
Correlation functions containing states in the twisted sectors have been 
discussed
in \cite{dfms87,dvv88}, and the  OPE between lwvs can be extracted from them.

Given a sequence $(R_i)_{i\in\NN}$ in $\RR^+$, we consider the sequence 
$(\CCC_{R_i}^{\ZZ_2},f_i^j)$ of
$\sone^1/\ZZ_2$-orbifold models such that on lwvs 
$f_i^j(|Q_{R_i};\ol Q_{R_i}\ket^{\ZZ_2})=
|Q_{R_j};\ol Q_{R_j}\ket^{\ZZ_2}$, 
$f_i^j(|\Theta_{R_i}\ket)=|\Theta_{R_j}\ket$,
$f_i^j(|\sigma_{R_i}^l\ket)=|\sigma_{R_j}^l\ket$, $f_i^j(|\tau_{R_i}^l\ket)=
|\tau_{R_j}^l\ket$.
This definition naturally extends to the descendants as in 
\req{torussequence}.

Then, as in the case of toroidal models, all correlation functions and the OPE
converge with respect to the $f_i^j$ if and only if $(R_i)_{i\in\NN}$
converges in $\RR^+$, $\lim_{i\rightarrow\infty}R_i= R_\infty>0$. 
In this case, $(\CCC_{R_i}^{\ZZ_2},f_i^j)$
is a fully convergent sequence of CFTs 
in the sense of Def.\ \ref{fulconv}, and $\NNN^\infty=\{0\}$, implying
the existence of correlation functions on $\PP^1$ 
(see Rem.\ \ref{simcon}.\ref{simconi}). No
degeneration occurs, which means that  correlation functions on surfaces
of positive genus  converge, too. Thus, in the limit we obtain
a full CFT, namely the $\sone^1/\ZZ_2$-model at radius $R_\infty$.

If $R_i\rightarrow 0$ or $R_i\rightarrow \infty$, our sequence of CFTs 
is convergent
in the sense of Def.\ \ref{convergent}. 
Indeed, all correlation functions between states with
convergent  weights converge, $\NNN^\infty=\{0\}$, 
and in the limit we obtain
a well-defined CFT on the sphere with degenerate vacuum sector. In the 
language of Sect.\ \ref{torex}, for $R_i\rightarrow\infty$ we can use
\req{latticedeco} with $M=\{0\}$ and 
$N=\left\{ \inv{\sqrt2}(m;m) \mid m\in\ZZ\right\}$,
$N^*=\left\{ \inv{\sqrt2}(n;-n) \mid n\in\ZZ\right\}$, and
$N\leftrightarrow N^*$ if $R_i\rightarrow0$.

By Prop.\ \ref{sptfromcft} we can associate a 
spectral pre-triple $(\wt\HH,H,\wt\AAA)$
to each orbifold model $\CCC_R^{\ZZ_2}$.
As mentioned after \req{trunc}, here we find $\HHH^\WWW\supsetneqq\ho$.
By \cite{dfms87,dvv88} the OPE-constants including twisted 
ground states are given by
\beqn\el{twistco}
C^R\!\left(|\sigma_R^k\ket^*,|Q_{m,n};\ol Q_{m,n}\ket^{\ZZ_2},
|\sigma_R^l\ket\right)
&\stack[\mbox{\scriptsize\req{cc},\req{trimetric}}]{=}&
C^R\!\left((|Q_{m,n};\ol Q_{m,n}\ket^{\ZZ_2})^*,|\sigma_R^k\ket,
|\sigma_R^l\ket\right)\nonumber\\
&=&{2\,(-1)^{ml}\delta_{n+l,k}\over 2^{Q_{m,n}^2+\ol Q_{m,n}^2}},
\eeqn
with notations as in \req{u1chargelattice}. Hence the 
$I_\WWW(\sigma^k_R,\sigma^l_R)$ used in \req{trunc} are infinite.
On the other hand, $\wt\AAA$ contains a subalgebra given by
the $\ZZ_2$-invariant
part $\wt\AAA^\prime:=\CC_\eps[\Gamma]^{\ZZ_2}$
of the respective algebra of the 
underlying circle theory, c.f.\ \eq{grpalg}. $|\Theta_R\ket$ acts on 
$\wt\AAA^\prime$ as a second order differential operator, and $\wt\AAA$ is an
$\wt\AAA^\prime$-module. 
Thus, $\wt\AAA$ can be regarded as the
space of sections of a sheaf over the non-commutative space associated to 
the restricted spectral pre-triple $(\wt\HH^\prime,H^\prime,\wt\AAA^\prime)$.

If $R_i\rightarrow\infty$, the zero mode algebra 
$\AAA^\infty\cong\CC[\ZZ]^{\ZZ_2}$ 
is generated by the lwvs 
represented by $|\inv[m]{\sqrt2R_i};\inv[m]{\sqrt2R_i}\ket^{\ZZ_2}$, 
$m\in\ZZ$. It is the algebra of $\ZZ_2$-symmetric
functions on the circle, i.e.\ the functions on $\sone^1/\ZZ_2$. In fact,
the $|m\ket_\infty^{\ZZ_2}:=
f_i^\infty\left( |\inv[m]{\sqrt2R_i};\inv[m]{\sqrt2R_i}\ket^{\ZZ_2}\right)$ 
are 
characterized by the recursion relation
$$
|m+1\ket_\infty^{\ZZ_2} 
\,=\,  |m\ket_\infty^{\ZZ_2}\boxast|1\ket_\infty^{\ZZ_2}
\,-\,|m-1\ket_\infty^{\ZZ_2},
$$
which agrees with the recursion relation for the (rescaled)
\textsc{Chebyshev polynomials of the first kind}, see e.g.\ \cite{he82}:
$$
T_m(\cos x) := 2 \cos( m x ), \quad
\mbox{ for } m\in\NN,\; x\in[0,\pi].
$$
Hence $|m\ket_\infty^{\ZZ_2}$ should be identified with the function
$x\mapsto T_m(\cos x)$.
This is not surprising, since the lwvs $|m\ket^{\ZZ_2}_\infty$ are 
$\ZZ_2$-symmetric combinations of lwvs in the underlying circle theories,
which in turn correspond to exponential functions.

Indeed, $\left\{ T_0/\sqrt2,\, T_1,\,T_2,\,\dots\right\}$ 
is an orthonormal basis 
of $L^2([0,\pi],\dvol_g)$ with $\dvol_g=dx/2\pi$, i.e.\ with the
flat standard metric $g$ on $[0,\pi]\cong\sone^1/\ZZ_2$. Hence the methods
of Sect.\ \ref{sptr} yield $\ol{\HH^\infty} = L^2(\sone^1/\ZZ_2,dx/2\pi)$,
$\ol{\AAA^\infty}=C^\infty(\sone^1/\ZZ_2)$, 
which according to Def.\ \ref{limgeomint} for the limit gives
the expected 
geometric interpretation  on $\sone^1/\ZZ_2$ with the flat metric
$g$ induced from the standard metric on $\sone^1$ and a trivial dilaton 
$\Phi$. 
Note also that the $m^{\mb{\small{}th}}$ Chebyshev polynomial $T_m$  
is an eigenfunction of the Laplacian 
$\inv{2}\Delta_g=-\inv{2}{d^2\over dx^2}$ with eigenvalue $\inv{2}m^2$,
as expected from 
\beqns
H^\infty|m\ket_\infty^{\ZZ_2}=\lambda_m^2|m\ket_\infty^{\ZZ_2}
\\
\quad\mb{ with }\quad\lambda_m^2
&=&\lim_{i\rightarrow\infty}R_i^2(h^i_{|Q_{m,0}\,;\,\ol Q_{m,0}\ket}
+\ol h^i_{|Q_{m,0}\,;\,\ol Q_{m,0}\ket}) = \inv{2}m^2.
\eeqns
As for the toroidal CFTs discussed in Sect.\ \ref{torex}, $\AAA^\infty$
acts on the entire pre-Hilbert space $\HHH^\infty$ which can be regarded
as the space of sections of a sheaf over $\sone^1/\ZZ_2$. Let us
restrict the discussion  to
the states $|\sigma^l\ket_\infty:=f_i^\infty\left(|\sigma_{R_i}^l\ket\right)$. 
The action of  $\AAA^\infty$ on them
can be extracted from the OPE-coefficients \req{twistco}:
\beqns
|Q_{m,n}\,;\,\ol Q_{m,n}\ket^{\ZZ_2}\boxast|\sigma_R^l\ket
&=&{2\,(-1)^{ml}\over 2^{Q_{m,n}^2+\ol Q_{m,n}^2}}\,|\sigma_R^{l+n}\ket
\\
&\Longrightarrow&\quad\quad
|m\ket_\infty^{\ZZ_2}\boxast
|\sigma^l\ket_\infty
=2(-1)^{ml}\,|\sigma^{l}\ket_\infty.
\eeqns
It follows that the sections corresponding to $|\sigma^l_R\ket$ 
are peaked around 
the respective $\ZZ_2$-fixed points, i.e.\ the endpoints of the interval
$[0,\pi]$. 
In the limit their support in fact shrinks to these points. The same holds 
true for all
other states in the twisted sectors. They are sections of skyscraper sheaves
over the fixed points of the orbifold action. As expected, 
in the limit the OPE of two states in the twisted sectors vanishes, unless 
the corresponding sections  have common support. 
This gives a nice geometric interpretation of the twisted sectors. 
\section{The $m\rightarrow\infty$, $c\rightarrow1$ limit
of the unitary Virasoro minimal models $\MMM(m,m+1)$}\el{kinfinity}
The present section contains the main results of this
work: In Sect.\ \ref{c1ex}
we show that the techniques introduced in Sect.\ \ref{lim} for 
the study of limits and degeneration phenomena
also apply to the family of diagonal
unitary Virasoro minimal models $\MMM(m,m+1),\, m\in\NN-\{0,1\}$,
which  gives a fully convergent sequence of CFTs. In Sect.\ \ref{gi1} we
determine and study a geometric interpretation of its limit $\MMM_\infty$ as 
$m\rightarrow\infty$, and we discuss the inherent D-brane geometry.
\subsection{The unitary Virasoro minimal models 
$\MMM(m,m+1)_{m\rightarrow\infty}$}\el{c1ex}
Both outset and favorite example for our investigation are the  
unitary Virasoro minimal models $\MM_m:=\MMM(m,m+1),\, m\in\NN-\{0,1\}$
\cite{bpz84}, which correspond to the $(A,A)$ (left-right
symmetric) modular invariant partition functions
in the CIZ classification \cite{ciz87a,ciz87b}. 
In this section, we explain how a fully convergent sequence 
$(\CCC^m,\,f_m^j)$ with $\CCC^m=\MMM_m$ 
for $m\in\NN-\{0,1\}$ can be defined according to
Def.\ \ref{fulconv}. 
To our knowledge, such a construction
was first alluded to in \cite[\S6 and App.~B]{dofa84} as well as in 
\cite[\S6]{mose89}.
Our approach also allows us to determine a geometric interpretation
of the limit of this sequence
as $m\rightarrow\infty$, according to Def.\ \ref{limgeomint}.

Let us start by recalling some of the 
main properties of the CFT $\MM_m$. Since this model is 
diagonal, we can restrict our 
discussion to the action of  the holomorphic Virasoro algebra.
The pre-Hilbert space of $\MMM_m$ decomposes
into a finite sum of irreducible representations of the Virasoro
algebra $\vir_{c_m}$ with central charge
\beq\el{charge}
c_m=1-{6\over m(m+1)}\;.
\eeq
These irreducible representations  are labeled by
$\NNN_m=\{(r,s)|r,s\in\NN,\;1\leq r<m,\;1\leq s<m+1\}/\!\!\sim$ with
$(r,s)\sim(m-r,m+1-s)$, i.e.\ $r+s\sim  2m+1-r-s$,
such that by choosing appropriate 
representatives we can write
\beq\el{z2choice}
\NNN_m=\{(r,s)|1\leq r<m,\;1\leq s<m+1,\;r+s\leq 2m+1-r-s\}\,.
\eeq
Each irreducible Virasoro module
$\VV^{m}_{(r,s)},\, (r,s)\in\NNN_m,$ has an lwv
$|r,s\ket_m$ of conformal dimension
\beq\el{miniweight}
h_{(r,s)}^{m}={(r(m+1)-s m)^2-1\over 4m(m+1)} 
\quad\stackrel{m\rightarrow\infty}{\sim}\quad
{(r-s)^2\over 4} + {r^2-s^2\over 4m}+ {s^2-1\over 4m^2}+\cdots\;.
\eeq
We choose the $|r,s\ket_m$ to be orthonormal.

The $n$-point functions for  $\MMM_m$ are discussed in \cite{dofa84},
in particular all OPE-coefficients $C^m$  are determined in 
\cite{dofa84,dofa85a}, see App.\ \ref{structconst}. The calculations
make use of the Feigin-Fuks integral representation
\cite{fefu83a} of $n$-point functions, assuming that $\MM_m$
has a Coulomb-gas representation. That the latter is indeed true is 
shown in \cite{fe89}.

To construct a sequence of CFTs according to Def.\ \ref{series} we note that
there are well-defined embeddings\footnote{Our choice of embeddings  is 
quite natural and has been used already in \cite{za87c} in the context of
slightly relevant perturbations of $\MMM_m$. 
However, there are other choices, leading to 
different limits of CFTs.}
$$
\NNN_m \hookrightarrow \NNN_{m+1}\,,\quad (r,s)\longmapsto (r,s)\, .
$$
We will extend these embeddings to  vector space
homomorphisms $f_m^{m+1}$ between 
the corresponding irreducible Virasoro modules. To meet 
Cond.\ \ref{wprimaries} of Sect.\ \mb{\ref{seli}},
we must map lwvs to lwvs:
\beqn\el{modemb}
\VV^{m}_{(r,s)}&\hookrightarrow&\VV^{m+1}_{(r,s)}\\
P\left(L_{n}^{m}\right) |r,s\ket_m 
&\longmapsto&\widetilde P
\left(L_{n}^{m+1}\right) |r,s\ket_{m+1} \,,
\nonumber
\eeqn
similarly to \req{torussequence}.
Here, 
$P,\,\widetilde P$ are elements 
of the same degree in the
weighted polynomial ring $\C[x_1,x_2,\dots]$ with $\deg x_n=n$, and we
substitute $x_n=L_{n}^{m}$ or $x_n=L_{n}^{m+1}$
in lexicographical order (see Def.\ \ref{monomials}). 

To construct consistent maps of type \req{modemb},
recall from \cite{bpz84} that the characteristic feature of the
representation $\VV^m_{(r,s)}$ is the fact that 
the Verma module built by the action of the Virasoro
algebra $\vir_{c_m}$ on
$|r,s\ket_m$ with character 
$q^{1-c_m/24}\chi_{h_{(r,s)}}^{\,\mb{\scriptsize gen}}(q)$
and $\chi_h^{\,\mb{\scriptsize gen}}$ as in \req{c1gen}
contains a proper non-trivial submodule of 
\textsc{singular vectors},
that is of lwvs of $\vir_{c_m}$ at positive
level. 
The occurrence of these singular vectors, which have been quotiented
out to obtain $\VV^m_{(r,s)}$, makes our construction slightly
delicate. However, the very properties of direct limits allow us to solve this
problem.
For later convenience, we give the following technical
\bdefi{monomials}
Let $m\in\NN-\{0,1\}$ and $(r,s)\in\NNN_m$. For each $N\in\NN$ choose a set
$\PPP^m_{(r,s)}(N)$ of monomials with weighted degree $N$, such that
\beq\el{basis}
\left\{ P(L_n^{m})|r,s\ket_m \mid P\in \PPP^m_{(r,s)}(N), \, N\in\NN \right\}
\eeq
is a basis of $\VV^m_{(r,s)}$, 
where for $P\in\PPP^m_{(r,s)}(N)$, 
$P(x_n)=\smash{\prod\limits_n} x_n^{a_n}$ with
$a_i\in\NN$ and $\smash{\sum\limits_n} a_n\cdot n=N$,
$$
P(L_n^{m}) := (L_1^{m})^{a_1}\circ(L_2^{m})^{a_2}\circ\cdots\,.
$$ 
If $B_{(r,s)}^m\in\NN$ obeys
$$
\forall\, N,N^\prime\in\NN\colon\quad
N+N^\prime< B_{(r,s)}^m
\;\Longrightarrow\;
\PPP^m_{(r,s)}(N)\cdot \PPP^m_{(r,s)}(N^\prime)
\;\subset\; \PPP^m_{(r,s)}(N+N^\prime),
$$
then $B_{(r,s)}^m\in\NN$ is called an \textsc{energy bound} of 
$\PPP^m_{(r,s)}=(\PPP^m_{(r,s)}(N))_{N\in\N}$. A system $\PPP^m_{(r,s)}$
with maximal energy bound among all systems giving bases \mb{\req{basis}}
of $\VV^m_{(r,s)}$
is called a \textsc{basic monomial system of weight $h^m_{(r,s)}$}.
A sequence $(\PPP^m_{(r,s)})_{m\geq M}$ of monomial systems
is called \textsc{special} 
if for all $m\geq M$ and for all $N<B_{(r,s)}^m$: 
$\PPP^m_{(r,s)}(N)=\PPP^{m+1}_{(r,s)}(N)$, where $B_{(r,s)}^m$
are the respective energy bounds, and almost all $\PPP^m_{(r,s)}$ are basic.
\edefi
Note that the relations which
arise from the existence of singular vectors in the Verma module 
over $|r,s\ket_m$, up to a global
pre-factor $(m(m+1))^{-K}$ with $K\in\NN$, are linear with respect to all
monomials $P(L_n^m)$ of a given weighted degree $N$, with coefficients
$a_P\in\RR[m]$ of degree at most $2N$. Moreover, 
as follows from the explicit character formula \req{minichar},
the  singular vectors  which under the action of $\vir_{c_m}$
generate the submodules of singular vectors have
weights $h_{(r+m,-s+m+1)}^{m}$ and $h_{(r,-s+2(m+1))}^{m}$, i.e.
levels $rs$ and $(m-r)(m+1-s)$, respectively. We conclude that
for fixed $r,s\in\NN-\{0\}$, the energy bound of basic monomial systems
$\PPP^m_{(r,s)}$ of weights $h^m_{(r,s)}$ is monotonic increasing in $m$.
Moreover,
\blem{bms}
For every pair $r,s\in\NN-\{0\}$ with $(r,s)\in\NNN_M$ ($M$ minimal), we can
choose a special sequence $(\PPP^m_{(r,s)})_{m\geq M}$ of  monomial
systems according to Def.\ \mb{\ref{monomials}}, and the respective
energy bounds approach infinity as $m\rightarrow\infty$.
\elem
In the following, $(\PPP^m_{(r,s)})_{m\geq M}$ will always denote 
a fixed special sequence of  monomial
systems of weights $h^m_{(r,s)}$ as in Lemma \ref{bms}. Note that we can 
depict these monomial systems in terms of a convex polyhedron,
as is customary in toric geometry.
We then define
\beq\el{minihom}
\forall\, N\in\NN,\, \forall\, P\in \PPP^m_{(r,s)}(N):\quad
f_m^{m+1}\left[ P(L_n^{m}) |r,s\ket_m \right]
:= P(L_n^{m+1}) |r,s\ket_{m+1}.
\eeq
Finally,
we linearly extend the $f_m^{m+1}$ to vector space homomorphisms
$$
f_m^{m}:={\rm id}_{\VV^{m}_{(r,s)}};\;\;
f_m^{m+1}\!:\;
\VV^{m}_{(r,s)}\hookrightarrow \VV^{m+1}_{(r,s)}; \;\;
f_m^{j}:=f_{j-1}^{j}\circ\cdots\circ f_m^{m+1} 
\!:\;
\VV^{m}_{(r,s)}\hookrightarrow \VV^{j}_{(r,s)}.
$$
Then by construction,
\blem{minisequence}
The sequence $(\MMM_m,f_m^j)$ is a sequence of CFTs with stable
Virasoro algebra according to Defs.\ \mb{\ref{series}} and \mb{\ref{wstable}}.
\elem
In the following, we show that the sequence $(\MMM_m,f_m^j)$ is
fully convergent according to Def.\ \ref{fulconv}. Although above
we have made a lot of choices, we will argue that our
limit is independent of all choices, including the use of monomials
and lexicographical order for their interpretation. 

First note that by
\req{charge} and \req{miniweight},
\beq\el{minilimweight}
c_m\;\stackrel{m\rightarrow\infty}{\longrightarrow}\;c=1,
\quad\quad\quad
h_{(r,s)}^m\;\stackrel{m\rightarrow\infty}{\longrightarrow}\;
h_{(r,s)}=\inv[(r-s)^2]{4},
\eeq
i.e.\ all structure constants of the stable Virasoro algebras 
$\vir_{c_m}$ converge. 
Moreover, setting 
\beq\el{struc}
C^{(p\p,p)}_{(n\p,n),(s\p,s)}
:= C^m\left( (|p\p,p\ket_m)^*, |n\p,n\ket_m, |s\p,s\ket_m
\right)
\eeq
with respect to
orthonormal $|r,s\ket_m$ as in \cite{dofa85},
our calculations \req{strconst}-\req{e} imply
$$
C_{(n\p, n)(s\p, s)}^{(p\p, p)}
\stackrel{m\rightarrow\infty}{\sim} 
A_{(n\p, n)(s\p, s)}^{(p\p, p)}(m+1)^{-E_{(n\p, n)(s\p, s)}^{(p\p, p)}}
$$
with $A_{(n\p, n)(s\p, s)}^{(p\p, p)}\in\R$, and 
$E_{(n\p, n)(s\p, s)}^{(p\p, p)}\geq0$ for non-vanishing 
$A_{(n\p, n)(s\p, s)}^{(p\p, p)}$
by Lemma \ref{Egeq0}. Hence each OPE-constant 
$C^{(p\p,p)}_{(n\p,n),(s\p,s)}$ converges
to a finite limit as $m\rightarrow\infty$. 
In fact, the properties of basic monomial systems and \req{minihom}
directly imply
\blem{mini1&6}
For the sequence $(\MMM_m,f_m^j)$,
Conds.\ \mb{\ref{opeconv}} and \mb{\ref{bigrading}} -- \mb{\ref{modeconv}} 
of Sect.\ \mb{\ref{seli}} hold. 
\elem
To meet Def.\ \ref{fulconv}, we need the more general
\blem{mini4pt}
For the sequence $(\MMM_m,f_m^j)$, all $n$-point functions on $\PP^1$ 
converge with respect to the $f_m^j$ as real analytic functions away from 
the partial diagonals,
with the standard behaviour near the singularities \mb{(}see App. 
\mb{\ref{cfts}}\mb{)}.
\elem
\bpr
By  Lemma \ref{mini1&6}, all structure constants of the Virasoro algebra
converge as $m\rightarrow\infty$. 
It will therefore suffice to prove convergence
of those $n$-point functions which only contain  primaries $|r,s\ket_m$, 
since all others can be obtained from them  by
application of differential operators with coefficients depending polynomially
on the structure constants of the Virasoro algebra.
Let $V_{(r,s)}^{m}(z,\ol{z})$ denote the field which creates $|r,s\ket_m$
as in \req{in}. By \cite{dofa84,dofa85}, an $n$-point function 
$$
\bra0|V_{(r_1,s_1)}^m(z_1;\ol z_1)\cdots V_{(r_n,s_n)}^m(z_n;\ol z_n) |0\ket_m
$$
on $\PP^1$ is a bilinear combination of a finite 
($m$-independent) number of specific
conformal blocks (see \req{coulombblock}) with coefficients 
given by OPE-constants. Since by Lemma \ref{mini1&6} all
OPE-constants converge as $m\rightarrow\infty$, it remains to prove
that the conformal blocks converge. To this end we use their Feigin-Fuks 
integral representations for $\MMM_m$. In particular,
we employ the Coulomb-gas formalism, 
i.e.\ a BRST construction of the $\VV_{(r,s)}^{m}$
(see \cite{fe89}), which is
adequate since the OPE-constants in $\MM_m$ have been calculated by
this technique in the first place \cite{dofa85a}. In fact,
the correction \cite{fe89err} to \cite[(3.14)]{fe89} ensures that the BRST
charges remain well-defined operators as $m\rightarrow\infty$,
yielding the Coulomb-gas description valid in our limit.

Recall (see, e.g., \cite{fe89,ags89}) that in the Coulomb-gas formalism the 
$\VV_{(r,s)}^{m}$ are obtained by a BRST construction from 
\textsc{charged Fock spaces}, built by the action 
of the Heisenberg algebra on $|r,s\ket_m$.
In particular, primary fields of $\MM_m$ are given by BRST invariant
operators with screening charges, such that $U(1)$ representation theory
can be used to calculate the $n$-point functions. 
That is, in an $n$-point function the field $V_{(r,s)}^m(z,\ol z)$
can be represented in terms of products of  holomorphic
\textsc{screened vertex operators}
\beqn\el{screen}
V^{i,j}_{(r,s)}(z)&:=&
\oint du_1\cdots \oint du_i 
\oint dv_1\cdots \oint dv_j \\
&&\qquad V_{\alpha^m_{(r,s)}}(z)
V_{\alpha^m_+}(u_1)\cdots V_{\alpha^m_+}(u_i)
V_{\alpha^m_-}(v_1)\cdots V_{\alpha^m_-}(v_j)\,\nonumber
\eeqn
and their antiholomorphic counter-parts.
Here, each $V_{\alpha}$ denotes the holomorphic part of
a vertex operator of charge $\alpha$ as in 
Sect.\ \ref{torex}: $V_{|Q;\ol Q\ket}(z,\ol z)=V_Q(z)V_{\ol Q}(\ol z)$,
and
$$
\alpha^m_\pm=\pm\left(\inv[m]{m+1}\right)^{\pm 1/2},\qquad 
\alpha^m_{(r,s)}:=\inv{2}((1-r)\alpha^m_++(1-s)\alpha^m_-)\,.
$$
Each conformal block is proportional to some
\beqn\el{coulombblock}
\oint_{C_1}\! du_1\cdots \oint_{C_N}\! du_N 
\oint_{S_1}\! dv_1\cdots \oint_{S_M}\! dv_M 
\bra0| V_{\alpha^m_1}(z_1)\cdots V_{\alpha^m_n}(z_n)
V_{-2\alpha^m_0}(\infty)\qquad\quad\\
V_{\alpha^m_+}(u_1)\cdots V_{\alpha^m_+}(u_N)
V_{\alpha^m_-}(v_1)\cdots V_{\alpha^m_-}(v_M)|0\ket_{\PP^1}\nonumber
\eeqn
with $\alpha_i^m:=\alpha_{(r_i,s_i)}^m$. Here, $M$ and $N$ are determined by 
the $r_i$, $s_i$, only, such that
the explicit numbers of screening charges which have to be
introduced  is independent
of $m$. 
The choice of integration
contours $C_i$, $S_i\subset\PP^1\backslash\{z_1,\ldots,z_n\}$ 
in \req{coulombblock} 
determines the specific conformal block and is independent of $m$. 
This yields the description of conformal blocks by representations
with screened vertex operators valid in our limit. 
By \cite{dofa84,dofa85}
the contours can be chosen in such
a way that the minimal distance between them as well as the minimal
distance between the contours and the $z_i$ is bounded away from zero by a 
constant. Since the integrand of \req{coulombblock} is the well-known
$n$-point function of vertex operators for the free bosonic theory,
see \req{toruscorr},
it therefore converges uniformly on the
integration domain implying that limit and integration can be interchanged.
Hence the integral of the limit function is  well-defined because
the integration domain is compact and does not hit singularities of the 
integrand.
\epr
Combining the above results, we find
\bprop{miniconv}
The sequence $(\MMM_m,f_m^j)$ of unitary Virasoro minimal models converges 
fully to a  limit $\MMM_\infty$ according to Def.\ \mb{\ref{fulconv}}.
\eprop
\bpr
In view of Lemmas \ref{mini1&6} and \ref{mini4pt} and by Def.\ \ref{fulconv}
it only remains to be shown that Cond.\ \ref{truncconv} of Sect.\ \ref{seli}
holds. We set
$$
\fa r,s\in\NN-\{0\}\colon\quad\quad
|r,s\ket_\infty := f_m^\infty( |r,s\ket_m ).
$$
By \req{minilimweight} we have
\beq\el{minizero}
\HH^\infty = \HHH_{0,0}^\infty 
= \span_\CC \left\{ |r,r\ket_\infty \bigm| r\in\NN-\{0\} \right\}.
\eeq
Then by Lemma \ref{modeaction} for all $r,s\p,s\in\NN-\{0\}$
and $h=\ol h=(s\p-s)^2/4$, the OPE-constant
$C^\infty\left(\psi^*,|r,r\ket_\infty, |s\p,s\ket_\infty \right)$ must
vanish for every primary 
$\psi\in\HHH^\infty$ with $\psi\not\in\wt\HH^\infty_{h,\ol h}$.
This is directly confirmed 
by Lemma \ref{Egeq0}. Moreover, \req{minifusion} implies that
$C^m\left( (|p\p,p\ket_m)^*, |r,r\ket_m, |s\p,s\ket_m \right)$ vanishes for 
all $m$ unless $|r-s^{(\prime)}|+1\leq p^{(\prime)}\leq
\min\{r+s^{(\prime)}-1, 2m-1-r-s^{(\prime)}\}$ and
$p^{(\prime)}+r+s^{(\prime)}\equiv1(2)$. This restricts $p$ and $p\p$ to a 
finite number of possibilities
as $m\rightarrow\infty$, implying Cond.\ \ref{truncconv} of 
Sect.\ \ref{seli}. In fact,  a straightforward calculation using
\req{mutilde}-\req{e} shows
$$
C^{(p,p)}_{(r,r)(s,s)} = 1 \quad
\mbox{ for } |r-s|+1\leq p\leq r+s-1,\quad p+r+s\equiv1(2)\,.
$$
Therefore,
\beq\el{cheby}
\fa r,s\in\NN-\{0\}\colon\quad\quad
|r,r\ket_\infty\boxast |s,s\ket_\infty
=\sum_{\stack[p=|r-s|+1,]{p+r+s\equiv 1(2)}}^{r+s-1}|p,p\ket_\infty\,.
\vspace*{-1em}
\eeq
\epr
Note that although we have made many choices in our construction above, 
the actual structure of the limit $\MMM_\infty$ is independent of those 
choices. This is largely due to the fact that 
Conds.\ \mb{\ref{opeconv} -- \ref{truncconv}} of Sect.\ \ref{seli}
are rather restrictive. For example, recall the two basic singular 
vectors of levels $rs$ and $(m-r)(m+1-s)$ in the Verma module built
on the primary $|r,s\ket_m$. The latter state does not play a 
role in the limit, since its level becomes infinite as $m\rightarrow\infty$.
In the language of our basic monomial systems of Def.\ \ref{monomials}
it always lies above the energy bound. On the other hand, the singular
vector at level $rs$ has dimension ${1\over 4}(r+s)^2$ and implies
that there also is a polynomial $P_{rs}$ of degree $rs$ such that
$P_{rs}(L_n)|r,s\ket_\infty=0$. 
Since up to normalization, $P_{rs}$ is uniquely determined by $r,\, s$, and by
the structure constants
of $\vir_{c=1}$, the dependence
on the choice of the basic monomial system
$(\PPP_{(r,s)}^{m}(N))_{N\in\NN}$ 
drops out in the limit.
\brem{ninfty}
In contrast to the examples discussed in Sect.\ \ref{exes},
for the limit of $(\MMM_m,f_m^j)$ we obtain additional null vectors,
i.e.\ $\NNN^\infty\neq\{0\}$. This is due to the fact that the conformal
weights of lwvs $|r,s\ket_m$ converge to $(r-s)^2/4$,  
while the central charge converges to $1$ (see \eq{minilimweight}). By the
above discussion of singular vectors,
the characters of the limit Virasoro modules before quotienting out
the null vectors
are  given by
$$
{1\over\eta(q)}\left(q^{(r-s)^2/4}-q^{(r+s)^2/4}\right)\,.
$$
But at $c=1$ there are null vectors \eq{sing} in the Fock spaces
built on lwvs with conformal weight $h$, $2\sqrt h\in\NN$, and the limit
characters decompose into characters \eq{char} of irreducible representations
of the Virasoro algebra of central charge $c=1$,
$$
{1\over\eta(q)}\left(q^{(r-s)^2/4}-q^{(r+s)^2/4}\right)
=\sum_{k=0}^{{\rm min}\{r,s\}-1}\chi_{{1\over 4}(|r-s|+2k)^2}\,.
$$
Those submodules of $\KKK^\infty$, where $\KKK^\infty/\NNN^\infty=\HHH^\infty$ 
as in \req{kdef}, which correspond to
lwvs at positive levels consist
of limit-null vectors, whose norms converge to zero 
for $m\rightarrow\infty$. For instance, the norm of 
$f_m^\infty(L^m_1|r,r\ket_m)$, $r>1$, is given by the limit of
\beq\el{scale}
|L_1^m|r,r\ket_m|^2=
C^m((L_1^m|r,r\ket_m)^*,\Omega^m,L_1^m|r,r\ket_m)=2h^m_{(r,r)}
\stackrel{m\rightarrow\infty}{\sim}{r^2-1\over 2m^2}\,.
\eeq
Thus this vector and all its descendants are elements of $\NNN^\infty$.

As alluded to in Rem.\ \ref{simcon}.\ref{simconi}, 
the quotienting out by
additional null vectors in \req{kdef}
spoils the factorization properties
of the limit-correlation functions on $\PP^1$. However, 
as pointed out in \cite[Sect.~3.1.1]{grw01} it is possible
to modify the definition of the $f_m^j$ in such a way that
$\NNN^\infty=\{0\}$. This is achieved by \textsc{scaling up the additional
null vectors}. For example, we can set
$$
\wt{f}_m^j(L^m_1|r,r\ket_m):={|L^m_1|r,r\ket_m|\over|L^j_1|r,r\ket_j|}
f_m^j(L^m_1|r,r\ket_m)\,.
$$
Indeed, homomorphisms $\wt{f}_m^j$ can be constructed in such a way that 
$(\MMM_m,\wt{f}_m^j)$ is a sequence of CFTs with stable Virasoro algebras
according to Defs.\ \ref{series}, \ref{wstable}, 
which does not lead to additional
null vectors as $m\rightarrow\infty$. However, the modification 
$f_m^j\mapsto\wt{f}_m^j$ could destroy the 
convergence of correlation functions.
That this is not the case, and that in fact the sequence 
$(\MMM_m,\wt{f}_m^j)$ of 
CFTs is fully convergent follows from the Coulomb-gas formalism. 
In the proof of Lemma \ref{mini4pt} we have already pointed out
that the expressions obtained from the Coulomb-gas formalism remain 
well-defined as $m\rightarrow\infty$. Recall that the Fock space
representation of elements of $\VV_{(r,s)}^{m}$ in the Coulomb-gas
representation is formally obtained from an 
action of the positive modes of the Heisenberg algebra on $|r,s\ket_m$. 
Hence singular vectors with respect to the  action of $\vir_{c_m}$
on $|r,s\ket_m$ are automatically  zero, see Ex.\ \ref{c1example} for an 
illustration. 
Namely, each singular vector $\nu\in\KKK^\infty$ is of the form
$\nu=f_m^\infty(\nu^m)$ with
$\nu^{m} = Q^{m}|r,s\ket_m$, where
$Q^{m}$ is an operator on $\HHH^m$ which can formally be
written as polynomial in the positive modes of the Heisenberg algebra,
with each coefficient converging to zero as $m\rightarrow\infty$.
In fact, each coefficient is a power series in $\inv{m}$ with vanishing
constant term. Therefore, our rescaling yields
$\wt\nu=\wt f_m^\infty(\nu^m) = \wt Q |r,s\ket_\infty$
with an operator $\wt Q$ on $\HHH^\infty$ which again can formally be  obtained
as polynomial in the positive modes of the Heisenberg algebra.
Hence all correlation functions involving $\wt Q |r,s\ket_\infty$ also 
converge.

This way, we can obtain a limit of the A-series of Virasoro minimal 
models whose 
correlation functions on $\PP^1$ have the usual factorization properties.
As a model case, in Lemma \ref{rescale} we 
also show by direct calculation that no divergences are introduced
in $C(|p\p,p\ket_\infty^*, L_1|r,r\ket_\infty, |s\p,s\ket_\infty )$ when
the singular vectors $L_1|r,r\ket_\infty$ are scaled up.
\vspace*{-1em}
\erem
\bex{c1example}
As in Ex.\ \ref{circle} let
$\CCC_{R_i},\, i\in\NN$, denote the CFT with central charge $c=1$ 
that describes a boson compactified on a circle of radius $R_i$, here with
$R_i:=1+{\sqrt2\over i}$. See in particular \req{u1chargelattice}
for notations.
According to \req{c1gen} -- \req{char}
the Verma module built on each 
$|Q_{m,n}^i;\ol Q_{m,n}^i\ket_i,\, m,n\in\Z$,
by the action of the Virasoro algebra is irreducible 
if $(m,n)\neq(0,0)$ because all our $R_i^2$
are irrational.
We can therefore define a direct system $(\HHH^i,\widehat f_i^j)$ by
$$
\fa (m,n)\neq(0,0)\colon\quad
\widehat f_i^j\left( P(L_k^i;\ol L_{\ol k}^i)\, 
|Q_{m,n}^i;\ol Q_{m,n}^i\ket_i \right)
:= P(L_k^j;\ol L_{\ol k}^j)\, |Q_{m,n}^j;\ol Q_{m,n}^j\rangle_j,
$$
where $P$ denotes a polynomial in the $L_k,\ol L_{\ol k},\, k,\ol k>0$.
In the vacuum sector we use
$$
\widehat f_i^j\left( P(a_k^i;\ol a_{\ol k}^i)\, 
|0;0\ket_i \right)
:= P(a_k^j;\ol a_{\ol k}^j)\, |0;0\ket_j
$$
as in \req{torussequence}, where as usual $a_k^i,\,\ol a_{\ol k}^i$
denote the modes  of the generators $j,\, \ol\j$
of $\fu(1)\oplus \ol{\fu(1)}$ in $\CCC_{R_i}$.
One checks that this gives a convergent sequence $(\HHH^i,\widehat f_i^j)$
of CFTs, but the direct
limit $\widehat\KKK^\infty$ possesses null vectors in
$\widehat\NNN^\infty$, where 
$\widehat\HHH^\infty=\widehat\KKK^\infty/\widehat\NNN^\infty$.
For example,
\beq\el{nullex}
\nu^i := \left( L_2^i - (L_1^i)^2 \right) 
|Q_{1,0}^i;\ol Q_{1,0}^i\ket_i,
\eeq
where for $|Q^i_{1,0};\ol Q^i_{1,0}\ket_i$ we have
$Q^i_{1,0}=\ol Q^i_{1,0} 
= \inv{\sqrt2R_i} \stackrel{i\rightarrow\infty}{\longrightarrow} \inv{\sqrt2},
\, h_i=\ol h_i 
= \inv{4(R_i)^2} \stackrel{i\rightarrow\infty}{\longrightarrow} \inv{4}$,
which gives a null vector $\nu=f_i^\infty(\nu^i)$.

On the other hand, in Sect.\ \ref{torex} we have already constructed
a fully convergent sequence $(\HHH^i, f_i^j)$ of CFTs via
$$
f_i^j\left( P(a_k^i;\ol a_{\ol k}^i)\, 
|Q_{m,n}^i;\ol Q_{m,n}^i\ket_i \right)
:= P(a_k^j;\ol a_{\ol k}^j)\, |Q_{m,n}^j;\ol Q_{m,n}^j\ket_j
$$
with $P$ as above.
Now the limit is the $\fs\fu(2)_1$ WZW model, i.e.\ a full fledged well-defined
CFT. Note that in terms of the latter Fock space representation, $\nu^i$
in \req{nullex} is  given by
$$
\nu^i = \inv{2} (a_1^i)^2 \left( 1-2(a_0^i)^2 \right)
|Q_{1,0}^i;\ol Q_{1,0}^i\ket_i
= \inv{2} (a_1^i)^2 \left( 1-(R_i)^{-2} \right)|Q_{1,0}^i;\ol Q_{1,0}^i\ket_i.
$$
Hence $f_i^\infty(\nu^i)\stackrel{i\rightarrow\infty}{\longrightarrow}0$
in $\HHH^\infty=\KKK^\infty$.
The direct system $(\HHH^i,f_i^j)$ yields the vectors
$(a_1)^2  |\inv{\sqrt2};\inv{\sqrt2}\ket_\infty$, 
$a_2  |\inv{\sqrt2};\inv{\sqrt2}\ket_\infty$
as linearly independent elements of $\HHH^\infty_{2+{1\over4},{1\over4}}$,
where the combination $\left(a_2  -\sqrt2 (a_1)^2\right)  
|\inv{\sqrt2};\inv{\sqrt2}\ket_\infty$ is Virasoro
primary. For $(\HHH^i,\widehat f_i^j)$, the corresponding vectors
$(L_1)^2 |\inv{\sqrt2};\inv{\sqrt2}\ket_\infty , \, 
L_2 |\inv{\sqrt2};\inv{\sqrt2}\ket_\infty $  differ by
the null vector $\nu$ and are thus identified in 
$\widehat\HHH^\infty_{2+{1\over4},{1\over4}}$. However, the 
above directly implies how the $\widehat f_i^j$ can be redefined 
by scaling up the additional null vectors, and then both limits give the
same well-defined CFT.
\eex
To approach the full limit structure
obtained on the pre-Hilbert space 
$\HHH^\infty$, recall that in the proof of Lemma \ref{mini4pt}
and in Rem.\ \ref{ninfty} we have argued that the correlation functions
in $\MMM_\infty$ are adequately described in terms of the 
Coulomb-gas formalism. A closer study of this formalism also shows that
it should be possible to represent the operator product algebra of the 
limit within the $\fs\fu(2)_1$
WZW model\footnote{This is in accord with \cite[p.~655]{dvv88}, where it
is stated that the $\fs\fu(2)_1$ WZW model ``in some sense can be regarded
as the limit $c\rightarrow1$ of the discrete unitary series''.}. Namely,
as follows from performing the limit in \req{screen},
the operator corresponding to  $|r,s\ket_\infty$ in a 
given correlation function
should be represented by a combination of the 
left-right symmetric  $\fu(1)$ vertex 
operator $V_{|Q={r-s\over\sqrt2};\ol Q={r-s\over\sqrt2}\ket}(z,\ol z)$ 
of the circle model $\CCC_{R=1}$ and 
the zero modes $Q_\pm$ of the holomorphic fields $J_\pm(z)$ which
create $|Q;\ol Q\ket=|\pm\sqrt2;0\ket$ as in \req{zero}, along with their
antiholomorphic counterparts.
\subsection{Geometric interpretation of 
$\MMM(m,m+1)_{m\rightarrow\infty}$}\el{gi1}
Note that by \eq{minilimweight} the limit $\MMM_\infty$ of the
sequence of unitary Virasoro minimal models
has an infinite degeneracy of every
energy level. This means that we cannot  interpret
$\MMM_\infty$ as  part of a well-defined CFT. However, 
the degeneration of the vacuum
sector allows us to apply the techniques introduced in 
Sect.\ \ref{geomintlim} and to find a geometric interpretation of the limit.
Indeed, in Prop.\ \ref{minigeomint} below we 
identify the algebra $\AAA^\infty$ obtained from 
$\HH^\infty$ by \req{cheby} with the algebra generated by 
the \textsc{Chebyshev
polynomials of the second kind}, i.e.\ with 
the algebra of continuous functions 
on an interval: 
\blem{chebylem}
For every $r\in\NN-\{0\}$, let $U_r$ denote the $r^{\mb{\small{}th}}$ 
\textsc{Chebyshev polynomial of the second 
kind}:
\beq\el{chebyshev}
U_r(\cos x) := {\sin(x r)\over \sin x}\,,\quad\quad
x\in[0,\pi]\,.
\eeq
Then $U_r(t=\cos x)$ is a polynomial of degree $r-1$ in $t\in[-1,1]$,
and the $U_r(t)$ form an orthonormal system of polynomials with
respect to the scalar product
\beq\el{scp}
\bra f,g\ket_\omega := \int_{-1}^1 \ol{f(t)}g(t)\, \omega(t) dt, \quad
\omega(t):= \inv[2]{\pi} \sqrt{ 1-t^2 }\,.
\eeq
Moreover, the 
Chebyshev polynomials of the second kind  obey the recursion relation
\beq\el{chrec}
\fa r,s\in\NN-\{0\},\,\fa t\in[-1,1]\colon\quad\quad
U_r(t) U_s(t)
=\sum_{\stack[p=|r-s|+1,]{p+r+s\equiv 1(2)}}^{r+s-1}U_p(t)\,.
\eeq
\elem
The proof of Lemma \ref{chebylem} is a straightforward calculation,
see e.g. \cite[Problems 3.1.10(a)]{he82}. 
Note in particular that this lemma implies
\beq\el{delta}
\fa t\in[-1,1],\,\fa x\in[0,\pi]\colon\qquad
\delta_{\cos x}(t)
=\sum_{p\p=1}^\infty U_{p\p}(t)U_{p\p}(\cos x)\,\inv[2]{\pi}\!\sin x\,.
\eeq
We are now in the position to give a geometric interpretation for our limit
according to Def.\ \ref{limgeomint}:
\bprop{minigeomint}
The limit $\MMM_\infty$ of the sequence $(\MMM_m,f_m^j)$ of unitary
Virasoro minimal models has a geometric interpretation on the interval
$[0,\pi]$ equipped with the dilaton-corrected 
metric $g(x)=\inv[4]{\pi^2}\sin^4\!x$ and dilaton $\Phi$ such that
$e^{2\Phi(x)}=\inv[2]{\pi}\sin^2\!x$
for $x\in[0,\pi]$. 
\eprop
\bpr
As a first step, we need to construct a spectral pre-triple
$(\HH^\infty, H^\infty, \AAA^\infty)$ from our limit $\MMM_\infty$
according to Def.\ \ref{limgeomint}. In fact, by Def.\ \ref{zeromodes},
$\HH^\infty$ is given in \req{minizero}, and $\AAA^\infty$ is the 
associated zero-mode algebra specified in \req{cheby}. Moreover,
\req{miniweight} shows that on $\HH^\infty$, according to 
Def.\ \ref{limgeomint}, 
we need to set
\beqn\el{hinfty}
\fa r\in\NN-\{0\}\colon\quad
\lambda_r^2 := \lim_{m\rightarrow\infty} 
m^2 \left( h^m_{(r,r)} + \ol h^m_{(r,r)} \right) &=& \inv[r^2-1]{2}, 
\nonumber\\
H^\infty |r,r\ket_\infty &:=& \inv[r^2-1]{2}\,|r,r\ket_\infty.
\eeqn
Comparison of \req{cheby} with
\req{chrec} shows that $\AAA^\infty$ agrees with the algebra generated by the
Chebyshev polynomials of the second kind. Here, similarly to the discussion
of Chebyshev polynomials of the first kind at the end of Sect.\ \ref{orbex},
we view the $U_r$ as functions $x\mapsto U_r(\cos x)$ with $x\in[0,\pi]$.
Therefore, $\ol{\AAA^\infty}$ 
can be identified with the algebra of smooth functions
on $[0,\pi]$, and
\req{scp} shows that $[0,\pi]$ is equipped with the dilaton-corrected
metric $g$ with 
$\dvol_g=\sqrt{g(x)}dx=\inv[2]{\pi}\sin^2\!x\,dx$ as claimed.
By the discussion in Sect.\ \ref{sptr} it therefore remains to identify
$H^\infty$ in \req{hinfty} with the generalized Laplacian
$H$ as defined in \req{genlaplace} and to read off the dilaton $\Phi$.
To this end we use the
characterization \req{lapl}, that is, for all $f,h\in C^\infty([0,\pi])$
we must have $\bra f, 2H h\ket_\omega=\bra f^\prime,h^\prime\ket_\omega$.
Since
\beqns
\bra f, 2H h\ket_\omega \stackrel{!}{=} \bra f^\prime,h^\prime\ket_\omega
&\reeq{scp}&
\int_0^\pi \ol{ f^\prime(x) } h^\prime(x) \inv[2]{\pi} \sin^2\!x\,dx\\
&=& -  \int_0^\pi \ol{ f(x) } 
\inv[d]{dx}\left(\sin^2\!x\, h^\prime(x) \right)\,\inv[2]{\pi} dx\,,
\eeqns
we deduce that $2H=-\sin^{-2}\!x\inv[d]{dx}\sin^2\!x\inv[d]{dx}$, and
thus $\wt{g}(x)\equiv1$ and $e^{2\Phi(x)}=\inv[2]{\pi}\sin^2\!x$.
With \req{chebyshev} one now checks that $H U_r(\cos x)$ 
$=\inv[r^2-1]{2}\,U_r(\cos x)$,
in perfect agreement with \req{hinfty}.
\vspace*{-1em}
\epr
\brem{minimetric}
The distance functional, which is associated to the dilaton-corrected metric 
$g(x)=\inv[4]{\pi^2}\sin^4\!x$ on the 
interval $[0,\pi]$ determined in Prop.\ \mb{\ref{minigeomint}}, is
$$
\fa a,b\in[0,\pi]\colon\qquad
d(a,b) = \inv{2\pi} \left| \xi(a)-\xi(b) \right|\quad
\mbox{ with }\quad 
\xi(\tau):=2\tau-\sin(2\tau).
$$
Here, $\xi(\tau/2)$ is the $x$-coordinate of a regular cycloid in 
Cartesian
coordinates. That is, if we consider a unit wheel which roles horizontally 
at unit speed, then
$2\pi\, d(0,\tau/2)$ measures the distance that the
point $(2,0)$ on the wheel travels horizontally within the time $\tau$.
\vspace*{-1em}
\erem
\brem{sigmamodel}
On the level of topological manifolds, our geometric interpretation
of $\MMM_\infty$ on an interval could have been predicted from the 
discussion in \cite[\S3.3]{frga93}. Namely, the unitary Virasoro 
minimal model $\MMM_m$ can be obtained by an $\fs\fu(2)$-coset 
construction:
$$
\MM_m \quad\longleftrightarrow \quad
{\fs\fu(2)_{m-2}\oplus \fs\fu(2)_1\over \fs\fu(2)_{m-1}}\;.
$$
In this language, the labels $r$ and $s$ in $|r,s\ket_m$ refer to the
relevant representations of $\fs\fu(2)_{m-2}$ and $\fs\fu(2)_{m-1}$,
respectively. Loosely speaking, since only states with $r=s$ enter
in our zero-mode algebra, our geometric interpretation can be expected
to yield a semiclassical limit of the coset WZW model
$\fs\fu(2)_m/\fs\fu(2)_m$ as
$m\rightarrow\infty$. That is, by \cite[(3.25)-(3.26)]{frga93} the
limit should have a geometric interpretation on the space 
${\rm SU}(2)/{\rm Ad}({\rm SU}(2))\simeq T/W$ with $T$ the Cartan subgroup
and $W$ the Weyl group of ${\rm SU}(2)$. Indeed, with $T=U(1)$,
$W=\ZZ_2$ we obtain $T/W\simeq[0,\pi]$.
An analogous observation was already made in \cite{ruwa02}. 
There, it was
also pointed out that\footnote{according to J.~Cardy} the geometric 
interpretation of $\MMM_\infty$ on the interval fits nicely with an analysis
of the qualitative Landau-Ginzburg description for the minimal models
$\MMM_m$ \cite{za86a}:
As $m\rightarrow\infty$, the Landau-Ginzburg potential approaches a 
square well with walls at $X=\pm1$, forcing the scalar field $X$ of
the Landau-Ginzburg theory to take values on the interval $[-1,1]$.

The sigma model metric, in principle, could  also be obtained by a 
gauged WZW model construction as was done in \cite{mms01} in the case of 
$\fs\fu(2)_k/\fu(1)$.
\erem
\vspace*{-1em}
\brem{watts}
Apart from the direct limit construction
studied above, one can introduce  
other sensible limits for the family $\MMM_m$
as $m\rightarrow\infty$, similarly to Rem.\ \ref{decomplim}. 
In particular, if there is a system of epimorphisms
$f_m\colon\,\wt\HHH^\infty\longrightarrow\HHH^m$ such that all
limits
$$
\bra 0|\varphi_1(z_1,\ol z_1)\ldots \varphi_n(z_n,\ol z_n)|0\ket
:=\lim_{m\rightarrow\infty}
\bra 0|f_m(\varphi_1)(z_1,\ol z_1)\ldots f_m(\varphi_n)(z_n,\ol z_n)|0\ket_m
$$
of $n$-point functions exist, then $\wt\HHH^\infty$ can be interpreted
as pre-Hilbert space of a limit theory $M_\infty$. We believe that this
is the structure underlying the ideas of \cite{grw01,ruwa01,ruwa02}.
Indeed, there
the authors find a limiting pre-Hilbert space of the form 
$$
\wt\HHH^\infty 
= \bigoplus_{r\in\RR^+-\NN} \VV_{r^2/4}^{\,\mb{\scriptsize gen}},
$$
where for $h\in\RR^+$ with $2\sqrt h\not\in\NN$, 
$\VV_{h}^{\,\mb{\scriptsize gen}}$ denotes the generic representation of the
Virasoro algebra $\vir_{c=1}$ with character \req{c1gen}. Analogously to
the situation in Rem.\ \ref{decomplim}, no degeneration phenomena occur
in this procedure, and the limit $M_\infty$
is conjectured to be part of a well-defined
non-rational CFT with central charge $c=1$, which has an interesting
resemblance to Liouville theory. Evidence for this conjecture is given in
\cite{ruwa01}, where in particular crossing symmetry is proven in some 
model cases. 

It seems that the two limits $\MMM_\infty$ and $M_\infty$ are complementary
in many respects: For instance, the representation content of $\wt\HHH^\infty$
is complementary to the one we have found in $\HHH^\infty$,
see \req{minilimweight}. Moreover, while the limit $M_\infty$ seems
to be a well-defined CFT, $\MMM_\infty$ shows the degeneration phenomena
discussed above, which allow to extract a geometric interpretation
from the limit structures.

A third approach to limiting processes is taken in \cite{fusc96}. There,
limits of WZW models at infinite level are introduced by means of 
\textsc{inverse limits} instead of direct limits. While our direct limit
construction takes advantage of those structures which the pre-Hilbert spaces
of minimal models $\MMM_m$ share
at $m\gg0$ and for sufficiently low conformal
dimensions, the inverse limit construction of \cite{fusc96} allows to interpret
the collection of fusion rings of $\fg-$WZW models as a category and to
identify a projective system in it. Clearly, as mentioned above, we cannot
view the family $(\MMM_m)_{m\in\NN-\{0,1\}}$ of minimal models as
direct system of CFTs with the natural ordering induced by $\NN$. The same
is true already on the level of $\fg-$WZW models; however, in \cite{fusc96}
a suitable non-standard partial ordering is found for the latter. Whether
geometric interpretations of $(\MMM_m)_{m\rightarrow\infty}$
with the expected properties arise from this construction remains to be seen.

We have not worked out the details of an application of our techniques to
$\fg$-WZW models at infinite level. However, we expect that the results
of \cite{frga93} should tie in naturally thus leading to a direct limit
construction with the expected geometric interpretation on the group manifold
$G$.
\erem
The results of Prop.\ \ref{minigeomint} and Rem.\ \ref{sigmamodel}
imply that under the coordinate change $t=\cos x$,
our limit $\MMM_\infty$ has a geometric interpretation  on the unit 
interval. By the ideas of \cite{frga93} this also means that
each unitary Virasoro minimal model $\MM_m$ with $m\gg0$
can be regarded as 
sigma model on the unit interval. We therefore expect to gain some 
insight\footnote{strictly speaking, after extending our constructions
of Sect.\ \ref{lim} to the boundary sector}
into the shape of 
the D-branes in this bulk-geometry by considering the bulk-boundary 
couplings for $m\gg0$. 

Recall that for each $\MMM_m$ we use the diagonal, that is the
charge conjugation invariant
partition function. Hence the Ishibashi
states $|p\p,p\ket\!\ket_m$ are labeled by $(p\p,p)\in\NNN_m$ with
$\NNN_m$ as in \req{z2choice}. Moreover, each $(r,s)\in\NNN_m$
labels a boundary condition. Its bulk-boundary coupling
with respect to $|p\p,p\ket\!\ket_m$  is 
given by
\beq\el{bbc}
\begin{array}{rcl}
\ds
B_{(r,s)}^{(p\p,p)}
&=&\ds{S_{(r,s)(p\p,p)}\over \sqrt{S_{(1,1)(p\p,p)}}}\\[4pt]
&=&\ds(-1)^{(r+s)(p\p+p)}
\left(\inv[8]{m(m+1)}\right)^{\!\!{1\over 4}}
{\sin\left({\pi r p\p\over m}\right)\sin\left({\pi s p\over m+1}\right)
\over\sqrt{\sin\left({\pi p\p\over m}\right)
\sin\left({\pi p\over m+1}\right)}}\,.
\end{array}
\eeq
In order to investigate  the geometry of the D-branes, we can restrict to the 
couplings of the bulk-fields $(p\p,p\p)$ which by Prop.\ \ref{minigeomint}
correspond  to the Chebyshev polynomials $U_{p\p}$ of the second kind. This
means that we will focus on the bulk-boundary couplings $B_{(r,s)}^{(p\p,p\p)}$
and the
bulk-boundary coupling support functions
$$
f_{(r,s)}^m(t)
:=
\inv[2]{\pi}\left(\inv[8]{m(m+1)}\right)^{\!\!\!-{1\over 4}}\;
\sum_{p\p=1}^{m-1}U_{p\p}(t)B_{(r,s)}^{(p\p,p\p)}.
$$
In the above definition of $f_{(r,s)}^m$ we have introduced the appropriate
pre-factor corresponding to the rescaling  in \req{hinfty} by
hand. In order to analyze $f_{(r,s)}^m(t)$ for $m\gg0$, we  
use $t=\cos x$ as before, and divide the domain of definition of $x$,
the interval $[0,\pi]$, equidistantly. That is, we set
$$
\fa (r,s)\in\NNN_m\colon\qquad
x_r:=\inv[r\pi]{m},\quad \wt x_s:=\inv[s\pi]{m+1}.
$$
Note the following useful reformulation of \req{chebyshev} for all
$p,r\in\NN-\{0\}$:
\beq\el{trick}
U_r(\cos(x_p))\sin(x_p)
= \sin(r x_p)=\sin(p x_r) = U_p(\cos(x_r))\sin(x_r),
\eeq
and analogously for $\wt x_p,\, \wt x_r$. Using $x_r\approx\wt x_r$
for $m\gg0$, we therefore find:
\beqns
f^m_{(r,s)}(t)
&\reeq{bbc}&
\inv[2]{\pi}\sum_{p\p=1}^{m-1}
U_{p\p}(t)\, {\sin(r x_{p\p})\sin(s\wt x_{p\p})\over 
\sqrt{\sin(x_{p\p})\sin(\wt x_{p\p})}  }\\
&\reeq{trick}&
\inv[2]{\pi}\sum_{p\p=1}^{m-1}
U_{p\p}(t)\, U_{r}(\cos(x_{p\p}))\, U_{s}(\cos(\wt x_{p\p}))\,
\sqrt{\sin(x_{p\p})\sin(\wt x_{p\p})}\\
&\stackrel{m\rightarrow\infty}{\sim}&
\inv[2]{\pi}\sum_{p\p=1}^\infty 
U_{p\p}(t)\; U_{r}(\cos(x_{p\p}))\, U_{s}(\cos( x_{p\p}))\,\sin(x_{p\p})\\
&\reeq{chrec}&
\sum_{\stack[p=|r-s|+1,]{p+r+s\equiv 1(2)}}^{r+s-1}\;
\inv[2]{\pi}\sum_{p\p=1}^\infty 
U_{p\p}(t)\; U_{p}(\cos(x_{p\p}))\, \sin(x_{p\p})\\
&\reeq{trick}&
\sum_{\stack[p=|r-s|+1,]{p+r+s\equiv 1(2)}}^{r+s-1}\;
\inv[2]{\pi}\sum_{p\p=1}^\infty 
U_{p\p}(t)\; U_{p\p}(\cos(x_{p}))\, \sin(x_{p})\\
&\reeq{delta}&
\sum_{\stack[p=|r-s|+1,]{p+r+s\equiv 1(2)}}^{r+s-1}\;
\delta_{\cos(x_p)}(t)\,.
\eeqns
We interpret this calculation in form of
\brem{dbranes}
For the unitary Virasoro minimal models $\MMM_m$ at $m\gg0$,
the D-branes corresponding to stable
boundary states labeled by $(r,1)$ which are elementary 
in the sense of 
\cite{rrs00} and the D-branes corresponding to the unstable 
boundary states  $(1,s)$
can  be interpreted as being localized in the points 
$t=\cos(x_r)=\cos({\pi r\over m})$ and
$t=\cos(\wt x_s)=\cos\left({\pi s\over m+1}\right)$
on the interval $[-1,1]$, respectively. On the other hand,
D-branes corresponding to the unstable boundary states $(r,s)$
with $r\neq1$, $s\neq1$ are supported on a union of these points.
In view of Rem.\ \ref{sigmamodel} this is in accord with the general shape
of D-branes in coset models \cite{ga01,frsc01}.
\erem
\vspace*{-1em}
\section{Discussion}\el{disc}
To conclude, let us address some open questions arising from our 
investigations. Of course, there are several interesting 
unsolved problems concerning the degenerating limit $\MMM_\infty$
of the A-series of 
unitary Virasoro minimal models of Sect.\ \ref{kinfinity}. 
For example,
it would be interesting to gain more insight into the representation 
of  this limit within  the $\fs\fu(2)_1$ WZW model, as mentioned at the
end of Sect.\ \ref{c1ex}. In particular, there are two fusion 
closed subsectors in $\MMM_\infty$, corresponding to the states
$|r,1\ket_\infty,\, r\in\NN-\{0\}$, and $|1,s\ket_\infty,\,s\in\NN-\{0\}$, 
respectively. We 
expect them to have a comparatively simple description in terms of the 
$\fs\fu(2)_1$ WZW model, because no additional null vectors occur in the
corresponding Verma modules. Moreover, by acting with the zero
mode algebra $\AAA^\infty$ on one of these subsectors, one can generate
the entire limit pre-Hilbert space $\HHH^\infty$. Thus an understanding
of these subsectors should also allow some insight into the geometry of
the entire $\AAA^\infty$ module $\HHH^\infty$, for instance the 
fiber structure of the corresponding sheaf. Finally, one could try to
extract the non-commutative geometries from 
the Virasoro
minimal models at finite level which at infinite level reduce to
the limit geometry on the interval determined in Prop.\ \ref{minigeomint}. 

Next, a generalization of our discussion in Sect.\ \ref{kinfinity}
to WZW models and their cosets in general would be nice, e.g. to
the families of unitary super-Virasoro minimal models.

More generally, for all limits of degenerating sequences of CFTs,
it would be interesting to understand the compatibility of the
limit structures with the action of the zero mode algebra $\AAA^\infty$. 
In particular, 
the limit OPE-constants are $\AAA^\infty$ homogeneous and therefore
should be 
induced by a corresponding fiberwise structure on the sheaf with $\HHH^\infty$
as space of sections. It is likely that the entire limit can be
understood in terms of such fiberwise structures together with the 
$\AAA^\infty$ action. This is in accord with the results of
\cite{koso00}.

In fact, the zero mode algebra  would be an interesting object
to study in its own right, 
not least because there seems to be a relation to Zhu's algebra as 
mentioned in Sect.\ \ref{cfttripel}. 

Finally, it would be natural to extend our constructions to the 
boundary sector. This could allow a more conceptual understanding
of geometric interpretations of D-branes, for example in terms of the
K-theory of $\AAA^\infty$. 
\def\thesection{\Alph{section}}
\setcounter{section}{0}
\section{Properties of conformal field theories}\el{cfts}
In this Appendix, we collect some properties of CFTs   that are used
in the main text. Recall the Virasoro algebra $\vir_c$
at central charge $c$,
with generators $L_n,\, n\in\Z$,
\beq\el{vir}
\forall\,m,\,n\in\ZZ:\quad
[L_m,L_n]= (n-m) L_{m+n} + \inv[c]{12} (n^3-n) \delta_{m+n,0}.
\eeq
In a given CFT $\CCC=(\HHH,\,\ast,\,\Omega,\,T,\,\ol T,\,C)$,
the vacuum $\Omega\in\HHH$ and its dual $\Omega^*\in\check\HHH^*$
are characterized by
\beq\el{vacuum} 
\ast(\Omega)=\Omega;\quad
\Omega^*(\Omega)=1;\quad
\fa n\leq1: \;
L_n\Omega=\ol L_n\Omega=0;\;\;
L_n^\dagger(\Omega^*)=\ol L_n^\dagger(\Omega^*)=0.
\eeq
The map $\HHH\rightarrow\check\HHH^*,\;\psi\mapsto\psi^*$ of 
\req{adj} can be explained by the relation between 
our OPE-coefficients $C$ and the 
\textsc{$n$-point functions}
\beq\el{npt}
\HHH^{\otimes n}\ni\varphi_1\otimes\cdots\otimes\varphi_n
\quad\longmapsto\quad
\bra0| \varphi_1(z_1,\ol{z}_1)\ldots\varphi_n(z_n,\ol{z}_n)|0\ket_\Sigma
\eeq
of a CFT. Here, $\Sigma$ is a conformal surface, and the right hand
side of \req{npt} denotes a real analytic function
$\Sigma^n\backslash D\rightarrow\CC$ outside the partial diagonals
$D=\cup_{i,j} D_{i,j}$ with 
$D_{i,j}:=\{(z_1,\ldots,z_n)\in\Sigma^n\,|\,\,z_i=z_j\}$.
Moreover, the right hand side of \req{npt} possesses expansions
around the partial diagonals $D_{i,j}$:
\beq\label{fpsing}
\sum_{(r,\ol{r})\in R_{i,j}} 
a_{r\ol{r}}(z_1,\ol z_1;\ldots;z_{i-1},\ol z_{i-1};
z_{i+1},\ol z_{i+1};\ldots;z_n,\ol z_n)\;
(z_i-z_j)^r(\ol{z}_i-\ol z_j)^{\ol{r}}.
\eeq
Here, $R_{i,j}\subset\RR^2$ is countable, and only finitely many 
$a_{r\ol{r}}$ are non-zero for $r<0$ or $\ol{r}<0$. Furthermore, the 
$a_{r\ol{r}}$ themselves are linear combinations of $(n-1)$-point functions 
with OPE-coefficients as linear factors. Finally, 
the right hand side of
\req{npt} is invariant under permutation of the $\varphi_i(z_i,\ol z_i)$.
One says that the correlation functions constitute
a \textsc{representation of the OPE}.

It is a basic feature of CFTs that each state $\psi\in\HHH$ possesses an
\textsc{adjoint} $\psi^\dagger\in\HHH$ such that two-point functions
on the sphere $\Sigma=\CC\cup\{\infty\}=\PP^1$ encode the metric on $\HHH$:
\beq\el{adjoint}
\fa\chi,\psi\in\HHH:\quad
\bra\psi|\chi\ket 
= \lim_{w,\zeta\rightarrow0} 
\bra0|\psi^\dagger( \ol w^{\,-1},w^{-1})\chi(\zeta,\ol \zeta)|0\ket_{\PP^1}.
\eeq
Using conformal invariance one can determine $\psi^\dagger(z,\ol z)$ as
the image of $\ast\psi(z,\ol z)$ under the transformation
$f:z\mapsto 1/z,\; \ol z\mapsto1/\ol z$. In particular, if
$\varphi\in\HHH_{h,\ol h}$ is real and 
\textsc{quasi-primary} (e.g. $\varphi=T$), then we can write
\beq\el{qpadj}
\varphi^\dagger(z,\ol z) 
= \varphi(\ol z^{\,-1}, z^{-1}) \ol z^{\,-2h} z^{-2\ol h}.
\eeq
As an abbreviation, one defines \textsc{in- and out-states} by setting
\beqn\el{in}
\fa\chi,\psi\in\HHH:\quad
\bra\psi|&:=& \lim_{w\rightarrow0} \,\bra0|\psi^\dagger( \ol w^{\,-1},w^{-1}),
\nonumber\\
|\chi\ket&:=& \lim_{\zeta\rightarrow0}\,\chi(\zeta,\ol \zeta)|0\ket_{\PP^1}.
\eeqn
Now the OPE-coefficients $C$ can be recovered as
\beqn\el{C3}
\fa\varphi,\chi,\psi\in\HHH:\quad
C(\psi^*,\varphi,\chi)
&=& \bra\psi|\varphi(1,1)|\chi\ket\\
&=& \lim_{w,\zeta\rightarrow0} 
\bra0|\psi^\dagger( \ol w^{\,-1},w^{-1})\varphi(1,1)
\chi(\zeta,\ol \zeta)|0\ket_{\PP^1}.\nonumber
\eeqn
Similarly, with $\varphi_x,\chi_x\in\HHH_{h_x,\ol h_x}$, four-point functions
can be brought into the form 
$$
\bra \varphi_a|\varphi_b(1)\varphi_c(z,\ol{z})|\varphi_d\ket
:=
\lim_{w,\zeta\rightarrow0} 
\bra0|\varphi_a^\dagger( \ol w^{\,-1},w^{-1})\varphi_b(1,1)\varphi_c(z,\ol z)
\varphi_d(\zeta,\ol \zeta)|0\ket_{\PP^1}.
$$
They have the following expansion around $z=0$:
\beqn\label{fpexp}
&&\bra \varphi_a|\varphi_b(1)\varphi_c(z,\ol{z})|\varphi_d\ket\\
&&\hphantom{\varphi_b(1)\varphi_c(z,\ol{z})}
=\sum_j C(\varphi_a^*,\varphi_b,\psi_j)
C(\psi_j^*,\varphi_c,\varphi_d) 
z^{h_j-h_c-h_d}\ol z^{\ol h_j-\ol h_c-\ol h_d},\nonumber
\eeqn
where $\{\psi_j\}_j$ denotes a suitable orthonormal basis of 
$\HHH$.

Using the above characterization of $\psi^\dagger$, conformal invariance,
and \req{qpadj},
one finds 
\beq\el{cc}
\fa\varphi,\chi,\psi\in\HHH\colon\quad
\ol{C(\psi^*,\varphi,\chi)} = C(\chi^*,\varphi^\dagger,\psi)
\stackrel{
\stackrel{\mbox{\small{}if $\varphi$ is quasi-}}{\mbox{\small{}primary}}
}{=}
C(\chi^*,\ast\varphi,\psi).
\eeq
Note that the OPE-coefficients involving only real states
$\ast\varphi=\varphi,\,\ast\chi=\chi,\,\ast\psi=\psi$ are always real.
Moreover, using \req{qpadj} one shows
\beq\el{viradj}
\fa n\in\ZZ:\quad
L_n^\dagger = L_{-n};
\quad\quad
\fa\chi,\psi\in\HHH:\quad
(L_n\psi)^*\chi = \psi^*(L_{-n}\chi).
\eeq
Since up to possible phases,
$n$-point functions \eq{npt} are invariant under permutations of the
$\varphi_i(z_i,\ol z_i)$, the second and third arguments in 
$C(\cdot,\cdot,\cdot)$ can be interchanged, up to a phase and contributions
of descendants to the OPE. However, the characterization \req{prime}
of primaries together with \req{viradj} ensures
that every primary state is orthogonal to each descendant. Hence,
\beqn\el{trimetric}
&&
\fa \varphi,\chi,\psi\in\HHH\;\mbox{ with }\;
\varphi\in\HHH_{h_\varphi,\ol h_\varphi},
\chi\in\HHH_{h_\chi,\ol h_\chi},
\psi\in\HHH_{h_\psi,\ol h_\psi}\cap\HHH^{\vir}:\nonumber\\
&&\quad\quad\quad\quad
C(\psi^\ast,\varphi,\chi)
= (-1)^{h_\chi-\ol h_\chi+h_\varphi-\ol h_\varphi-h_\psi+\ol h_\psi}\;
C(\psi^\ast,\chi,\varphi).
\eeqn
To define \textsc{modes} associated to each $\varphi\in\HHH$, note that
for all $h,\,\ol h,\,\mu,\,\ol\mu$, the space $\HHH_{h+\mu,\ol h+\ol\mu}$
is finite dimensional by \eq{cftgrowth}, so we can set
\beqn\el{mode}
&&
\forall\,\varphi\in\HHH,\;
\forall\,\mu,\,\ol\mu, \, h,\,\ol h\in\RR, \;
\forall\,\chi\in\HHH_{h,\ol h}:\\
&&\quad
\varphi_{\mu,\ol\mu}\chi \in \HHH_{h+\mu,\ol h+\ol\mu}\;
\mbox{ s. th. } \;
\forall\,\psi \in \HHH_{h+\mu,\ol h+\ol\mu}:
\;
\psi^* (\varphi_{\mu,\ol\mu}\chi) = C(\psi^*, \varphi, \chi). \nonumber
\eeqn
If $\varphi\in\HHH_{h,\ol h}$, then 
$\varphi_{h,\ol h}$ obeys
$\varphi=\varphi_{h,\ol h}\Omega$. This gives
$$
[L_0,\varphi_{h,\ol h}] 
=
h\varphi_{h,\ol h}, \quad
[\ol L_0,\varphi_{h,\ol h}] 
=
\ol h\varphi_{h,\ol h}.
$$
In general, all three-point functions in a CFT can be obtained 
as linear combinations of three-point functions of the
primaries, acted on by differential operators. For example,
if $\varphi\in\HHH_{h_\varphi,\ol h_\varphi},\,
\chi\in\HHH_{h_\chi,\ol h_\chi},\,
\psi\in\HHH_{h_\psi,\ol h_\psi}$, then
\beq\el{l1action}
C(\psi^*, L_1\varphi, \chi) 
= (h_\psi-h_\varphi-h_\chi)\,C(\psi^*, \varphi, \chi),
\eeq
and analogously for $\ol L_1$.
On the other hand, analogously to \req{fpexp},
all $n$-point functions of a CFT can be recovered 
from its OPE-constants. This 
imposes many consistency conditions on the latter. An important example 
for this is
\textsc{crossing symmetry} \req{cross}
of four-point functions on the sphere.

Before discussing crossing symmetry, let us introduce  \textsc{W-algebras},
since we will use them to 
rewrite \eq{fpexp} in a slightly different way. Namely, for 
$\varphi\in\ker(\ol L_0)$ and
$\chi\in\HHH_{h,\ol h}$, $\varphi_{\mu,\ol\mu}\chi\neq0$ implies
$(\mu,\ol\mu)=(n,0)$ with $n\in\ZZ$, and similarly for elements
of $\ker(L_0)$ with $\mu,\ol\mu$ interchanged. 
The modes associated to states in  $\ker(\ol{L}_0)$, 
$\ker(L_0)$ generate a \textsc{holomorphic} or 
\textsc{antiholomorphic W-algebra} 
$\WWW^*\supset \vir_c,\,\ol\WWW^*\supset \ol\vir_c$ defined by
\beqn\el{walg}
\WWW^*&:=&\span_\CC\left\{ \varphi_{n,0} \bigm| n\in\ZZ, \; 
\varphi\in \ker(\ol L_0)\right\}\\
&=&\bigoplus_{n\in\ZZ}\WWW^*_n, \quad 
\WWW^*_n:=\left\{ w\in\WWW^*\bigm| [L_0,w]=n w\right\},\nonumber
\eeqn
and analogously for $\ol\WWW^*$ or any subalgebra $\WWW$ of 
$\WWW^*\oplus\ol\WWW^*$.
We suppose that $\HHH$ decomposes into a sum of tensor products 
of irreducible lowest weight representations 
$\VV_a^{\WWW^*}$, $\ol{\VV}_{\ol a}^{\ol\WWW^*}$ of 
the 
holomorphic and antiholomorphic W-algebras,
$$
\HHH=\bigoplus_{(a,\ol{a})\in\III}
\VV_a^{\WWW^*}\otimes\ol{\VV}_{\ol{a}}^{\ol\WWW^*}\,.
$$
Moreover, the OPE determines the commutative associative product
on the representation ring of $\WWW^*\otimes\ol\WWW^*$ which is known as
\textsc{fusion}:
$$
\left[\varphi_a\right]\bullet\left[\varphi_b\right] 
= \sum_c N_{ab}^c \left[\varphi_c\right]
$$
for conformal families $\left[\varphi_a\right]$ with 
$\varphi_a\in\VV_a$ etc.

We now consider an
orthonormal basis $\left\{ \psi_j \right\}_{j\in\NN}$ of primaries
of a given CFT with respect to a subalgebra $\WWW$ of the 
holomorphic and antiholomorphic W-algebra 
as in  \eq{prime}. 
Without loss of generality we can assume
that  $\ast\psi_j=\psi_j$ and 
$\psi_j\in\HHH_{h_j,\ol h_j}$ for all 
$j\in\NN$. Moreover, let $\{\psi_j^{\{k,\ol{k}\}}\}_{k\in K,\ol{k}\in \ol{K}}$
with $K,\ol K\subset\oplus_p \NN^p$ denote a basis
of the descendants of $\psi_j$, which is $(L_0,\ol{L}_0)$-homogeneous,
with bi-degree of $\psi_j^{\{k,\ol{k}\}}$ 
given by $(h_j+|k|,\ol{h}_j+|\ol{k}|)$,
$|k|,|\ol{k}|\in\NN$ for all $k\in K$, $\ol{k}\in\ol{K}$.
For $a,b,j\in\NN$ we set
$$
C_{ab}^j := C(\psi_j^*, \psi_a, \psi_b), \quad\quad
\psi_j^{\{(),()\}}:=  \psi_j.
$$
Then, there are constants 
$\beta^{j\{k^\prime\}}_{ab}, \ol{\beta}^{j\{\ol{k}^\prime\}}_{ab}\in\RR$, 
such that
\beqn\el{beta}
&&
\fa j\in\NN,\;\fa k\in K,\,\ol{k}\in\ol{K}:\\[3pt]
&&\quad
C\left( (\psi_j^{\{k,\ol{k}\}})^*, \psi_a,\psi_b\right)
=  \sum_{k^\prime,\ol{k}^\prime} C_{ab}^j\, \beta^{j\{k^\prime\}}_{ab}\,
\ol{\beta}^{j\{\ol{k}^\prime\}}_{ab}\,
C\left( (\psi_j^{\{k,\ol{k}\}})^*,\Omega, \psi_j^{\{k^\prime,\ol{k}^\prime\}}
\right).\quad\quad\nonumber
\eeqn
Here, $\beta^{j\{()\}}_{ab}=\ol{\beta}^{j\{()\}}_{ab}:=1$. Now, for all 
$a,b,c,d,j\in\NN$ the \textsc{conformal blocks}
are given by
\beqn\el{confbl}
\f{j}{a}{b}{c}{d}(z)
&:=& \sum_k {\beta^{j\{k\}}_{ab} \over \sqrt{C^d_{c j}}}\,
C\left(\psi_d^*, \psi_c, \psi_j^{\{k,()\}}\right)\, z^{h_{j}-h_a-h_b+|k|},\\
\olf{j}{a}{b}{c}{d}(\ol{z})
&:=& \sum_{\ol{k}} {\ol{\beta}^{j\{\ol{k}\}}_{ab} \over \sqrt{C^d_{c j}}}\,
C\left(\psi_d^*, \psi_c, \psi_j^{\{(),\ol{k}\}}\right)\, 
\ol{z}^{\ol{h}_{j}-\ol{h}_a-\ol{h}_b+|\ol{k}|}.\nonumber
\eeqn
Up to factors $z^{h_{j}-h_a-h_b}$ ($\ol{z}^{\ol{h}_{j}-\ol{h}_a-\ol{h}_b}$),
the conformal blocks are (anti-)me\-ro\-mor\-phic functions on $\CC$ with 
poles at $0,\, 1,\, \infty$. 
They encode the four-point functions of primaries by
$$
\langle \psi_d | \psi_c(1,1) \psi_a(z,\ol z) | \psi_b \rangle
= \sum_j C^j_{ab} C^d_{c j}\; \f{j}{a}{b}{c}{d}(z)\;\;\olf{j}{a}{b}{c}{d}(\ol{z}),
$$
and \textsc{crossing symmetry} reads: for all $a,b,c,d\in\NN$
\beqn\el{cross}
&&\sum_j C^j_{ab} C^d_{c j}\; 
\f{j}{a}{b}{c}{d}(z)\;\;\olf{j}{a}{b}{c}{d}(\ol{z})
\\
&&\qquad\qquad\quad
=\sum_j C^j_{ad} C^b_{c j}\; 
\f{j}{a}{d}{c}{b}(z^{-1})\;\;
\olf{j}{a}{d}{c}{b}(\ol{z}^{-1}) z^{-2h_a}\ol z^{-2\ol h_a}.
\qquad\qquad\nonumber
\eeqn
\section{$c=1$ Representation theory}\label{c1}
In this Appendix, let $\CCC=(\HHH,\,\ast,\,\Omega,\,T,\,\ol T,\,C)$
denote a unitary conformal field theory with $c=1$. We
recall some basic facts about its representation content;
see also \cite{ga02}.

Since all known unitary conformal field theories at $c=1$ can be constructed
with   energy momentum tensor
$T={1\over2}\!\!\!:\!\!\!j j\!\!\!:$ and $j$ a $\fu(1)$ current (which not
necessarily is a field of the theory), it is convenient to use  
the Heisenberg algebra
\beq\el{heisenberg}
j(z)=\sum_{n=-\infty}^\infty a_n z^{n-1},\quad \mbox{ where }\quad
[a_n,a_m]=m\delta_{n+m,0}.
\eeq
Then all states in the pre-Hilbert space of every known theory $\CCC$
with central charge $c=1$
are obtained from the Fock space that we construct 
from appropriate polynomials in the $a_n, n>0$, acting on an 
appropriate subset of all lwvs of the Virasoro algebra. 
To build the latter it 
suffices to take states
\begin{eqnarray}\el{u1norm}
\hwv[h]{Q}, \quad\mbox{ such that }\quad
L_0 \hwv[h]{Q} &=&  h\hwv[h]{Q}, \mbox{ with } h={\textstyle{Q^2\over2}},
\nonumber\\
a_0 \hwv[h]{Q} &=& Q \hwv[h]{Q},\\
\ast\left(\hwv[h]{Q}\right) &=&  \hwv[h]{-Q},\nonumber
\end{eqnarray}
as well as
so-called twisted ground states with $h=\overline h\leq1/16$,
which we will not make use of in the following, however.
We will always normalize the $\hwv[h]{Q}$ such that 
\beq\el{u1primenorm}
C\left( \hwv[h]{Q}^\ast,\Omega,\hwv[h]{Q}\right )
\reeq{adj} \bra h,-Q \hwv[h]{Q}
=1.
\eeq
In a consistent theory, all left and right charges 
$(Q;\overline Q)$ are contained in a charge lattice. Namely,  for every 
theory $\CCC$
there is a fixed $R\in\R^+$ such that all $(Q;\overline Q)$ that may
occur are given by
\beq\el{circlespec}
(Q;\overline Q)=\inv{\sqrt2} \left( m R+ \inv[n]{R}\,;\,m R-\inv[n]{R}\right), 
\quad m,n\in\Z.
\eeq
In a so-called \textsc{circle theory at radius $R$}, 
the pre-Hilbert space is just the
entire Fock space built on the set of vacua 
$|Q;\ol Q\ket:=
\hwv[{{Q^2\over2}}]{Q}\otimes\hwv[{{\overline Q^2\over2}}]{\overline Q}$ 
with all allowed values of $(Q;\overline Q)$. 
The $\fs\fu(2)_1$ WZW-model  is the circle theory at radius
$R=1$. All $|Q;\ol Q\ket$ are simple currents, and
the leading terms in the OPE are given by
\beq\el{circleope}
C\left( |Q+Q^\prime;\ol Q+\ol Q^\prime\ket^\ast, 
|Q;\ol Q\ket, |Q^\prime;\ol Q^\prime\ket \right)
= (-1)^{(Q+\ol Q)(Q^\prime-\ol Q^\prime)/2},
\eeq
with all other OPE-constants vanishing.
Equivalently, 
\beqn\el{circletrunc}
|Q;\ol Q\ket\boxast |Q^\prime;\ol Q^\prime\ket
&=& \eps\left( (Q;\ol Q), (Q^\prime;\ol Q^\prime) \right)
|Q+Q^\prime;\ol Q+\ol Q^\prime\ket\nonumber\\
&=& (-1)^{(Q+\ol Q)(Q^\prime-\ol Q^\prime)/2}\;
|Q+Q^\prime;\ol Q+\ol Q^\prime\ket
\eeqn
with  notations as in \eq{trunc}. The \textsc{cocycle factor}
$\eps$ introduces the appropriate phases.

For central charge $c=1$, the   character of a Virasoro irreducible 
representation with lowest
weight vector of dimension $h$ generically  is
\beq\el{c1gen}
\chi_h^{gen}(q) = \inv{\eta(q)} q^h .
\eeq
But for $n=2\sqrt h\in\N$, the representation contains a null vector at level 
$n+1$,  namely \cite[(8.34)]{dms96}
\beq\el{sing}
S_n\,{\ts\left|{n^2\over 4}\right\rangle}:=
\sum_{\stackrel{\scriptstyle p_i\geq1,}{p_1+\cdots+p_k=n+1}}\ts
{(-1)^{n+1+k} \over 
\prod\limits_{l=1}^{k-1}(p_1+\cdots+p_l)(n+1-p_1-\cdots-p_l) }
L_{p_1}\cdots L_{p_k} \left|{n^2\over 4}\right\rangle\,,
\eeq
where $|{n^2\over 4}\ket$ denotes the lowest weight vector of 
conformal weight $h={n^2\over 4}$, $n\in\NN$. 
Hence the character reduces to
\beq\el{char}
\chi_{{1\over4}n^2}=\inv{\eta(q)}(q^{n^2/4}-q^{(n+2)^2/4}) .
\eeq
In the following, we restrict attention to the holomorphic side only.
The  generic W-algebra $\WWW$ of   circle theories is generated
by the $\fu(1)$ current $j$. The  $\hwv{Q}$ are just the
lowest weight vectors of irreducible
representations $\VV_Q^{\fu(1)}$ of $\WWW$ with characters 
$$
X_{\sqrt2 Q}= \inv{\eta(q)}q^{{Q^2\over2}},
$$
regardless of the value of $Q$. In particular, if $\sqrt2 Q=n\in\Z$,
by \req{char}
$$
X_n = \sum_{k=0}^\infty \chi_{{1\over 4}(|n|+2k)^2},
$$
and the Fock space built on  $\hwv{Q}$ contains infinitely many
Virasoro irreducible representations with lowest weight vectors
$\hwv[h]{Q}, h={Q^2\over2}+N, N=k(\sqrt2|Q|+k),\,k\in\N$. 
Let
$$
\hw{m}:=\ts\left|{n^2\over4},{m\over\sqrt2}\right\rangle,
$$
and let $\ver{m}$ denote the
space of states in the irreducible representation of the Virasoro algebra 
with lwv $\hw{m}$ of norm $1$.
Note that   e.g.\  for the circle theory at $R=1$ (the $\fs\fu(2)_1$ WZW model)
each positive eigenvalue
of $L_0$ is highly degenerate since this theory has an enhanced $\fs\fu(2)_1$ 
Kac-Moody algebra the zero modes of whose generators commute with $L_0$.
More precisely,  
$$
\mathcal V_{h={n^2\over4}}=
\bigoplus_{m=-n, m\equiv n(2)}^n \ver{m} .
$$
All the representations $\ver{m}$ with $|m|\leq n$, $m\equiv n(2)$ have
the same character $\chi_{{1\over 4}n^2}$ as in \req{char}.
Let $J_\pm(z)$ denote the holomorphic fields creating 
$|Q;\ol Q\ket = |\pm\!\sqrt2;0\ket$ as in \req{in}.
Then we define
\beq\el{zero}
Q_\pm := \int dz J_\pm(z), \quad
\mbox{ i.e. } [Q_+,Q_-]=\sqrt2 a_0 =: 2J_0,\quad
[J_0,Q_\pm]=\pm Q_\pm,
\eeq
the zero modes of $J_\pm,J$ 
in the enhanced $\fs\fu(2)_1$ Kac-Moody algebra of
the circle model at radius $R=1$. Since $[L_n,Q_\pm]=0$ for all 
$n\in\Z$, from \req{zero} together with \eq{u1norm}
it follows that
$\hwv[h]{Q}=$ $\kappa Q_\mp\hwv[h]{Q\pm\sqrt2}$ for some $\kappa\in\CC^*$ if 
$\hwv[h]{Q\pm\sqrt2}$ exists. More precisely, \req{zero} inductively shows
$$
Q_+^l Q_-^l \hw{n} = \inv[l!n!]{(n-l)!} \hw{n},
$$
if the left hand side does not vanish.
From our normalization \eq{u1primenorm} it now follows that
$$
\begin{array}{rcl}
\mbox{for } m,l\in\N,\quad\quad\quad\\[3pt]
\hw[n=m+2l]{m}
&=& \sqrt{\inv[(n-l)!]{l!n!}}\, Q_-^l \hw{n},
\mbox{ and } Q_{\pm}\hw{\pm n}=0.
\end{array}
$$
In particular, $Q_-^l \hw{n}=0$ for $l>n$.
\section{The unitary Virasoro minimal models, their structure constants,
and their $c\rightarrow1$ limit}\el{structconst}
The unitary diagonal Virasoro minimal model   
$\MM_m:=\MMM(m,m+1)$ with $m\in\NN-\{0,1\}$ has central charge
$c_m$ given by \req{charge}. Its irreducible representation $(r,s)$ of the
Virasoro algebra has an lwv $|r,s\ket_m$ with weight
\req{miniweight}, and character
\beq\el{minichar}
\begin{array}{rcl}\ds
\chi_{(r,s)}^{m}(q)
&=&\ds {q^{-{c_m\over24}}\over \prod\limits_n(1-q^n)}
\left[ 
q^{h_{(r,s)}^{m}} - \sum_{k=1}^\infty\left\{
q^{h_{(r+(2k-1)m,-s+m+1)}^{m}} + q^{h_{(r,2k(m+1)-s)}^{m}}\right.\right.
\\[3pt]
&&\ds \left.\left.\hphantom{{q^{-c_m/24}\over \prod_n(1-q^n)\sum\sum}
q^{h_{(r,s)}^{m}} }
-q^{h_{(r+2km,s)}^{m}} - q^{h_{(r,2k(m+1)+s)}^{m}}
\right\}\right].
\end{array}
\eeq
Fusion reads
\beq\el{minifusion}
\VV_{(n\p,n)}^{m}\bullet \VV_{(s\p,s)}^{m}
= \bigoplus_{\stack[p\p=|n\p-s\p|+1,]{p\p+n\p+s\p\equiv 1(2)}}^{\min\{
n\p+s\p-1,2m+1-n\p-s\p\}}\,\,
\bigoplus_{\stack[p=|n-s|+1,]{p+n+s\equiv 1(2)}}^{\min\{
n+s-1,2m-1-n-s\}}\VV_{(p\p,p)}^{m}.
\eeq
The structure constants  as in \req{struc} are 
given by \cite{dofa85}
\beq\el{strconst}
\begin{array}{rcl}
&&C_{(n\p, n)(s\p, s)}^{(p\p, p)}
=\mu_{l\p, l}\sqrt{{a_{n\p, n}a_{s\p, s}\over a_{p\p, p}}}\\[12pt]
&&\quad
\times\prod\limits_{i=0}^{l\p -2}{\Gamma(s-s\p+1+i-y\p(s\p-1-i))
\Gamma(n-n\p+1+i-y\p(n\p-1-i))\Gamma(p\p-p+1+i+y\p(p\p+1+i))\over
\Gamma(s\p-s-i+y\p(s\p-1-i))\Gamma(n\p-n-i+y\p(n\p-1-i))
\Gamma(p-p\p-i-y\p(p\p+1+i))}\\[12pt]
&&\quad\times\prod\limits_{j=0}^{l -2}{\Gamma(s\p-s+2+j-l\p+y(s-1-j))
\Gamma(n\p-n+2+j-l\p+y(n-1-j))\Gamma(p-p\p+2+j-l\p-y(p+1+j))\over
\Gamma(s-s\p-1-j+l\p-y(s-1-j))\Gamma(n-n\p-1-j+l\p-y(n-1-j))\Gamma(p\p-p-1-j+l\p+y(p+1+j))}
\\[12pt]
&&\quad\quad\quad\quad=:\mu_{l\p, l}\sqrt{{a_{n\p, n}a_{s\p, s}
\over a_{p\p, p}}}\tilde{C}_{(n\p, n)(s\p, s)}^{(p\p, p)}\,,
\end{array}
\eeq
with
$$
\begin{array}{rcl}
y&:=&{1\over m+1}\,,\quad y\p:={1\over m}\,,\quad
l:={1\over 2}(s+n-p+1)\,,\quad l\p:={1\over 2}(s\p+n\p-p\p+1)\,,\\[12pt]
\mu_{l\p, l}&:=&
(1-y)^{4(l\p-1)(l-1)}\prod\limits_{i=1}^{l\p-1}\prod\limits_{j=1}^{l-1}{1
\over (i-j+y j)^2}\prod\limits_{i=1}^{l\p-1}{\Gamma(i+i y\p)\over
\Gamma(1-i-i y\p)}\prod\limits_{j=1}^{l-1}{\Gamma(j-j y)\over\Gamma(1-j+j y)}
\\
&=&(-1)^{(l-1)(l\p-1)}
(1-y)^{4(l\p-1)(l-1)}\prod\limits_{i=1}^{l\p-1}{\Gamma(i+i y\p)\over
\Gamma(1-i-i y\p)}\prod\limits_{j=1}^{l-1}{\Gamma(j-l\p+1-j y)\over\Gamma(l\p-j+j y)}\,,\\
[12pt]
a_{n\p, n}&:=&\prod\limits_{i=1}^{n\p-1}\prod\limits_{j=1}^{n-1}\!
\left({i-j+y(1+j)\over i-j+y j}\right)^{\!\!2}\,
\prod\limits_{i=1}^{n\p-1}\!
{\Gamma(i+i y\p)\Gamma(1-i-y\p(1+i))\over\Gamma(1-i-i y\p)\Gamma(i+y\p(1+i))}
\prod\limits_{j=1}^{n-1}\!
{\Gamma(j-j y)\Gamma(1-j+y(1+j))\over\Gamma(1-j+y j)
\Gamma(j-y(1+j))}\\
&=&
\prod\limits_{i=1}^{n\p-1}
{\Gamma(i+i y\p)\Gamma(1-i-y\p(1+i))\over\Gamma(1-i-i y\p)\Gamma(i+y\p(1+i))}
\prod\limits_{j=1}^{n-1}{\Gamma(j-n\p+1-j y)\Gamma(n\p-j+y(1+j))\over\Gamma(n\p-j+y j)
\Gamma(j-n\p+1-y(1+j))}\,.
\end{array}
$$
Note that $\mu_{l\p,l}$, $a_{n\p, n}$ and 
$\tilde{C}_{(n\p, n)(s\p, s)}^{(p\p, p)}$
are products of expressions
$$
\begin{array}{rclrl}
G(N,M,\epsilon)&:=&
{\Gamma(1+N-M\epsilon)\over\Gamma(-N+M\epsilon)}
&=&(-1)^N\,\Gamma^2(1+N-M\epsilon)\,{\sin(\pi M\epsilon)\over\pi}
\\
&&&=&\left((-1)^N\,\Gamma^2(-N+M\epsilon)\,{\sin(\pi M\epsilon)\over\pi}
\right)^{-1}\,,
\end{array}
$$
where $N,M\in\ZZ$ and $\epsilon\in\{y,y\p\}$.
We also have the following expansions 
for $m\rightarrow\infty$, to lowest order in $y=y\p+O(y^2)$:
$$\ts
\begin{array}{rcl}
G(N,M,\epsilon)
&\stackrel{y\rightarrow0}{\sim}& 
\left((-1)^N M y\,
\Gamma^2\left(\Big|{1+\si(N)\over 2}+N\Big|\right)\right)^{\si(N)}
=:y^{\si(N)}e(N,M)\,,\\[5pt]
{\Gamma(N+M\epsilon)\over\Gamma(N+M\p\epsilon\p)}
&\stackrel{y\rightarrow0}{\sim}&
\left\{\begin{array}{ll}
{M\p\over M}&\mbox{ if }N\leq 0\\ 1&\mbox{ if }N>0\end{array}\right.\,,
\end{array}
$$
where $\si(N)=1$ for $N\geq 0$, $\si(N)=-1$ for $N<0$. Hence we obtain
the lowest order expansions 
\beqn\el{mutilde}
\mu_{l\p, l}&\stackrel{y\rightarrow0}{\sim}&\ts
y^{|l-l\p |}{(l-1)!(l\p-1)!\over ((\min\{l,l\p\}-1)!)^2}
\prod\limits_{i=1}^{l\p-1}\prod\limits_{\stack[j=1,]{j\neq i}}^{l-1}\!
{1\over(i-j)^2}
\prod\limits_{i=1}^{l\p-1}\!(-1)^i\Gamma(i)^2
\prod\limits_{j=1}^{l-1}\!(-1)^{j+l\p}\Gamma(j)^2 \nonumber\\
&=:&y^{|l-l\p|}\tilde{\mu}_{l\p, l},\\
a_{n\p, n}&\stackrel{y\rightarrow0}{\sim}&\ts{\min\{n\p,n\}\over 
\max\{n\p,n\}},\nonumber\\ 
\tilde{C}_{(n\p, n)(s\p, s)}^{(p\p, p)} 
&\stackrel{y\rightarrow0}{\sim}&  
\tilde{A}_{(n\p, n)(s\p, s)}^{(p\p, p)}
y^{\tilde{E}_{(n\p, n)(s\p, s)}^{(p\p, p)}},\nonumber
\eeqn
where
\beq\el{etilde}
\tilde{E}_{(n\p, n)(s\p, s)}^{(p\p, p)} 
:= k(s\p -s,l\p-2,l-2)+k(n\p-n,l\p-2,l-2)+k(p-p\p,l\p-2,l-2), 
\eeq
\beqns
k(x,a,b)&:=&d(x,a)+d(-x,b)-2g(x,a,b),\\ 
d(x,a)&:=&\max\{\min\{a+1,a+1-2x\},-(a+1)\} \\
&=&\ts{1\over 2}\left(-x-|x|+|2a+2-x-|x||\right), \\
g(x,a,b)&:=&\ts
(\min\{a-{x\over 2},b+{x\over 2}\}-{|x|\over 2}+1)\Theta(a-x)\Theta(b+x)
 \\
&=&\ts{1\over 2}(-|x|+a+b+2-|a-b-x|)\Theta(a-x)\Theta(b+x) ,
\eeqns
so
\beq\el{kdefb}
k(x,a,b) = |a-b-x|-|x| \quad \mbox{for } a,b\geq-1.
\eeq
Moreover,
\beqns
\tilde{A}_{(n\p, n)(s\p, s)}^{(p\p, p)}&=&\ts 
\prod\limits_{i=0}^{l\p-2}\left\{e(s-s\p+i,s\p-1-i)\,e(n-n\p+i,n\p-1-i)\right.\qquad
\qquad\qquad\\
\ts &&\left. e(p\p-p+i,-p\p-1-i)\right\}
\prod\limits_{j=0}^{l-2}\left\{ e(s\p-s+1+j-l\p,-s+1+j)\right.\\
\ts &&
\left. e(n\p-n+1+j-l\p,-n+1+j)\,e(p-p\p+1+j-l\p,p+1+j)\right\}
\nonumber
\eeqns
Thus in the limit $m\rightarrow\infty$ we have
\beq\el{climit}
C_{(n\p, n)(s\p, s)}^{(p\p, p)}
\stackrel{y\rightarrow0}{\sim} 
A_{(n\p, n)(s\p, s)}^{(p\p, p)}y^{E_{(n\p, n)(s\p, s)}^{(p\p, p)}}
\eeq
with
\beqn\el{e}
A_{(n\p, n)(s\p, s)}^{(p\p, p)}
&=&
\ts\left({\min\{n\p,n\}\min\{s\p,s\}\max\{p\p,p\}\over
\max\{n\p,n\}\max\{s\p,s\}\min\{p\p,p\}}\right)^{1/2}
\tilde{\mu}_{l\p, l}\tilde{A}_{(n\p, n)(s\p, s)}^{(p\p, p)}\,,\nonumber\\
E_{(n\p, n)(s\p, s)}^{(p\p, p)}
&=&|l-l\p|+\tilde{E}_{(n\p, n)(s\p, s)}^{(p\p, p)}\,.
\eeqn
Note that $A_{(n\p, n)(s\p, s)}^{(p\p, p)}$ never vanishes in the
allowed regime $p\p+n\p+s\p\equiv p+n+s\equiv1(2)$,
$|n^{(\prime)}-s^{(\prime)}|<p^{(\prime)}<n^{(\prime)}+s^{(\prime)}$.
These  constants obey
\blem{Egeq0}
Given $(p\p,p),(n\p,n),(s\p,s)$ such that 
$A_{(n\p, n)(s\p, s)}^{(p\p, p)}\neq 0$, we have
$E_{(n\p, n)(s\p, s)}^{(p\p, p)}\geq 0$. More precisely,
with $\nu:=n\p-n,\sigma:=s\p-s,\pi:=p\p-p$,
$$
E_{(n\p, n)(s\p, s)}^{(p\p, p)}=0 \quad\Longleftrightarrow\quad
|\pi|\in\left[ \vphantom{\sum}
\min\{ |\sigma+\nu|,|\sigma-\nu| \},
\max\{ |\sigma+\nu|,|\sigma-\nu| \} \right] .
$$
\vspace*{-1em}
\elem
\bpr
Since $l\p-l={1\over2}(\sigma+\nu-\pi)$, from \req{etilde}, \req{kdefb}, 
\req{e} we find 
\beqns
E_{(n\p, n)(s\p, s)}^{(p\p, p)}
&=&
\inv{2}|\sigma+\nu-\pi| + \inv{2}|-\sigma+\nu-\pi|
+ \inv{2}|\sigma-\nu-\pi| + \inv{2}|\sigma+\nu+\pi|\nonumber\\
&&-|\sigma|-|\nu|-|\pi| \nonumber\\
&=& \max\{|\sigma+\nu|,|\pi|\} + \max\{|\sigma-\nu|,|\pi|\}
-|\sigma|-|\nu|-|\pi|.
\eeqns
Therefore,
$$
E_{(n\p, n)(s\p, s)}^{(p\p, p)}
=\left\{
\begin{array}{ccl}
2\max\{|\sigma|,|\nu|\}-|\sigma|-|\nu|-|\pi|
 &>&0, \\
\hphantom{|\pi|  -|\nu| -|\nu| -|\nu|-|\nu|}
\mbox{ if }|\pi|&<&\min\{ |\sigma+\nu|,|\sigma-\nu| \},\\
\max\{ |\sigma+\nu|,|\sigma-\nu| \}-|\sigma|-|\nu|&=&0, \\
\hphantom{|\pi|   -|\nu|-|\nu| -|\nu|-|\nu|}
\mbox{ if }|\pi|&\in& \left[ \vphantom{\sum}
\min\{ |\sigma+\nu|,|\sigma-\nu| \},\right.\\
&&\left.\vphantom{\sum}\quad\max\{ |\sigma+\nu|,|\sigma-\nu| \} \right],\\
|\pi|-|\sigma|-|\nu|
&>&0, \\
\hphantom{|\pi|  -|\nu| -|\nu| -|\nu|-|\nu|}
\mbox{ if }|\pi|&>&\max\{ |\sigma+\nu|,|\sigma-\nu| \},
\end{array}\right.
$$
which proves the lemma.
\epr
In Rem.\ \ref{ninfty} we explain how additional null vectors in the
limit $\MMM_\infty$ of unitary Virasoro minimal models can be scaled
up without introducing divergences in three-point functions. In fact,
Lemma \ref{Egeq0} can be used  in order to extend the example 
of scaling up null vectors given in \cite[Sect.~3.1.1]{grw01} by
a direct calculation:
\blem{rescale}
All vectors $L_1|r,r\ket_\infty,\, r>1,$ can be scaled up to non-vanishing norm
without introducing divergences in the OPE-constants
$C(|p\p,p\ket_\infty^*, L_1|r,r\ket_\infty, |s\p,s\ket_\infty )$.
\elem
\bpr
By \req{scale}, a normalization of $L_1|r,r\ket_\infty$ 
to non-vanishing but finite
norm is given by
$$
D_{r,r,1}^m:=\lim_{m\rightarrow\infty} (m+1) \;L_1|r,r\ket_m,
$$
i.e.\ we set 
$$
\wt f_m^j(D_{r,r,1}^m):=D_{r,r,1}^j.
$$
Note that for finite $m$, \req{minifusion} shows that
$C^{(p\p,p)}_{(r,r)(s\p,s)}$ is only
non-vanishing if 
$\inv{2}(r+s^{(\prime)}-1-p^{(\prime)})\in\{0,\dots,
\min\{r,s^{(\prime)}\}-1\}$, hence we restrict
to such $p,\,p\p$. 
By \req{l1action} we find
\begin{eqnarray*}
C^{(p\p,p)}_{D_{r,r,1}(s\p,s)}
&\stackrel{m\rightarrow\infty}{\sim}&
(m+1)\left( h^{m}_{(p\p,p)}-h^{m}_{(r,r)}-h^{m}_{(s\p,s)}\right)
C^{(p\p,p)}_{(r,r)(s\p,s)}\\ 
&\stackrel{\req{climit}}{\sim}&
(m+1)^{1-E^{(p\p,p)}_{(r,r)(s\p,s)}}\;
\left( h^{m}_{(p\p,p)}-h^{m}_{(r,r)}-h^{m}_{(s\p,s)}\right)
A^{(p\p,p)}_{(r,r)(s\p,s)}.
\end{eqnarray*}
Therefore, if $E^{(p\p,p)}_{(r,r)(s\p,s)}\geq1$, 
the assertion follows directly from the convergence of each term in the 
latter expression.

On the other hand,   for $p\p,p$ in the range given above, by Lemma \ref{Egeq0}
we have 
$E^{(p\p,p)}_{(r,r)(s\p,s)}=0$ iff $|p\p-p|=|s\p-s|$. Hence in this case
\begin{eqnarray*}
C^{(p\p,p)}_{D_{r,r,1}(s\p,s)}
&\stackrel{m\rightarrow\infty}{\sim}&
(m+1)\left( h^{m}_{(p\p,p)}-h^{m}_{(r,r)}-h^{m}_{(s\p,s)}\right)
C^{(p\p,p)}_{(r,r)(s\p,s)}\\ 
&\stackrel{\mbox{\scriptsize\req{miniweight}}}{=}&
\left\{\inv[m+1]{4m} \left( (p\p)^2-p^2-(s\p)^2+s^2 \right)
+ {\cal O}\left(\inv{m}\right)\right\}A^{(p\p,p)}_{(r,r)(s\p,s)}
\end{eqnarray*}
remains finite, too.
\vspace*{-1.5em}
\epr 
%
%
%
\def\polhk#1{\setbox0=\hbox{#1}{\ooalign{\hidewidth
  \lower1.5ex\hbox{`}\hidewidth\crcr\unhbox0}}}
\providecommand{\bysame}{\leavevmode\hbox to3em{\hrulefill}\thinspace}

\end{document}